\documentclass[final,3p,times,twocolumn,authoryear]{elsarticle}

\usepackage[breaklinks=true]{hyperref}
\usepackage{xspace}
\usepackage{amssymb}
\usepackage{amsmath}
\usepackage{amsfonts}
\usepackage{aas_macros}
\usepackage{textcmds}
\usepackage{longtable}

\journal{New Astronomy}

\newcommand{\ergs}{\ensuremath{\mathrm{erg\,s}^{-1}}\xspace}

\newcommand{\ergscm}{\ensuremath{\mathrm{erg\,s}^{-1}\mathrm{cm}^{-2}}\xspace}

\newcommand{\Msol}{\ensuremath{\mathrm{M}_{\odot}}\xspace}
\newcommand{\Rsol}{\ensuremath{\mathrm{R}_{\odot}}\xspace}

\newcommand{\Ps}{\ensuremath{P_\mathrm{s}}\xspace}
\newcommand{\Pb}{\ensuremath{P_\mathrm{b}}\xspace}

\newcommand{\Athena}{\textsl{Athena}\xspace}
\newcommand{\intg}{\textsl{INTEGRAL}\xspace}
\newcommand{\xmm}{\textsl{XMM-Newton}\xspace}
\newcommand{\swift}{\textsl{Swift}\xspace}
\newcommand{\chandra}{\textsl{Chandra}\xspace}
\newcommand{\nustar}{\textsl{NuSTAR}\xspace}
\newcommand{\suzaku}{\textsl{Suzaku}\xspace}
\newcommand{\xte}{\textsl{RXTE}\xspace}

\newcommand{\vela}{Vela~X-1\xspace}
\newcommand{\lsi}{\mbox{LS\,I\,+61$^{\circ}$\,303}\xspace}

\newcommand{\bexrb}{BeXRB\xspace}
\newcommand{\bexrbs}{BeXRBs\xspace}
\newcommand{\sghmxb}{sgHMXB\xspace}
\newcommand{\sghmxbs}{sgHMXBs\xspace}

\newcounter{tabref}
\newcommand{\tbrefcite}[2]{\refstepcounter{tabref}[\thetabref]\label{#1}~\protect{\citet{#2}}}

\makeatletter
\newcommand{\tbsmartref}[1]{[\ref{#1}\checknextarg}
\newcommand{\checknextarg}{\@ifnextchar\bgroup{\gobblenextarg}{]}}
\newcommand{\gobblenextarg}[1]{,\ref{#1}\@ifnextchar\bgroup{\gobblenextarg}{]}}

\long\def\@@address[#1]#2{\g@addto@macro\elsaddress{%
    \def\baselinestretch{1}%
     \refstepcounter{affn}
     \xdef\@currentlabel{\theaffn}
     \elsLabel{#1}%
     }}

\newcommand{\printaffil}[2]{%
\noindent\textsuperscript{\elsRef{#1}}\,\textit{\footnotesize #2}\par}
\makeatother

\begin{document}

\begin{frontmatter}



\title{Advances in Understanding High-Mass X-ray Binaries with \intg and Future Directions}

\author[esac]{Peter Kretschmar}
\author[esac-quasar]{Felix F{\"u}rst}
\author[inaf-iasf]{Lara Sidoli}
\author[isdc]{Enrico Bozzo}
\author[cab]{Julia Alfonso-Garz{\'o}n}
\author[gcsu]{Arash Bodaghee}
\author[aim,apc]{Sylvain Chaty}
\author[dcu,dias]{Masha Chernyakova}
\author[isdc]{Carlo Ferrigno}
\author[sharjah-daa,saasst]{Antonios Manousakis}
\author[dfapl-ua]{Ignacio Negueruela}
\author[sternberg,kazan]{Konstantin Postnov}
\author[inaf-iasf]{Adamantia Paizis}
\author[iafrt,uoc]{Pablo Reig}
\author[dfests-ua,iufacit-ua]{Jos\'e Joaqu{\'{\i}}n Rodes-Roca}
\author[turku,iki]{Sergey Tsygankov}
\author[soton]{Antony J. Bird}
\author[ecap]{Matthias Bissinger n{\'e} K{\"u}hnel}
\author[viu]{Pere Blay}
\author[esac-aurora]{Isabel Caballero}
\author[soton]{Malcolm J. Coe} 
\author[cab]{Albert Domingo}
\author[iaat,iki]{Victor Doroshenko}
\author[isdc,iaat]{Lorenzo Ducci}
\author[issi]{Maurizio Falanga}
\author[iki]{Sergei A. Grebenev}
\author[iaat]{Victoria Grinberg}
\author[mit-kavli]{Paul Hemphill}
\author[remeis,ecap]{Ingo Kreykenbohm}
\author[franz-ludwig,remeis]{Sonja Kreykenbohm n{\'ee} Fritz}
\author[desy]{Jian Li}
\author[iki]{Alexander A. Lutovinov}
\author[ifca]{Silvia Mart\'inez-N\'u\~nez}
\author[cab]{J. Miguel Mas-Hesse}
\author[inaf-oas,uabello]{Nicola Masetti}
\author[saao,iuidia,iaudev]{Vanessa A. McBride}
\author[apc,isdc]{Andrii Neronov}
\author[cresst,gsfc]{Katja Pottschmidt}
\author[aim]{J{\'e}r{\^o}me Rodriguez}
\author[inaf-oab]{Patrizia Romano}
\author[cass]{Richard E. Rothschild}
\author[iaat]{Andrea Santangelo}
\author[inaf-oas]{Vito Sguera}
\author[iaat]{R{\"u}diger Staubert}
\author[ssl]{John A. Tomsick}
\author[dfests-ua,iufacit-ua]{Jos{\'e} Miguel Torrej{\'on}}
\author[ieec,icrea]{Diego F. Torres}
\author[isdc]{Roland Walter}
\author[remeis,ecap]{J{\"o}rn Wilms}
\author[msfc]{Colleen A. Wilson-Hodge}
\author[ihep]{Shu Zhang}


\address[esac]{European Space Astronomy Centre (ESA/ESAC), Operations Department, E-28692 Villanueva de la Ca\~nada (Madrid), Spain}
\address[esac-quasar]{Quasar Science Resources S.L for ESA, European Space Astronomy Centre (ESA/ESAC), Operations Department, E-28692 Villanueva de la Ca\~nada (Madrid), Spain}
\address[inaf-iasf]{INAF -- IASF, Istituto di Astrofisica Spaziale e Fisica Cosmica, Via A. Corti 12, I-20133 Milano, Italy}
\address[isdc]{Department of Astronomy, ISDC, University of Geneva, Chemin d'Ecogia 16, CH-1290 Versoix, Switzerland}
\address[cab]{Centro de Astrobiolog{\'\i}a -- Departamento de Astrof{\'\i}sica (CSIC-INTA), E-28692 Villanueva de la Ca\~nada (Madrid), Spain}
\address[gcsu]{Department of Chemistry, Physics and Astronomy, Georgia College and State University, Milledgeville, GA 31061, USA}
\address[aim]{AIM, CEA, CNRS, Universit\'e Paris-Saclay, Universit\'e de Paris, F-91191 Gif-sur-Yvette, France}
\address[apc]{Universit\'e de Paris, CNRS, Astroparticule et Cosmologie, F-75006 Paris, France}
\address[dcu]{School of Physical Sciences and CfAR, Dublin City University, Dublin 9, Ireland}
\address[dias]{Dublin Institute for Advanced Studies, 31 Fitzwilliam Place, Dublin 2, Ireland}
\address[sharjah-daa]{Department of Applied Physics \& Astronomy, University of Sharjah, Sharjah, UAE}
\address[saasst]{Sharjah Academy of Astronomy, Space Sciences, and Technology (SAASST), Sharjah, UAE}
\address[dfapl-ua]{Department of Applied Physics, University of Alicante, E-03080 Alicante, Spain}
\address[sternberg]{Sternberg Astronomical Institute, Moscow State University, 119234, Moscow, Russia}
\address[kazan]{Kazan Federal University, Kazan, Russia}
\address[iafrt]{Institute of Astrophysics, Foundation for Research and Technology-Hellas, 71110 Heraklion, Crete, Greece}
\address[uoc]{University of Crete, Physics Department \& Institute of Theoretical \& Computational Physics, 70013 Heraklion, Crete, Greece}
\address[dfests-ua]{Department of Physics, Systems Engineering and Signal Theory, University of Alicante, E-03690 Alicante, Spain}
\address[iufacit-ua]{University Institute of Physics Applied to Sciences and Technologies, University of Alicante, E-03690 Alicante, Spain}
\address[turku]{Department of Physics and Astronomy, FI-20014 University of Turku, Finland}
\address[iki]{Space Research Institute of the Russian Academy of Sciences, Profsoyuznaya Str. 84/32, Moscow 117997, Russia}
\address[soton]{School of Physics and Astronomy, Faculty of Physical Sciences and Engineering, University of Southampton, Southampton SO17 1BJ, UK}
\address[ecap]{Erlangen Centre for Astroparticle Physics (ECAP), Erwin-Rommel-Strasse 1, 91058 Erlangen, Germany}
\address[viu]{Universidad Internacional de Valencia - VIU, C/Pintor Sorolla 21, 46002, Valencia, Spain}
\address[esac-aurora]{Aurora Technology B.V. for ESA, European Space Astronomy Centre (ESA/ESAC), Operations Department, E-28692 Villanueva de la Ca\~nada (Madrid), Spain}
\address[iaat]{Institut f{\"u}r Astronomie und Astrophysik, Universit\"at T\"ubingen, Sand 1, 72076 T{\"u}bingen, Germany}
\address[issi]{International Space Science Institute (ISSI), Hallerstrasse 6, CH-3012 Bern}   
\address[mit-kavli]{MIT Kavli Institute for Astrophysics and Space Research, Massachusetts Institute of Technology, Cambridge, MA 02139, USA}
\address[remeis]{Dr. Karl Remeis-Sternwarte, Friedrich-Alexander-Universität Erlangen-N{\"u}rnberg, Sternwartstr. 7, D-96049 Bamberg, Germany}
\address[franz-ludwig]{Franz-Ludwig-Gymnasium, Franz-Ludwig-Straße 13, 96047 Bamberg}
\address[desy]{Deutsches Elektronen Synchrotron DESY, D-15738 Zeuthen, Germany}
\address[ifca]{Instituto de F{\'\i}sica de Cantabria (CSIC-Universidad de Cantabria), E-39005, Santander, Spain}
\address[inaf-oas]{INAF -- OAS, Osservatorio di Astrofisica e Scienza dello Spazio, via Gobetti 93/3, I-40129 Bologna, Italy}
\address[uabello]{Departamento de Ciencias F{\'\i}sicas, Universidad Andr{\'e}s Bello, Fern\'andez Concha 700, Las Condes, Santiago, Chile}
\address[saao]{South African Astronomical Observatory, Observatory Road, Observatory, 7925, Cape Town, RSA}
\address[iuidia]{Inter-University Institute for Data-Intensive Astronomy, Department of Astronomy, University of Cape Town, Private Bag X3, Rondebosch 7701, South Africa}
\address[iaudev]{IAU-Office of Astronomy for Development, P.O. Box 9, 7935 Observatory, South Africa}
\address[cresst]{CRESST, Department of Physics, and Center for Space Science and Technology, UMBC, Baltimore, MD 21250, USA}
\address[gsfc]{NASA Goddard Space Flight Center, Greenbelt, MD 20771, USA}
\address[inaf-oab]{INAF -- Osservatorio Astronomico di Brera, Via E. Bianchi 46, I-23807 Merate, Italy}
\address[cass]{Center for Astrophysics and Space Sciences, University of California, San Diego, 9500 Gilman Drive, La Jolla, CA 920093-0424, USA}
\address[ssl]{Space Sciences Laboratory, 7 Gauss Way, University of California, Berkeley, CA 94720-7450, USA}
\address[ieec]{Institute of Space Sciences (ICE, CSIC), Campus UAB, Carrer de Can Magrans s/n, E-08193 Barcelona, Spain; Institut d'Estudis Espacials de Catalunya (IEEC), Gran Capità 2-4, E-08034 Barcelona, Spain}
\address[icrea]{Instituci{\'o} Catalana de Recerca i Estudis Avan\c{c}ats (ICREA), E-08010 Barcelona, Spain}
\address[msfc]{ST12 Astrophysics Branch, NASA Marshall Space Flight Center, Huntsville, AL 35812, USA}
\address[ihep]{Key Laboratory of Particle Astrophysics, Institute of High Energy Physics, Chinese Academy of Sciences, 19B Yuquan Road, Shijingshan District, 100049, Beijing, China}

\begin{abstract}
High mass X-ray binaries are among the brightest X-ray sources in the Milky Way, as well as in nearby Galaxies. Thanks to their highly variable emissions and complex phenomenology, they have attracted the interest of the high energy astrophysical community since the dawn of X-ray Astronomy. In more recent years, they have challenged our comprehension of physical processes in many more energy bands, ranging from the infrared to very high energies. 

In this review, we provide a broad but concise summary of the physical processes dominating the emission from high mass X-ray binaries across virtually the whole electromagnetic spectrum. These comprise the interaction of stellar winds with the high gravitational and magnetic fields of compact objects, the behaviour of matter under extreme magnetic and gravity conditions, and the perturbation of the massive star evolutionary processes by presence in a binary system.

We highlight the role of the INTEGRAL mission in the discovery of many of the most interesting objects in the high mass X-ray binary class and its contribution in reviving the interest for these sources over the past two decades. We show how the INTEGRAL discoveries have not only contributed to significantly increase the number of high mass X-ray binaries known, thus advancing our understanding of the population as a whole, but also have opened new windows of investigation that stimulated the multi-wavelength approach nowadays common in most astrophysical research fields.

We conclude the review by providing an overview of future facilities being planned from the X-ray to the very high energy domain that will hopefully help us in finding an answer to the many questions left open after more than 18 years of INTEGRAL scientific observations. 
\end{abstract}

\end{frontmatter}

\onecolumn
\begin{center}
\noindent%
\printaffil{esac}{European Space Astronomy Centre (ESA/ESAC), Operations Department, E-28692 Villanueva de la Ca\~nada (Madrid), Spain}
\printaffil{esac-quasar}{Quasar Science Resources S.L for ESA, European Space Astronomy Centre (ESA/ESAC), Operations Department, E-28692 Villanueva de la Ca\~nada (Madrid), Spain}
\printaffil{inaf-iasf}{INAF -- IASF, Istituto di Astrofisica Spaziale e Fisica Cosmica, Via A. Corti 12, I-20133 Milano, Italy}
\printaffil{isdc}{Department of Astronomy, ISDC, University of Geneva, Chemin d'Ecogia 16, CH-1290 Versoix, Switzerland)}
\printaffil{cab}{Centro de Astrobiolog{\'\i}a -- Departamento de Astrof{\'\i}sica (CSIC-INTA), E-28692 Villanueva de la Ca\~nada (Madrid), Spain}
\printaffil{gcsu}{Department of Chemistry, Physics and Astronomy, Georgia College and State University, Milledgeville, GA 31061, USA}
\printaffil{aim}{AIM, CEA, CNRS, Universit\'e Paris-Saclay, Universit\'e de Paris, F-91191 Gif-sur-Yvette, France}
\printaffil{apc}{Universit\'e de Paris, CNRS, Astroparticule et Cosmologie, F-75006 Paris, France}
\printaffil{dcu}{School of Physical Sciences and CfAR, Dublin City University, Dublin 9, Ireland}
\printaffil{dias}{Dublin Institute for Advanced Studies, 31 Fitzwilliam Place, Dublin 2, Ireland}
\printaffil{sharjah-daa}{Department of Applied Physics \& Astronomy, University of Sharjah, Sharjah, UAE}
\printaffil{saasst}{Sharjah Academy of Astronomy, Space Sciences, and Technology (SAASST), Sharjah, UAE}
\printaffil{dfapl-ua}{Department of Applied Physics, University of Alicante, E-03080 Alicante, Spain}
\printaffil{sternberg}{Sternberg Astronomical Institute, Moscow State University, 119234, Moscow, Russia}
\printaffil{kazan}{Kazan Federal University, Kazan, Russia}
\printaffil{iafrt}{Institute of Astrophysics, Foundation for Research and Technology-Hellas, 71110 Heraklion, Crete, Greece}
\printaffil{uoc}{University of Crete, Physics Department \& Institute of Theoretical \& Computational Physics, 70013 Heraklion, Crete, Greece}
\printaffil{dfests-ua}{Department of Physics, Systems Engineering and Signal Theory, University of Alicante, E-03690 Alicante, Spain}
\printaffil{iufacit-ua}{University Institute of Physics Applied to Sciences and Technologies, University of Alicante, E-03690 Alicante, Spain}
\printaffil{turku}{Department of Physics and Astronomy, FI-20014 University of Turku, Finland}
\printaffil{iki}{Space Research Institute of the Russian Academy of Sciences, Profsoyuznaya Str. 84/32, Moscow 117997, Russia}
\printaffil{soton}{School of Physics and Astronomy, Faculty of Physical Sciences and Engineering, University of Southampton, Southampton SO17 1BJ, UK}
\printaffil{ecap}{Erlangen Centre for Astroparticle Physics (ECAP), Erwin-Rommel-Strae 1, 91058 Erlangen, Germany}
\printaffil{viu}{Universidad Internacional de Valencia - VIU, C/Pintor Sorolla 21, 46002, Valencia, Spain}
\printaffil{esac-aurora}{Aurora Technology B.V. for ESA, European Space Astronomy Centre (ESA/ESAC), Operations Department, E-28692 Villanueva de la Ca\~nada (Madrid), Spain}
\printaffil{iaat}{Institut f{\"u}r Astronomie und Astrophysik, Universit\"at T\"ubingen, Sand 1, 72076 T{\"u}bingen, Germany}
\printaffil{issi}{International Space Science Institute (ISSI), Hallerstrasse 6, CH-3012 Bern}   
\printaffil{mit-kavli}{MIT Kavli Institute for Astrophysics and Space Research, Massachusetts Institute of Technology, Cambridge, MA 02139, USA}
\printaffil{remeis}{Dr. Karl Remeis-Sternwarte, Friedrich-Alexander-Universit\"at Erlangen-N{\"u}rnberg, Sternwartstr. 7, D-96049 Bamberg, Germany}
\printaffil{franz-ludwig}{Franz-Ludwig-Gymnasium, Franz-Ludwig-Strasse 13, 96047 Bamberg}
\printaffil{desy}{Deutsches Elektronen Synchrotron DESY, D-15738 Zeuthen, Germany}
\printaffil{ifca}{Instituto de F{\'\i}sica de Cantabria (CSIC-Universidad de Cantabria), E-39005, Santander, Spain}
\printaffil{inaf-oas}{INAF -- OAS, Osservatorio di Astrofisica e Scienza dello Spazio, Area della Ricerca del CNR, via Gobetti 93/3, I-40129 Bologna, Italy}
\printaffil{uabello}{Departamento de Ciencias F{\'\i}sicas, Universidad Andr{\'e}s Bello, Fern\'andez Concha 700, Las Condes, Santiago, Chile}
\printaffil{saao}{South African Astronomical Observatory, Observatory Road, Observatory, 7925, Cape Town, RSA}
\printaffil{iuidia}{Inter-University Institute for Data-Intensive Astronomy, Department of Astronomy, University of Cape Town, Private Bag X3, Rondebosch 7701, South Africa}
\printaffil{iaudev}{IAU-Office of Astronomy for Development, P.O. Box 9, 7935 Observatory, South Africa}
\printaffil{cresst}{CRESST, Department of Physics, and Center for Space Science and Technology, UMBC, Baltimore, MD 21250, USA}
\printaffil{gsfc}{NASA Goddard Space Flight Center, Greenbelt, MD 20771, USA}
\printaffil{inaf-oab}{INAF -- Osservatorio Astronomico di Brera, Via E. Bianchi 46, I-23807 Merate, Italy}
\printaffil{cass}{Center for Astrophysics and Space Sciences, University of California, San Diego, 9500 Gilman Drive, La Jolla, CA 920093-0424, USA}
\printaffil{ssl}{Space Sciences Laboratory, 7 Gauss Way, University of California, Berkeley, CA 94720-7450, USA}
\printaffil{ieec}{Institute of Space Sciences (ICE, CSIC), Campus UAB, Carrer de Can Magrans s/n, E-08193 Barcelona, Spain; Institut d'Estudis Espacials de Catalunya (IEEC), Gran Capit\`a 2-4, E-08034 Barcelona, Spain}
\printaffil{icrea}{Instituci\'o Catalana de Recerca i Estudis Avan\c{c}ats (ICREA), E-08010 Barcelona, Spain}
\printaffil{msfc}{ST12 Astrophysics Branch, NASA Marshall Space Flight Center, Huntsville, AL 35812, USA}
\printaffil{ihep}{Key Laboratory of Particle Astrophysics, Institute of High Energy Physics, Chinese Academy of Sciences, 19B Yuquan Road, Shijingshan District, 100049, Beijing, China}
\end{center}
\twocolumn

\tableofcontents

\section{Introduction}
\label{Sec:Intro}

In High-Mass X-ray Binaries (HMXBs) a compact object --  most frequently a neutron star (NS) -- accretes matter from a binary companion star with a mass above $\sim$10\Msol. They form a sub-class of X-ray Binaries (XRBs), hosting very massive donors ($M_\mathrm{donor} \geq 10\,\Msol$).
Galactic HMXBs were among the earliest sources detected by the new field of X-ray astronomy in the 1960's \citep[see, e.g.,][]{giacconi62a,chodil67,schreier72,webster72} and have remained an intense subject 
of study ever since. 
These systems can form during the joint evolution of a pair of massive stars in a sequence involving mass transfer between the companions also before the first supernova explosion \citep{vdHeuvel72,tauris17}.  Massive stars influence their environment significantly, through their strong, ionizing ultraviolet radiation, as well as by their
strong winds and final explosions which provide a significant input of energy and chemically enriched matter into the interstellar medium \citep{kudritzki2002}. Some of these systems may develop further into double compact objects, i.e., future sources of gravitational wave events
\citep[see, e.g.,][and references therein]{vdHeuvel2019}.

Accretion in HMXBs can occur in different ways, as described in Section~\ref{Sec:PhysObs}. Observationally, one usually distinguishes between disc-fed systems, showing Roche-Lobe overflow, wind-accreting Supergiant X-ray Binaries (\sghmxbs), with stellar type O or B mass donors -- including the sub-class of peculiar Supergiant Fast X-ray Transients (SFXTs) -- and Be X-ray Binaries (BeXRBs), in which accretion is driven by the interaction between the compact object and the decretion disc around the mass donor. 
Another, special class are the Wolf-Rayet X-ray Binaries with only 7 known examples, 6 of which are in other galaxies \citep{Esposito2015}. The very well-known, but peculiar X-ray binary Cyg~X-3 is the only example in our Milky Way and has been studied frequently by \intg \citep[e.g.][]{Beckmann:2007CygX-3,Zdziarski2012a,Zdziarski2012b}.

\begin{figure}[tb]
\begin{center}
\includegraphics[angle=0,width=\columnwidth]{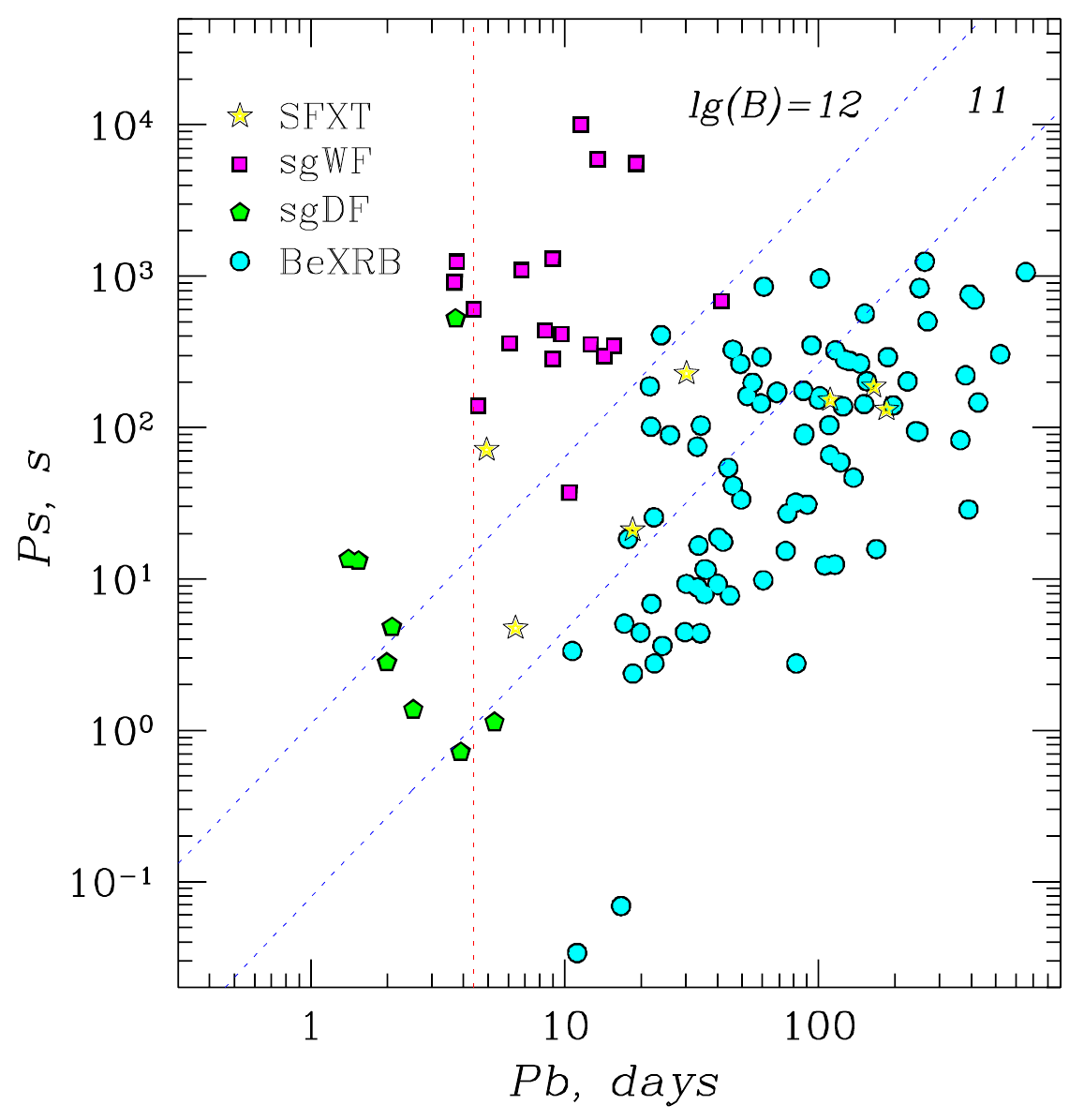}
\caption{Spin-period (\Ps) over binary (\Pb) period diagram (``Corbet diagram'') for HMXBs. The different classes are distinguished by colour and symbol shape: SFXTs, disc-fed (sgDF) supergiant binaries (including ultra-luminous X-ray sources (ULXs)), wind-fed (sgWF) supergiant binaries, as well as Be X-ray Binaries (BeXRB). The vertical line indicates the binary period at which a 20~\Rsol, 22~\Msol supergiant fills its Roche lobe. Below the blue lines quasi-spherical accretion from the stellar wind for two different dipole magnetic field strengths $B$ 
is inhibited by the centrifugal barrier (see Section~\ref{Sec:magnetosphere}), assuming a wind speed of 800~km ~s$^{-1}$. Updated from Fig.~4 in \citet{Grebenev:2009SFXT}.}
\label{fig:Corbet}
\end{center}
\end{figure}

Due to the large field of view of the \intg instruments and their sensitivity, especially in the hard X-ray band, as well as an observing programme concentrating in particular 
on regions in the Galactic disc, the number of known HMXBs has grown very significantly with \intg and new types of HMXBs have been identified. 
Specifically, \intg observations led to the detection of very highly absorbed sources (see Sect.~\ref{Sec:absorbed}) and to the identification of SFXTs as a
new class (see Sect.~\ref{Sec:SFXT}). In addition, \intg observed almost all major outbursts from BeXRBs and discovered eight new such systems (see Sect.~\ref{Sec:BeX-INT}). Studies of accreting X-ray pulsars with long pulse periods, have also benefited from the long, uninterrupted observations possible due to the highly-elliptic orbit of \intg.
The relatively high fraction of \sghmxbs among the sources found with \intg (see also Table~\ref{tab:obs}) has led to a more even distribution of known source types and allowed us to fill the parameter space in the spin-period (\Ps) over binary (\Pb) period diagram (Figure~\ref{fig:Corbet}), often referred to as ``Corbet-Diagram''.  

While for obvious reasons detections of new HMXB systems were especially frequent in the early years of \intg, when certain areas of the sky where covered in depth for the first time by its instruments (see Table~\ref{tab:obs} and its references), the number of identified sources 
has been steadily growing, also due to the transient nature of many systems. For catalogues of \intg surveys see \citet{Revnivtsev:2004Survey,Bird:2006,Bird:2007,Bird:2010,Krivonos:2010Survey,Krinovos:2012Survey}, while results from the Optical Monitor Camera have been published by \citet{Alfonso-Garzon:2012}. An extensive review of HMXBs in the Milky Way and \intg{'s} contribution was published by \citet{walter15a}.

In the following, we first summarize in Section~\ref{Sec:PhysObs} the physics and observable phenomena which are behind the rich phenomenology seen in HMXBs by \intg and other satellites. Section~\ref{Sec:HMXB-INT} summarizes key results obtained for the different classes of sources and specific example cases. In Section~\ref{Sec:Pop} we discuss the distribution of HMXB systems in the Galaxy and their contribution to the overall luminosity of the Galaxy in these wavelengths. Finally, Section~\ref{Sec:Future} focuses on the future of HMXB observations with \intg and other observatories.

\section{Physics and observable phenomena}
\label{Sec:PhysObs}


In this Section, we give a short overview of the most important observable phenomena and their physical interpretation. We discuss flux variability, which is linked to variations in the accretion rate (Sects. \ref{Sec:windRL}, \ref{Sec:RLOF}, \ref{Sec:BeX}), super-orbital variability (Sect.~\ref{Sec:SuperOrb}), interactions at the magnetosphere (Sect.~\ref{Sec:magnetosphere}), X-ray continuum emission properties of neutron stars (Sect.~\ref{Sec:continuum}), the formation of Cyclotron Resonant Scattering Features (CRSFs, Sect.~\ref{Sec:CRSF}), and X-rays from black hole systems (Sect.~\ref{Sec:BH}).

\subsection{X-ray binaries accreting from stellar winds}
\label{Sec:windRL}

In the majority of \sghmxbs, the compact object accretes directly from the stellar wind of its companion. Therefore the physical state, structure, and density of the wind has an immediate effect on the accretion rate and hence on the observed X-ray luminosity. A large fraction of the observed aperiodic variability can be qualitatively explained by accretion from a ``clumpy'' stellar wind; however, quantitative calculations are more difficult to obtain, as outlined below.

\subsubsection{Acceleration of stellar winds}
\label{Sec:windRL:acc}

The winds of hot luminous stars are mainly driven and accelerated by ultra-violet (UV) resonance lines (like ions of C, N, O, Fe, etc), therefore they are known as line driven winds. The theory of radiatively driven stellar winds was first developed  by \citet{castor75a} (CAK hereafter) and refined in many subsequent publications, see e.g. \citet{windsreview00,Puls08}. 
The acceleration depends on the ionization, excitation, and chemical composition of the stellar wind \citep[see e.g., Eq. 7 in][]{castor75a}. The amount of ionization is strongly affected by the stellar parameters (e.g. effective temperature, $T_\text{eff}$) and in X-ray binary systems also by the X-ray emission of the compact object \citep[see e.g.,][for recent studies]{ASander+18,Krticka+Kubat+Krtickova:2018}.


From the phenomenological point of view, the stellar wind can be characterized by two parameters: i) the mass loss rate ($\dot{M}$) per unit of time and ii) the terminal velocity ($\upsilon_{\infty}$) at large distances from the star, where wind acceleration becomes insignificant. Typical mass loss rates of massive stars are of the order of $\dot{M}\sim 10^{-6}\:\mathrm{M}_{\odot}$\,yr$^{-1}$. Typical values of the  terminal velocity in massive stars are of the order of $\upsilon_{\infty}\sim$500\,--\,2000\,km\,s$^{-1}$ \citep{windsreview00,Puls08}. 

Assuming a spherically symmetric (non-rotating) stellar wind, the mass loss rate can be derived from  $\dot{M}=4\pi r^{2}\,\rho(r) \,\upsilon(r)$, where $r$ is the distance from the center of the star, $\rho(r)$ is the density and $\upsilon(r)$ is the velocity at that distance. 
The radial density profile is commonly parametrized  assuming a so-called $\beta$-velocity law (CAK): 
\begin{equation} \label{eq:sw_betalaw}
\upsilon(r)=\upsilon_{\infty}\,\big(1-R_{\star}/r\big)^{\beta},
\end{equation}
where $R_{\star}$ is the stellar radius and $\beta$ describes the steepness of the wind velocity profile. Both wind terminal velocity ($\upsilon_{\infty}$) and the $\beta$-parameter are obtained mainly through spectral fitting of optical/UV data. Typical values of $\beta$ parameters are in the range between $0.5$ and $1$ \citep{windsreview00}. 
\subsubsection{Inhomogeneous (clumpy) winds}

It was recognized early \citep{Lucy+Solomon:1970,Lucy+White:1980} that line driven wind acceleration is likely to be unstable, leading to the formation of shocks and inhomogeneous regions in the wind, commonly referred to as clumps \citep{OwockiRybicki84,owocki88a}.
This phenomenon is usually described as line-deshadowing (or just line-driven) instability (LDI), which will lead to a very unstable outer wind, but which may be damped close to the stellar surface, although \citet{SundqvistOwocki13} also found an unstable wind in near photospheric layers.
For most of the wind mass, the dominant overall effect of the instability is to concentrate material into dense clumps, leading to a density contrast up to $10^{4\text{--}5}$ \citep{Puls08}. 

Accounting for clumpy winds can affect the interpretation of observable quantities for the stellar wind (e.g.\ mass-loss rates derived from spectral line fits) by large factors. LDI simulations tend to favour relatively small clump sizes and masses, compared to some values assumed in studies of X-ray binaries, see \citet{martinez17a} for a detailed discussion.

\begin{figure}[ht!]
  \centering
  \includegraphics[angle=0,width=0.75\columnwidth]{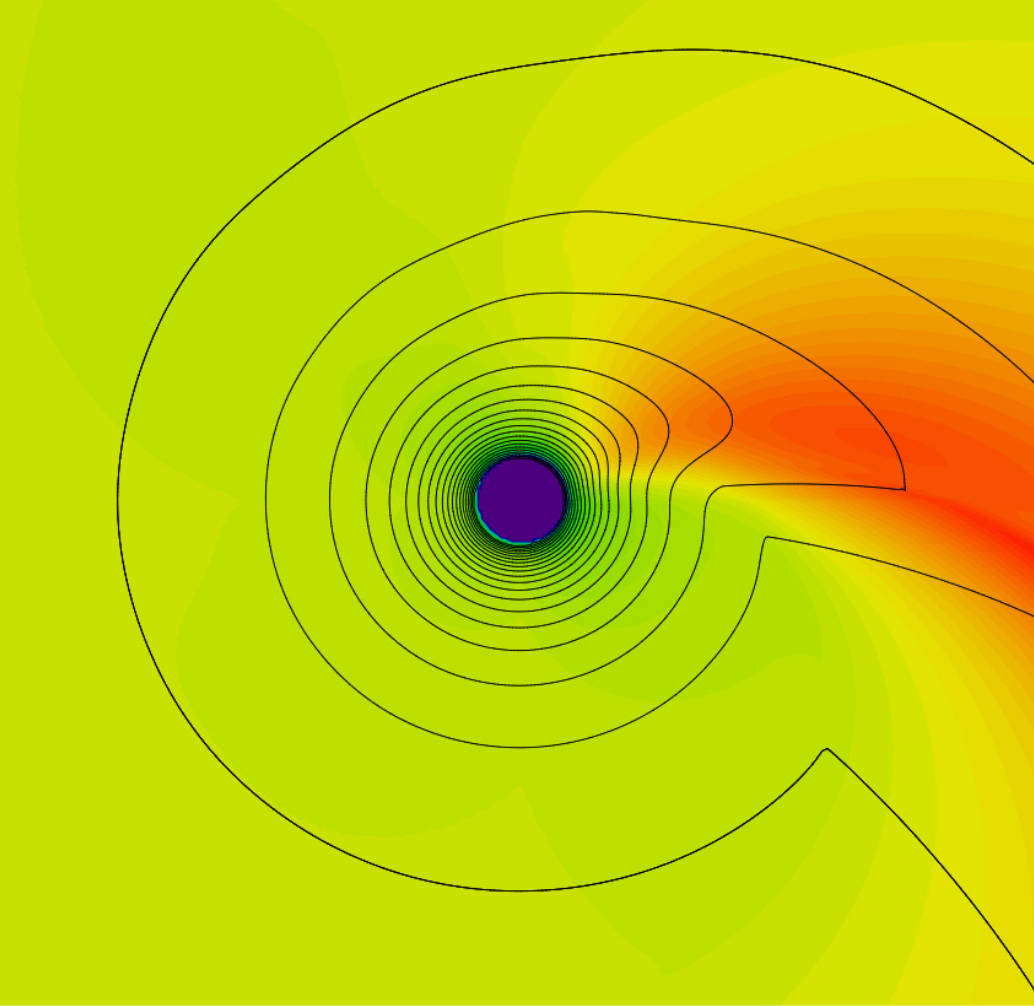}
  \includegraphics[angle=0,width=0.75\columnwidth]{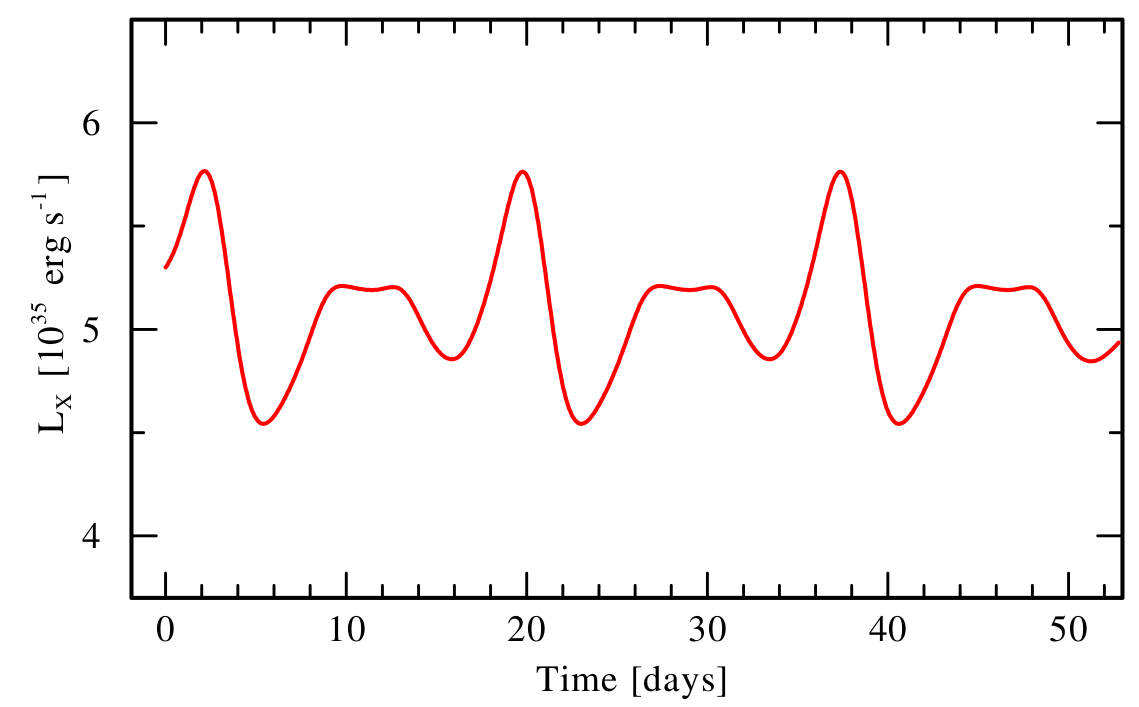}
  \caption{Top: an example of a single-arm CIR presented by 
  \citet{2017A&A...608A.128B} and obtained from the hydrodynamic code developed by \citet{lobel}. In this case, the CIR is generated by a bright spot on the stellar surface and it has a period of 10.3~days. The colors code the wind density of the CIR model relative to the density of the supergiant star unperturbed, smooth wind. The maximum over-density in the CIR is by a factor 1.22 (red colors). The black solid lines show 20 overplotted contours of equal radial velocity in the hydrodynamic rotating wind model. Bottom: Simulated long-term light curve of IGR~J16493$-$4348, assuming that the superorbital period of $\sim$20~days is produced by a single-arm CIR in the top figure.}
\label{fig:cirb}
\end{figure}

\subsubsection{Corotating interaction regions}
\label{Sec:windRL:CIR}

The existence of large structures in OB supergiant winds, beyond the typical size expected for the clumps, was suggested already in the 1980s by \citet{mullan1984}, and confirmed later by the observational detection of  so-called discrete absorption components \citep[DACs; see, e.g.,][and references therein]{underhill1975}.
These DACs are understood to be generated by corotating interaction regions (CIRs) induced by irregularities on the stellar surface, related either to dark/bright spots, magnetic loops, or non-radial pulsations \citep{cranmer1996}. 
CIRs are spiral-shaped density and velocity perturbations in the stellar wind that can extend outward up to several tens of stellar radii, and the physical properties of such structures (thickness, inclination, velocity and density profiles, number of spiral arms, rotational period) can be determined from the characteristics assumed for the irregularity(ies) from which they originate (see also Fig.~\ref{fig:cirb}). 
Unfortunately, there is a paucity of observational data on DACs in supergiant stars, with all evidence obtained with the IUE satellite \citep{boggess1978}, due to the lack of an UV facility with comparable or better sensitivity. But their existence is further supported by modulations of the X-ray emission observed in single OB stars \citep{oskinova2001, naze2013, massa2014}.

\begin{figure}[hbt]
\begin{center}
\includegraphics[angle=0,width=\columnwidth]{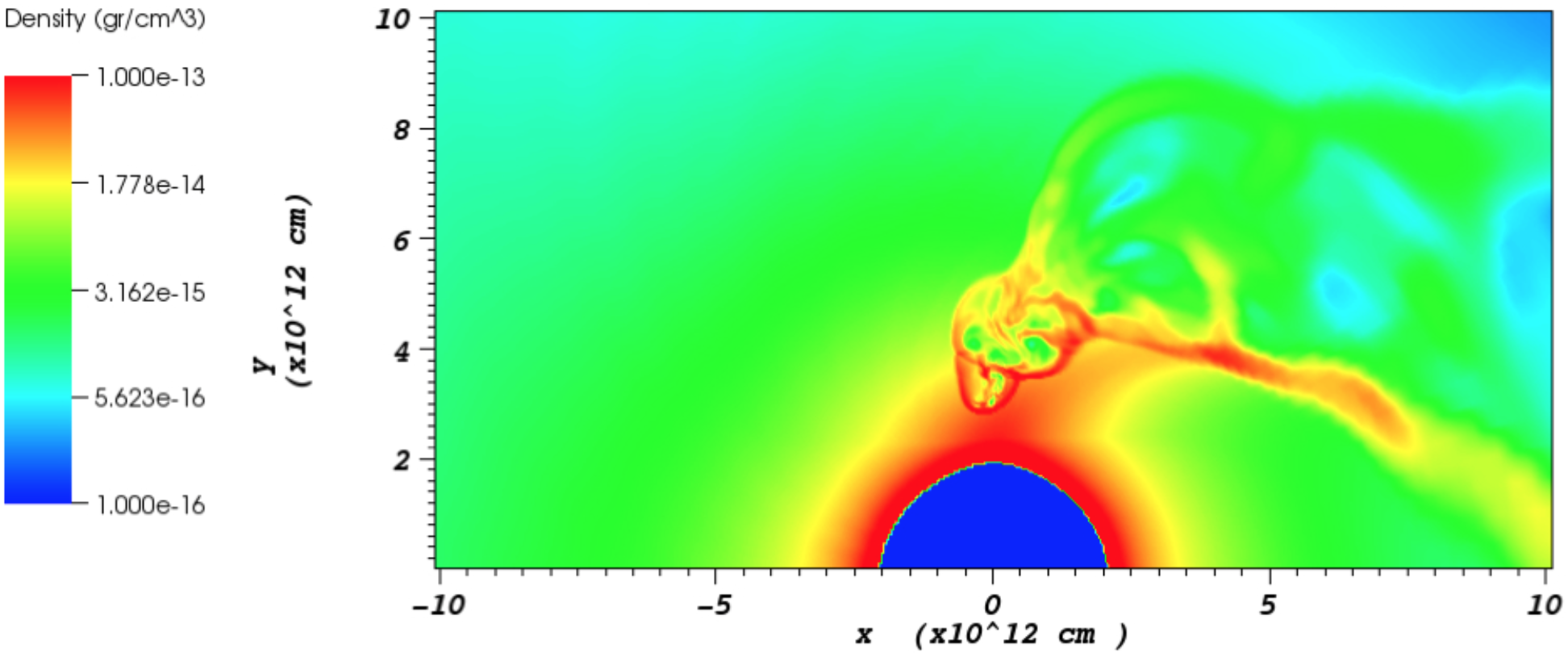}
\caption{Snapshot of a hydrodynamic simulation of a model of the Vela~X-1 system using the 
VH1 
code developed by John Blondin. Originally Fig.~7.5 in \citet{manousakis11a}.}
\label{Fig:VelaHydromodel}
\end{center}
\end{figure}

\subsubsection{The influence of the compact object}
\label{Sec:windRL:influence}
In wind accreting HMXBs, the presence of the compact object moving through the dense stellar wind close to the massive star significantly influences the wind flow. The gravity of the compact object will focus the wind towards the orbital plane and the position of the accretor; the orbital movement of the companion will lead to the formation of a bow shock and a trailing accretion wake; and finally the X-ray emission will ionize the local environment, changing the availability of resonance lines for acceleration. The ionization state of the wind is defined as ($\xi\left(r\right)={L_\mathrm{X}}/{N r^2}$), where $L_\mathrm{X}$ is the average X-ray luminosity and $N$ is the gas density at the distance $r$ from the NS \citep{Tarter69}. X-ray photoionization and heating can lead to the formation of a Str\"omgren sphere, where the wind is not accelerated anymore. At the boundary of this a shock is formed and sheets of gas trailing the X-ray source result \citep{FranssonFabian}. In extreme cases, wind accretion might be inhibited, leading to a feedback process \citep[e.g.][]{Blondin+90,Krticka+Kubat+Krtickova:2018}.

In general, in a \sghmxb system an accretion wake develops around and behind 
the compact object and evolves with time, showing fluctuations. Even when assuming a smooth wind as starting condition, the wind is likely to be heavily disrupted due to the hydrodynamical effects, accounting for high X-ray variability even in the absence of clumps \citep{manousakis15a}. 
First steps to model the accretion of clumps from a realistic clumpy wind model \citep{Sundqvist+Owocki+Puls:2018} were done by \citet{ElMellah18}, but without including the ionizing feedback effect on the incoming matter.

We note that most studies relating wind density and X-ray emission effectively assume immediate Bondi-Hoyle accretion of the matter reaching the vicinity of the NS. As shown in Section~\ref{Sec:magnetosphere}, the detailed physics of accretion close to or at the magnetosphere will further modulate the variability, e.g., by ``magnetic gating'', possibly dampening or enhancing the intrinsic density fluctuations. Simulations combining all these multi-scale, multi-physics effects are still in the future.

\subsection{Potential Roche-lobe Overflow systems}
\label{Sec:RLOF}

While wind accretion is the main mechanism to feed X-ray emission in HMXBs, for a limited subset of these sources the presence of an accretion disc is inferred from their substantially higher luminosity, their position in the spin period over orbital period (Corbet) diagram \citep{Corbet:86}, marked trends in their spin periods, or their optical lightcurves. 
Well-known examples are LMC X-4 \citep{Heemskerk+vanParadijs:89}, Cen X-3 \citep{Tjemkes:86} and SMC X-1 \citep{Hutchings:77}.
Other systems in which transient discs have been reported and/or strongly suggested due to observationally indirect evidences include OAO~1657$-$415 \citep{Jenke:2012,sidoli18a}, 4U~0114+650 \citep{Hu:2017}, GX~301$-$2 \citep{2019A&A...629A.101N}, and possibly  IGR~J08408$-$4503 \citep{Ducci:2019IGRJ08408}; see also \citet{Taani:2019} for a recent compilation of possible evidence of
disc accretion in HMXBs.

In the spin-period (\Ps) over binary (\Pb) period diagram (a.k.a. ``Corbet diagram'', Figure~\ref{fig:Corbet}) a group of bright X-ray binaries occupies
a distinct position in the lower left part of the diagram 
and these 
are generally considered to be accreting from a disc. The very high and persistent X-ray flux has an effect on the mass donors, which are ``bloated'' -- their luminosity class and brightness is higher than expected for their mass (which has been accurately measured, unlike in almost all other HMXBs). There may even be X-ray heating of the stellar surface, and this may have an effect on wind acceleration, connected to, but not quite the same as, the effects discussed in Section~\ref{Sec:windRL:influence} above for fainter X-ray sources.

The disc formation has often been ascribed to Roche-lobe Overflow (RLO). Fully-developed RLO will lead to mass transfer on the thermal timescale of the donor, at a rate of about $10^{-3}\:\mathrm{M}_{\odot}$\,yr$^{-1}$, which will completely extinguish the X-ray source. However, it was shown by \citet{Savonije:78,Savonije:83} that prior to filling the Roche lobe, already a phase of
beginning atmospheric RLO may set in, which in the case of a core-hydrogen-burning donor may last many thousands of years and supply mass-transfer rates below $10^{-8}\:\mathrm{M}_{\odot}$\,yr$^{-1}$. In addition, \citet{Quast:2019} showed that a stable RLO phase with mass-transfer rates below $10^{-5}\:\mathrm{M}_{\odot}$\,yr$^{-1}$ -- as observed in ULXs -- may last for $>$200\,000 yrs, before full thermal-timescale RLO sets in.
On the other hand, recent theoretical publications \citep{ElMellah:2019,Karino:2019} indicate that under certain conditions, especially for slower wind speeds than assumed in the past,
also accretion discs may form at least temporarily in wind-accreting systems without the need to invoke Roche-lobe Overflow.

We only briefly mention the well-known X-ray binary SS~433, discussed in its own review as part of this collection \citep{Cherepashchuk:2019NewAst}. The accreting X-ray pulsar Her~X-1 \citep[e.g.][]{Staubert:2017}, an intermediate mass system and disc accretor, is also discussed in another review in this volume \citep[Sect.~6.1]{Sazonov:2020arXiv}.

\subsection{Be X-ray Binaries and their outbursts}
\label{Sec:BeX}

When the optical counterpart in an HMXB is not an evolved (supergiant) star but a Be star, then the system is called Be/X-ray binary \citep[\bexrb,][]{Maraschi:76Be,reig11,paul11}. The Be components are early-type B (earlier than B3) or late-type O (later than O8) stars with masses in the range 8\,--\,15\,\Msol whose most prominent property is the presence of a disc around the star's equator. The disc is formed by material ejected from the photosphere, due to the star's rapid rotation in addition to stellar activity \citep[e.g.,][]{Balona+Ozuyar:2020a} 
and not by accretion from an external source. Since material is transported outwards by the same viscosity mechanism that drags matter inwards in an accretion disc, the disc is referred to as a decretion disc. The disc provides the reservoir of material that is accreted by the compact object, in contrast to the stellar wind in \sghmxbs, and is ultimately responsible for the variability observed in these systems at all frequency bands. In the optical and infrared, Be stars show emission lines in their spectra, excess flux that increases with wavelength (with respect to a canonical B star of the same spectral type), and polarization. The emission lines and infrared excess are formed by recombination in the disc. Linear polarization results from Thomson scattering, when photons from the Be star scatter with electrons in the Be disc \citep{poeckert79,wood96,yudin01,halonen13,haubois14}. In the X-rays, \bexrbs show two distinct types of outbursting behaviour: an orbitally-modulated increase in X-ray flux, normally coincident with periastron passage (type I), and giant and long-lasting outbursts (type II). During type I outbursts, the X-ray luminosity is generally below a few $10^{37}$ erg s$^{-1}$, while in type II outbursts the luminosity may reach a few $10^{38}$ erg s$^{-1}$, close to the Eddington value \citep{stella86, okazaki01}

The Be phenomenon, i.e., the presence of a circumstellar disc, was first observed in isolated Be stars, without NS companions. In principle, one could turn to the vast amount of studies on classical Be stars to shed light on the properties of \bexrbs since the variability in \bexrbs is closely linked to the evolution of the decretion disc. The disc forms, grows and dissipates on time scales of years. However, it turns out that the presence of a compact companion affects the characteristics and evolution of the disc. Discs in \bexrbs are smaller and denser than in isolated systems \citep{reig97,zamanov01,okazaki02,reig16}. The reason is that discs in \bexrbs are truncated at the outer rim  \citep{okazaki01,okazaki02}. Disc truncation was also proposed as a natural explanation for the existence  of the two different types of outbursts \citep{okazaki01}.

Observational evidence for disc truncation stems from the various correlations of disc parameters with the size of the orbit, expressed as the orbital period \Pb, eccentricity $e$, or orbital separation $a$. Circumstellar discs in narrow-orbit systems are naturally more affected by the tidal torque exerted by the NS than systems with longer orbital periods. Thus, we expect faster and larger amplitude variations of the observables in systems with tighter orbits. The following correlations support the disc truncation idea:

\textsl{-- Orbital separation and disc size}. The correlation between the orbital period and the highest historical value of the equivalent width of the H$\alpha$ line (EW(H$\alpha$)) was the first to be suggested as evidence for disc truncation \citep{reig97} and confirmed in subsequent studies \citep{reig07a,antoniou09,coe2015a,reig16}. Because the equivalent width is directly related to the size of the disc \citep{quirrenbach97,grundstrom06},  these correlations imply that large discs can only develop in systems in which the two components are far apart. In systems with small orbital separation, the tidal torque exerted by the NS prevents the disc from expanding freely.

\textsl{-- Orbital period and variability}. Systems with small orbital periods are more variable both in the continuum and line optical emission \citep{coe05,reig16}. A similar result was found by analysing the variability in the X-ray band \citep{reig07a}. Because the discs in systems with short orbital periods suffer strong tidal torques exerted by their NS, they cannot reach a stable configuration over long timescales. 

\textsl{-- Disc recovery after dissipation}. Systems with shorter orbital periods display larger growth rates after a disc-loss episode. Owing to truncation, the disc becomes denser more rapidly in shorter orbital period systems, and so the equivalent width of the H$\alpha$ line increases faster. Not only the disc formation, but also the entire formation and dissipation cycle appears to be faster in systems with short orbital periods, while longer timescales are associated with longer orbital periods \citep{reig16}. 

Disc truncation has implications for theories that explain type II (giant) outbursts. X-ray outbursts are caused by the mass transfer from the Be star's disc to the NS. But if the disc is truncated, how can large amounts of matter be transferred to the NS? The current idea is that the giant outbursts occur when the NS captures a large amount of gas from a warped and eccentric Be disc, highly misaligned with respect to the orbital plane \citep{okazaki13,martin11,martin14}. The models show that highly distorted discs result in enhanced mass accretion when the NS gets across the warped part. Observational evidence for misaligned discs comes from optical spectra \citep{moritani11,moritani13} and polarization \citep{reig18}. 

In Be stars, the H$\alpha$ line can display very different shapes, from single peaked to double-peaked. These varied flavours in the emission line appearance are attributed to the different inclinations of the line of sight with respect to the circumstellar disc \citep{hanuschik95,hummel94}. Single-peaked profiles are seen (generally showing flank inflections due to non-coherent scattering, producing the wine-bottle profile) in low inclination systems. For intermediate inclinations, Doppler broadening gives rise to double-peaked profiles. For large inclination systems, the outer cooler regions of the disc intercept the line of sight and produce shell profiles (deep narrow absorption cores that go below the continuum level).  But, what if all the types of profiles described above are seen in the same source, as in the \bexrb 4U 0115+63 \citep{negueruela01,reig07b}? Because the spin axis of the Be star cannot change on time scales of days, the appearance of different profiles in the same star can only imply that the disc axis is changing direction. This phenomenon is then interpreted as evidence for a precessing and warped disc.
A similar interpretation has been used to explain the complex, three-peaked H$\alpha$ profiles shown by 1A~0535+262 and AX~J0049.4-7323  \citep{moritani11,Ducci:2019AXJ0049}. The warping of the disc may be caused by the tidal interaction with the NS \citep{martin11} or by radiation from the
central star \citep{porter98}

Further evidence for warped discs comes from polarization. The light coming from a Be star is polarized.  The net polarization is perpendicular to the scattering plane (the plane containing
the incident and scattered radiation). Since the photons that get scattered come from the Be star and the scattering medium is the disc, the polarization angle is expected to be perpendicular to the major elongation axis.  Therefore, if the disc precesses, we should expect the polarization angle to change.  Changes in the optical polarization angle on time scales comparable to the orbital period were reported for the first time during a giant X-ray outburst in the \bexrb 4U 0115+63 and were interpreted as variation in the orientation of the disc \citep{reig18}.

As with the supergiant systems, BeXRBs in the Milky Way are generally affected by heavy extinction. Advances in our understanding of the class have come through the detailed analysis of a few representative systems \citep[e.g.][and references therein]{reig07b, monageng17}. In contrast, the Small Magellanic Cloud (SMC) contains a very large number of BeXRBs (approaching 100), almost free of interstellar absorption (at a moderately large distance, though), which can be used to perform population studies \citep{coe2015a}. This large number of BeXRBs is unexpected in such a small galaxy, and likely related to a recent burst of star formation due to interaction with the Large Magellanic Cloud (LMC) \citep{antoniou10}. Thanks to these properties, the SMC has become the prime laboratory for the study of BeXRBs \citep{haberl16}. Optical properties \citep{mcbride08}
and X-ray properties \citep{galache08} can be studied statistically, providing valuable input for the investigation of the accretion process \citep[e.g.][]{yang17}, formation mechanisms \citep[e.g.][]{Townsend:2011b} or even strict constraints on models for the production of gravitational wave emitting systems \citep[e.g.][]{vinciguerra20}.

\subsection{Superorbital modulations}
\label{Sec:SuperOrb}

Superorbital modulations are periodic variations of the X-ray luminosity observed from several HMXBs on time-scales longer than their orbital period \citep[typically a factor of 3--10 longer; see, e.g.,][and references therein]{kotze12}. In a few disc-fed HMXBs, as SMC~X-1 and LMC~X-4, superorbital variations have been known for decades, and have been clearly detected in all long-term monitoring data collected with \xte, \intg, and \swift \citep[see, e.g.,][and references therein]{dage19}. These modulations can be interpreted as being caused by irradiation from the X-rays emitted by the compact object onto a tilted and/or warped accretion disc, which is then forced to precess and periodically obscures the X-ray source \citep[see, e.g.,][]{pringle1996, ogilvie01}. 

Superorbital modulations have also been more recently discovered in several wind-fed \sghmxbs, mainly by using \swift data and then confirmed in a few cases also in the \intg and \xte data \citep[see][]{Corbet2013, corbet2018}; in some occurrences, only promising indications are found and confirmations are expected in the future when more data will be available. The interpretation of super-orbital modulations in wind-fed systems is less straightforward than in disc accreting binaries. It is unlikely that the presence of temporary accretion discs in wind-fed \sghmxbs could be the cause of the super-orbital modulations, as these periodicities require a mechanism stable over years to produce variations that are detected by folding the decade-long data-sets collected with \intg, \swift, and \xte. The interpretation put forward to explain the super-orbital variability in \sghmxbs involves either the variability of the mass-loss rate from the donor star induced by tidally-regulated oscillations of its outer layers \citep{koenigsberger2006}, or the presence of a third star in a hierarchical system \citep{chou01}. The problem with these two interpretations is that the first has been shown to work only for strictly circular orbits, while a stable triple hierarchical system implied by the second interpretation would require the presence of a third body in a very distant orbit that is not compatible with the fact that superorbital modulations in \sghmxbs have a period that is in general not longer than roughly three times the orbital period of these sources \citep{Corbet2013}. 

More recently, \citet{2017A&A...608A.128B} proposed an alternative idea according to which superorbital modulations could be related to the interaction between the CIRs of the supergiant (see Sect.~\ref{Sec:windRL}) and the NS orbiting the companion. When the NS encounters the CIR, the different velocity and density of this structure compared to the surrounding stellar wind produces the required long-term variation of the mass accretion rate to give rise to a super-orbital modulation with the observed intensity. As the CIRs do not necessarily have the same rotational period of the supergiant star and their number as well as geometrical properties are not yet well known, different combinations of a single or multiple CIR arms can be invoked in the different \sghmxbs in order to obtain the observed super-orbital periods. 
Fig.~\ref{fig:cirb} shows an example of applying this idea to interpret the $\sim$20~days-long superorbital modulation observed from the 
\sghmxb IGR~J16493$-$4348 \citep[with an orbital period of 6.78~days; see, e.g.,][and references therein]{Corbet2013}. In this case, the superorbital modulation is produced by a single-arm CIR with a rotational period of 10.3~days. 

\citet{coley19} presented an attempt to observationally understand the nature of the superorbital modulation in IGR~J16493$-$4348. They combined  an analysis of long-term intensity changes traced via \swift BAT with a broad-band spectral analysis combining two quasi-simultaneous \swift XRT and \nustar observation at the maximum and minimum phase of the superorbital modulation within a single 20~days cycle. They did not observe any significant differences between the spectral parameters of the two sets, apart from the overall flux change and could not firmly identify the mechanism causing the modulation. Despite the limitation of the short amount of time during which the broad-band spectral properties are measured, mechanisms where a significant change in the neutral hydrogen column density would be expected were considered unlikely. 
More observations covering much longer integration time scales, e.g., folding data covering many superorbital cycles, are clearly needed in order to advance our understanding on the intriguing superorbital variability of wind-fed \sghmxbs.

\subsection{Interactions at the magnetosphere}
\label{Sec:magnetosphere}

In the case of accreting NSs their usually strong magnetic fields add another level of complexity to the accretion physics. Note that this is often happening at the scale of a single pixel within the grids of models describing the system at a whole and thus tends to be handled in a very simplified manner in the kind of models described previously.

Interaction of plasma accreting onto a magnetized NS is usually described using an ideal magneto-hydrodynamic (MHD) approximation. In this approximation, the accreting plasma flow is significantly disturbed by the NS magnetic field at the radius (the Alfv\'en radius, $R_A$) determined by the balance between accreting plasma pressure (thermal and dynamical) and magnetic pressure. For example, for a spherically symmetric flow characterized by a mass accretion rate $\dot M$ onto a NS with mass $M_{\mathrm{x}}$ and dipole magnetic moment $\mu$, one obtains \citep{1977ApJ...215..897E}:
\begin{equation}
R_{\mathrm{A}}=\left(\frac{\mu^2}{\dot M\sqrt{(2GM_{\mathrm{x}}\,)}}\right)^{2/7}
\label{Eq:AlfvenRadius}
\end{equation}
This is a convenient reference formula to which real values of the magnetospheric boundary  $R_{\mathrm{m}}$, different by different factors in each particular source, can be normalized. It is convenient to write $R_{\mathrm{m}}=\zeta R_{\mathrm{A}}$, where the coefficient $\zeta$ is generally a function of $\dot M, \mu$ and other parameters and geometry of the flow (see, e.g., \citealt{2014EPJWC..6401001L} but also the criticism to this approach expressed by \citealt{bozzo09, bozzo18}). Note that the dependence of the Alfv\'en radius on accretion rate $R_{\mathrm{A}}\propto \dot M^{-2/7}$ was indirectly checked by the analysis of aperiodic X-ray variability of bright accreting NSs \citep{2009A&A...507.1211R} and is confirmed by the disc accretion torque-luminosity dependence in transient X-ray pulsars with Be components \citep{Sugizaki:2017,2017AstL...43..706F}. 


Magnetospheric interaction differs for disc or quasi-spherical accretion. The type of accretion (disc or quasi-spherical) is determined by the specific angular momentum of captured matter $j_{\mathrm{m}}$ at the magnetospheric boundary $R_{\mathrm{m}}$. A disc is formed around the magnetosphere if $j_\mathrm{m}$ exceeds the specific Keplerian value at the magnetospheric boundary,  $j_{\mathrm{m}}(R_{\mathrm{m}})>j_{\mathrm{K}}(R_{\mathrm{m}})=\sqrt{GM_{\mathrm{x}}R_{\mathrm{m}}}$. This is always the case for Roche-lobe overflow, but more rarely occurs in wind-fed systems (see above in Section \ref{Sec:RLOF}). In the opposite case, the accretion flow arriving at the magnetosphere is quasi-spherical. 

\subsubsection{Disc accretion and propeller effect}
\label{Sec:disc-propeller}

In the case of disc accretion, the inner disc radius $R_{\mathrm{d}}=\zeta_{\mathrm{d}} R_{\mathrm{A}}$ is the key parameter directly related to observational phenomena (spin-up/spin-down transitions in HMXB X-ray pulsars, the propeller effect, etc.). Presently, there are a number of models effectively describing $R_{\mathrm{d}}$, which is determined by plasma microphysics and the model of plasma-magnetospheric interaction. For example, assuming a purely diamagnetic Shakura-Sunyaev $\alpha-$disc \citep{1973A&A....24..337S}, one readily finds $\zeta_{\mathrm{d}}\approx \alpha^{2/7}$ \citep{1980A&A....86..192A}. In other models
(see, e.g., \cite{1979ApJ...232..259G,1979ApJ...234..296G,1995MNRAS.275..244L,2007ApJ...671.1990K}, among many others), different effective values of $\zeta_{\mathrm{d}}$ are obtained \citep{bozzo09, 2014EPJWC..6401001L}.

From an observational point of view, the plasma-magnetospheric interaction can be probed by spin-up/spin-down studies of X-ray pulsar spin periods and by analysis of non-stationary phenomena (X-ray outbursts). The torques acting on a NS are usually split into spin-up ($K_{\mathrm{su}}$) and spin-down ($K_{\mathrm{sd}}$) parts so that the angular momentum balance implies $I\dot\omega_{\star}=K_{\mathrm{su}}-K_{\mathrm{sd}}$, where $\omega_{\star}=2\pi/P_{s}$ is the NS spin frequency, $I$ is the NS moment of inertia. The spin-up torque can be written as $K_{\mathrm{su}}=\dot M\omega_{\mathrm{m}} R_{\mathrm{m}}^2$, where $\omega_{\mathrm{m}}$ is the angular frequency of matter at the magnetospheric boundary. 
The spin-down torque includes the magnetic part $\sim \mu^2/R_{\mathrm{m}}^3$ (which generally may have different sign depending on the twisting of the magnetic field lines) and the part due to the possible mass outflow from the inner disc radius $\sim \dot M_{\mathrm{ej}} R_{\mathrm{m}}^2\omega_{\star}$. Note that adding the matter ejection part provides an explanation for 
the observed strong spin-down episodes in Her X-1 \citep{2009A&A...506.1261K}. 

A widely accepted approach is to consider, in the first approximation, an equilibrium spin period $\Ps^\mathrm{\,eq}$ obtained from the balance $K_{\mathrm{su}}=K_{\mathrm{sd}}$. Obviously, this is a model-dependent quantity, $\Ps^\mathrm{\,eq}(\dot M, \mu, \zeta_{\mathrm{d}},...)$. The notion of an equilibrium spin period is frequently used for indirect estimation of the NS magnetic field from observations of \Ps and the X-ray luminosity produced near the NS surface -- the latter is related to $\dot M$ as $L_{\mathrm{x}} \approx \dot M c^2$. 
This period is close (but not identical) to the critical NS spin period derived from the condition for accretion to be centrifugally allowed, which is obtained by equating the corotation radius to the inner disc radius: $R_{\mathrm{c}}=(GM_{\mathrm{x}}/\omega^{*2})^{1/3}=\zeta_{\mathrm{d}}R_{\mathrm{A}}$. In the first approximation by assuming $\zeta_{\mathrm{d}}=\mathrm{const}$, 
we get $\Ps^\mathrm{\,crit}\propto \dot M^{-3/7}$. In other words, with decreasing $\dot M$ the inner disc radius increases, and once at a given \Ps it reaches the corotation radius, accretion is centrifugally inhibited, and the NS enters the so-called `propeller' state \citep{1975A&A....39..185I}. At this stage, the matter can be centrifugally expelled along open magnetic field lines, and magnetically dominated Poynting jets can be formed \citep{2014MNRAS.441...86L}. 
Nevertheless, a residual, strongly reduced X-ray luminosity (compared to the accretion state), can still be sustained by an inefficient plasma entry rate into the magnetosphere caused by diffusion, cusp instabilities, etc., as discussed e.g. by \cite{1984ApJ...278..326E}, or can be due to thermal emission from the magnetospheric accretion with $L_{\mathrm{x,m}}\simeq GM_{\mathrm{x}}\dot M/R_{\mathrm{m}}$, as in the model developed for $\gamma$ Cas stars by \citet{2017MNRAS.465L.119P}. 

Variable accreting X-ray sources offer the possibility to probe the accretion-propeller transitions during rise and decay of outbursts. The propeller mechanism is frequently invoked to explain a variety of transient phenomena in HMXBs, including SFXT outbursts \citep{2007AstL...33..149G,2008ApJ...683.1031B}, luminosity changes in bright transient X-ray pulsars \citep{2016A&A...593A..16T,2017ApJ...834..209L,2018MNRAS.479L.134T} and in  ultra-luminous X-ray pulsars (ULXPs) \citep{2016MNRAS.457.1101T}. 

\subsubsection{Quasi-spherical accretion: supersonic (Bondi) vs subsonic (settling)}

In the case of quasi-spherical accretion, interaction of plasma at $R_{\mathrm{m}}$ can be responsible for different steady-state and non-stationary phenomena in wind-fed HMXBs (see Section \ref{Sec:SFXT} below). Here a new important parameter appears -- the plasma cooling time at the magnetospheric boundary, which determines the type of magnetosphere inflow and the torques that apply to the NS. 

It has long been recognized that plasma entry into the NS magnetosphere in accreting X-ray binaries occurs via an interchange instability -- Rayleigh-Taylor (RT) in the case of slowly rotating NSs \citep{1976ApJ...207..914A,1977ApJ...215..897E} or Kelvin-Helmholtz (KH) in rapidly rotating NSs  \citep{1983ApJ...266..175B}.
In the case of disc accretion, the plasma entry into the magnetosphere via the RT instability was compellingly demonstrated by multi-dimensional numerical MHD simulations \citep{2008MNRAS.386..673K}. However, global MHD simulations of large NS magnetospheres ($\sim 10^9$ cm) have not been performed as yet, and information about physical processes near NS magnetospheres should be inferred from observations.  

During quasi-spherical wind accretion onto slowly rotating NSs, there is a characteristic luminosity, $L^*\simeq 4\times 10^{36}~\ergs$, that separates two physically distinct accretion regimes: the free-fall Bondi-Hoyle supersonic accretion occurring at higher X-ray luminosity, when the effective Compton cooling time of infalling plasma $t_\mathrm{cool}$ is shorter than the dynamical free-fall time $t_\mathrm{ff}$ \citep{1984ApJ...278..326E}, and subsonic settling accretion at lower luminosities, during which a hot convective shell forms above the NS magnetosphere \citep{2012MNRAS.420..216S,2018ASSL..454..331S}. In the latter case, a steady plasma entry rate is controlled by plasma cooling (Compton or radiative) and is reduced compared to the maximum possible value determined by the Bondi-Hoyle-Littleton gravitational capture rate from the stellar wind of the optical companion, $\dot M_{\mathrm{B}}\simeq \rho_{\mathrm{w}} R_\mathrm{B}^2/v_{\mathrm{w}}^3$, by a factor $f(u)^{-1}\approx (t_\mathrm{cool}/t_\mathrm{ff})^{1/3}>2$. Here $\rho_{\mathrm{w}}$ and $v_{\mathrm{w}}$ are the stellar wind density and the velocity relative to the NS, respectively, and $R_{\mathrm{B}}=2GM_{\mathrm{x}}/v_{\mathrm{w}}^2$ is the Bondi gravitational capture radius. The necessary conditions for settling accretion are met at low-luminosity states in HMXBs \citep{2017arXiv170100336P}.

Settling accretion, unlike supersonic Bondi-Hoyle accretion, enables angular momentum transfer from the magnetosphere, which makes it possible to find an equilibrium NS spin period from the torque balance $K_{\mathrm{su}}=K_{\mathrm{sd}}$. However, unlike in the disc case, the equilibrium period for a standard NS magnetic field turns out to be proportional to the binary orbital period \Pb, and can be very long: $\Ps^\mathrm{\,eq}\approx \Pb(R_{\mathrm{m}}/R_{\mathrm{B}})^2\approx 1000\:{\mathrm{s}}\,P_{10}\, \mu_{30}^{12/11}\,L_{36}^{-4/11}\,v_8^4$. Here $P_{10}=\Pb/10\:{\mathrm{d}}$, $L_{36}=L_{\mathrm{x}}/(10^{36}\mathrm{erg\, s}^{-1})$, $v_8=v_{\mathrm{w}}/(1000\,\mathrm{km\, s}^{-1})$. This can explain the existence of very long-period X-ray pulsars without invoking a superstrong magnetic field for the NSs \citep{2011ApJ...742L..11M,2017MNRAS.469.3056S}. Further implications and population synthesis modeling of HMXBs at the settling accretion stage are discussed in \cite{2018arXiv181102842P}.

At the settling accretion stage, in a rather narrow X-ray luminosity range between $\sim$ a few $\times 10^{35}$ \ergs and $L^*$, Compton cooling is still effective enough to enable steady plasma entry at a rate $\dot M_x=f(u)_\mathrm{Comp}\dot M_B$ via the RT instability. The classical persistent X-ray pulsars Vela~X-1 and GX~301-2 provide suitable examples \citep{2012MNRAS.420..216S}. 
With decreasing X-ray luminosity, radiative cooling becomes more effective than the Compton one. However, it is unclear whether the RT-mediated plasma entry into the magnetosphere can be steadily sustained by radiative cooling. Indeed, observations of `off'-states in Vela X-1 show abrupt decreases in the observed X-ray flux by more than an order of magnitude during which X-ray pulsations are clearly visible, suggesting a temporary transition to the radiative cooling regime \citep{2013MNRAS.428..670S}. Similar changes were observed in the SFXT IGR J11215$-$5952 \citep{2007A&A...476.1307S}. The application of the settling accretion model to SFXTs is further discussed below in Section \ref{Sec:SFXT}. 

\subsubsection{Magnetic and centrifugal inhibition of accretion in wind-fed NS HMXBs}
\label{Sec:inhibition-wind}

\citet{2008ApJ...683.1031B} proposed a different approach to identify  the diverse accretion regimes in NS HMXB, especially SFXTs, expanding on \citet{1983ApJ...266..175B}, \citet{davies79}, and \citet{davies81}. Compared to the previous sub-section, this also considers the case of fast rotating NSs, for which plasma penetration into the magnetosphere through the KH instability can be more effective than the RT instability.  

This description is based on the relative sizes of three essential radii defined in wind acrretion \citep[see][for detailed equations and definitions]{2008ApJ...683.1031B}:
\begin{itemize}
\item the accretion radius: $R_\mathrm{a}$ is the distance at which the inflowing matter is gravitationally focused toward the NS;
\item the magnetospheric radius: $R_\mathrm{M}$, at which the pressure of the NS magnetic field ($\mu^2$/(8$\pi$$R_\mathrm{NS}^6$), with $\mu$ the NS magnetic moment) 
balances the ram pressure of the inflowing matter;
\item the corotation radius: $R_\mathrm{co}$, at which the NS angular velocity equals the Keplerian angular velocity.
\end{itemize}
Assuming typical values for the NS and the stellar wind, these three radii are all of the order of a few times $10^{10}$~cm.

Changes in the relative position of these radii result into transitions across different regimes for the NS. In particular, the accretion radius and magnetospheric radius depend on the wind parameters ($R_\mathrm{a}\propto v_\mathrm{w}^{-2}$, $R_\mathrm{a}\propto v_\mathrm{w}^{-1/6}$, with the wind velocity $v_\mathrm{w}$), which can vary on a wide range of timescales (from seconds to months) and are usually assumed to trigger the transition between the different regimes, together with the corresponding variations in the X-ray luminosity. 
Below the different regimes of a magnetic rotating NS, subject to a varying stellar wind, are summarised \citep[see also][]{bozzo09b}. \\

\noindent\emph{Outside the accretion radius: magnetic inhibition of accretion ($R_\mathrm{M} > R_\mathrm{a}$)}. \\ 

In systems with $R_\mathrm{M} > R_\mathrm{a}$ the mass flow from the companion star interacts directly with the NS magnetosphere without significant gravitational focusing, forming a bow shock at $R_\mathrm{M}$. The power released in this region $L_\mathrm{shock}$ is estimated to be relatively low luminosity -- a few times $10^{29}$ erg~s$^{-1}$ 
--  and is mainly radiated in the X-ray band.  
Two different regimes of magnetic inhibition of accretion can be distinguished:

\begin{enumerate}
\item \textit{The super-Keplerian magnetic inhibition regime\/} ($R_\mathrm{M} > R_\mathrm{a}, R_\mathrm{co}$):
In this case the magnetospheric radius is larger than both the accretion and corotation radii. Matter that is shocked and halted close to R$_\mathrm{M}$ cannot proceed further inward, due to the rotational drag of the NS magnetosphere which is locally super-Keplerian. 
Since magnetospheric rotation is also supersonic, the interaction between the NS magnetic field and matter at R$_\mathrm{M}$ results in rotational energy dissipation and thus, NS spin down. This process releases energy at a rate $L_\mathrm{sd}$, again of the order of a a few times $10^{29}$ erg~s$^{-1}$ for typical parameters,
which adds to the shock luminosity. 

\item \textit{The sub-Keplerian magnetic inhibition regime} ($R_\mathrm{a} < R_\mathrm{M} < R_\mathrm{co}$):
In this case the magnetospheric drag is sub-Keplerian and matter can penetrate the NS magnetosphere through the KH instability.  
The mass inflow rate across the magnetospheric boundary R$_\mathrm{M}$ resulting from this instability depends on the efficiency factor $\eta_\mathrm{KH}\sim 0.1$, the shear velocity $v_\mathrm{sh}$ at R$_\mathrm{M}$, and the densities $\rho_\mathrm{i}$ and $\rho_\mathrm{e}$ inside and outside R$_\mathrm{M}$, respectively. The luminosity released by accretion of this matter onto the NS is estimated differently \citep[see][for details]{2008ApJ...683.1031B}, depending on the choice of the post shock gas velocity or the rotational velocity of the NS magnetosphere, but is estimated to be of the order of a a few times $10^{34}$ erg~s$^{-1}$ for typical parameters.
\end{enumerate}

\noindent\emph{Inside the accretion radius:} $R_\mathrm{M} < R_\mathrm{a}$.\\ 

Once $R_\mathrm{M}$ is inside the accretion radius, matter flowing from the companion star is shocked adiabatically at $R_\mathrm{a}$ and halted at the NS magnetosphere.  
In the region between $R_\mathrm{a}$ and $R_\mathrm{M}$ this matter redistributes itself into an approximately spherical configuration (resembling an ``atmosphere''), whose shape and properties are determined by the interaction between matter and NS magnetic field at $R_\mathrm{M}$. A hydrostatic equilibrium ensues when radiative losses inside $R_\mathrm{a}$ are negligible; the atmosphere is stationary on dynamical timescales, and a polytropic law of the form p $\propto$ $\rho^{1+1/n}$ can be assumed for the pressure and density of the atmosphere. 
The value of the polytropic index $n$ depends on the conditions at the inner boundary of the atmosphere, and in particular on the rate at which energy is deposited there. Three different regimes can be distinguished: 

\begin{enumerate}
\item \textit{The supersonic propeller regime} ($R_\mathrm{co} < R_\mathrm{M} < R_\mathrm{a}$):  
In this case the rotational velocity of the NS magnetosphere at $R_\mathrm{M}$ is supersonic; the interaction with matter in the atmosphere leads to dissipation of some of the star's rotational energy and thus spin-down. Turbulent motions are generated at $R_\mathrm{M}$, which convect this energy up through the atmosphere, until it is lost at its outer boundary. 
Matter that is shocked at $\sim R_\mathrm{a}$, reaches the magnetospheric boundary at $R_\mathrm{M}$, where the interaction with the NS magnetic field draws energy from the NS's rotation. According to \citet{PringleRees:72}, this gives the largest contribution to the total luminosity in this regime, $L_\mathrm{sd}$, of the order of a a few times $10^{31}$ erg~s$^{-1}$ for typical parameters.      

\item \textit{The subsonic propeller regime} ($R_\mathrm{M} < R_\mathrm{a}, R_\mathrm{co}$, $\dot{M}_\mathrm{w} < \dot{M}_\mathrm{lim}$): 
The break down of the supersonic propeller regime occurs when the magnetosphere rotation is no longer supersonic with respect to the surrounding material. The structure of the atmosphere changes and  the transition to the subsonic propeller regime takes place. Since the rotation of the magnetosphere is subsonic, the atmosphere is roughly adiabatic ($n$=3/2). 
In the subsonic propeller regime, the centrifugal barrier does not operate because $R_\mathrm{M} < R_\mathrm{co}$, but the energy input at the base of the atmosphere is still too high for matter to penetrate the magnetosphere at the capture rate $\dot{M}_\mathrm{capt}$  at which it flows towards the magnetosphere. 
Nevertheless, a fraction of the matter inflow at $R_\mathrm{a}$ is expected to accrete onto the  NS, mainly due to the KH instability, leading to a luminosity $L_\mathrm{KH} > 10^{35}$ erg~s$^{-1}$ for typical parameters. 
The rotational energy dissipation at $R_\mathrm{M}$ gives a small contribution $L_\mathrm{sd}$ of order of 10$^{30}$ erg~s$^{-1}$ under the same assumptions.

The subsonic propeller regime applies until the critical accretion rate $\dot{M}_\mathrm{lim}$
is reached, at which the gas radiative cooling completely damps convective motions inside the atmosphere. 
If this cooling takes place, direct accretion at the rate $\dot{M}_\mathrm{capt}$ onto the NS surface becomes possible.  
 
\item \textit{The direct accretion regime} ($R_\mathrm{M} < R_\mathrm{a}, R_\mathrm{co}$, $\dot{M}_\mathrm{w} > \dot{M}_\mathrm{lim}$): 
If $R_\mathrm{M} <R_\mathrm{co}$ and matter outside the magnetosphere cools efficiently, accretion onto the NS takes place at the full capture rate $\dot{M}_\mathrm{capt}$. 
The corresponding luminosity 
\begin{equation}\begin{split}
L_\mathrm{acc}= G M_\mathrm{NS} \dot{M}_\mathrm{capt} / R_\mathrm{NS} \\ 
\simeq2\times10^{35}\dot{M}_{15}~\mathrm{erg ~s}^{-1}, 
\label{eq:lacc} 
\end{split}\end{equation}
where $\dot{M}_{15}$=$\dot{M}_\mathrm{capt}$/10$^{15}$ g s$^{-1}$. 
This is the standard accretion regime, identified in some of the previous sections as the Bondi-Hoyle accretion. 
\end{enumerate} 

\subsubsection{Accretion regimes in SFXTs}
\label{sec:SFXT-theory}
Supergiant Fast X-ray Transients (SFXT) have been established as a class by \intg and are discussed in detail in Section~\ref{Sec:SFXT} later.
Their behaviour is characterized by transient emission and a huge dynamical range during outbursts.
This suggests inhibition of accretion between the flares, which can be due to physically different mechanisms. Chronologically, the first model invoked the \textit{magnetospheric gating} due to the magnetic and/or centrifugal propeller effect in a wind-fed system discussed above in Section \ref{Sec:magnetosphere}  \citep{2007AstL...33..149G,2008ApJ...683.1031B}. Indeed, for a fixed value of the mass accretion rate $\dot M$, if a NS spins fast enough and/or if its magnetic field strength is sufficiently intense, the magnetospheric radius can end up being either larger than the corotation radius but inside the accretion radius, or larger than both the accretion and corotation radii. 
Under these circumstances, the system might end up either in the supersonic propeller regime or even in the super-Keplerian propeller regime, where the largest inhibition of accretion occurs due to the magnetic and centrifugal gate. In this regime, the system is expected to be characterized by a low luminosity state ($\lesssim$10$^{33}$~erg~s$^{-1}$). Temporary increases in $\dot M$, for example related to the clumps in the winds of OB-supergiants \citep{2016A&A...589A.102B,2017A&A...608A.128B}, can `open' the magnetopsheric/centrifugal gates and induce transitions to the other different accretion regimes introduced
earlier in this section. Among them, the subsonic propeller or the direct accretion regimes allow a much higher accretion rate onto the compact object to take place, thus explaining the brightest X-ray states observed from the SFXTs ($\gtrsim 10^{36}$--$10^{37}$~erg~s$^{-1}$). 

\begin{figure*}[ht!]
\centering
\includegraphics[scale=0.34]{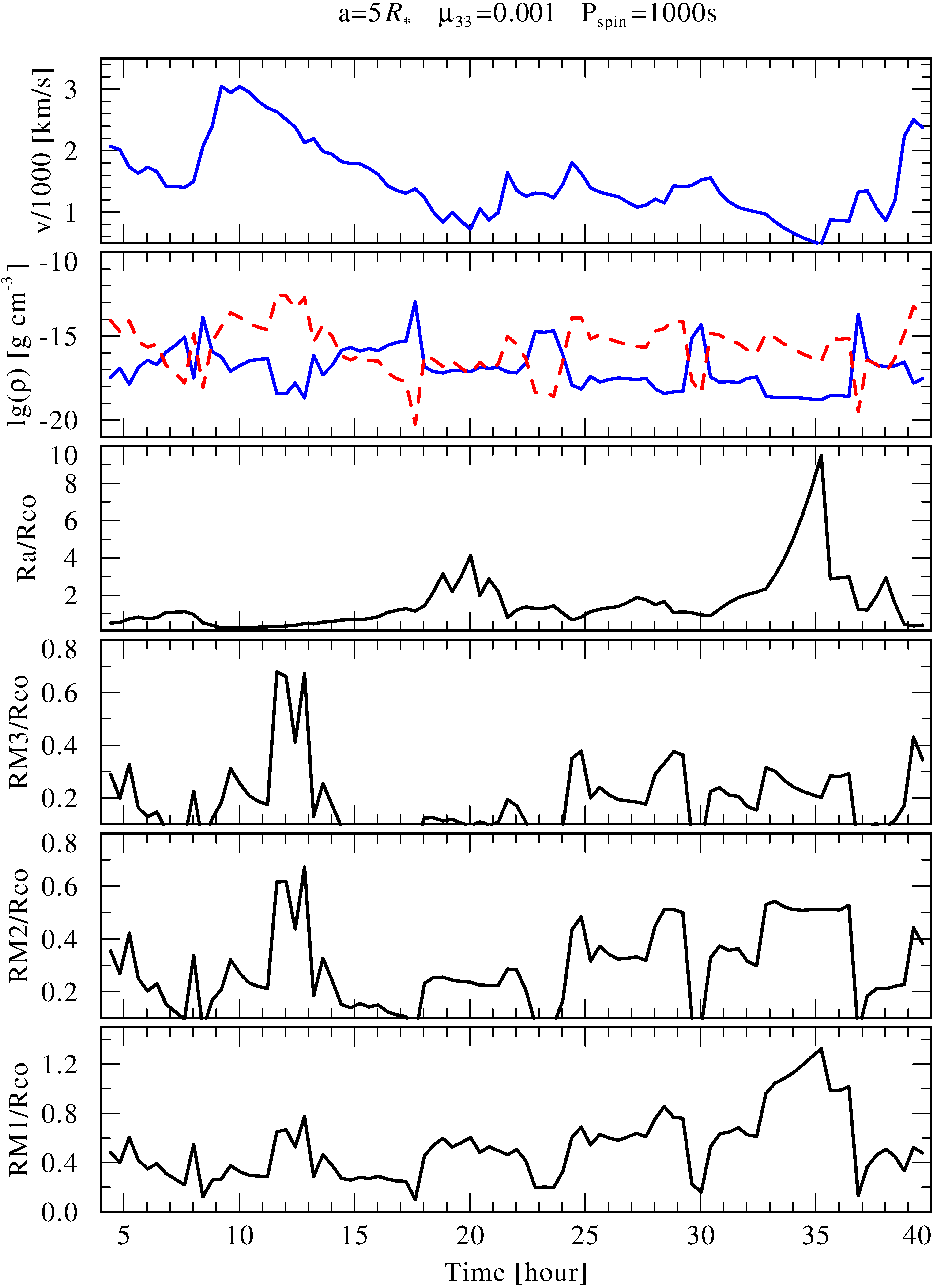}
\includegraphics[scale=0.34]{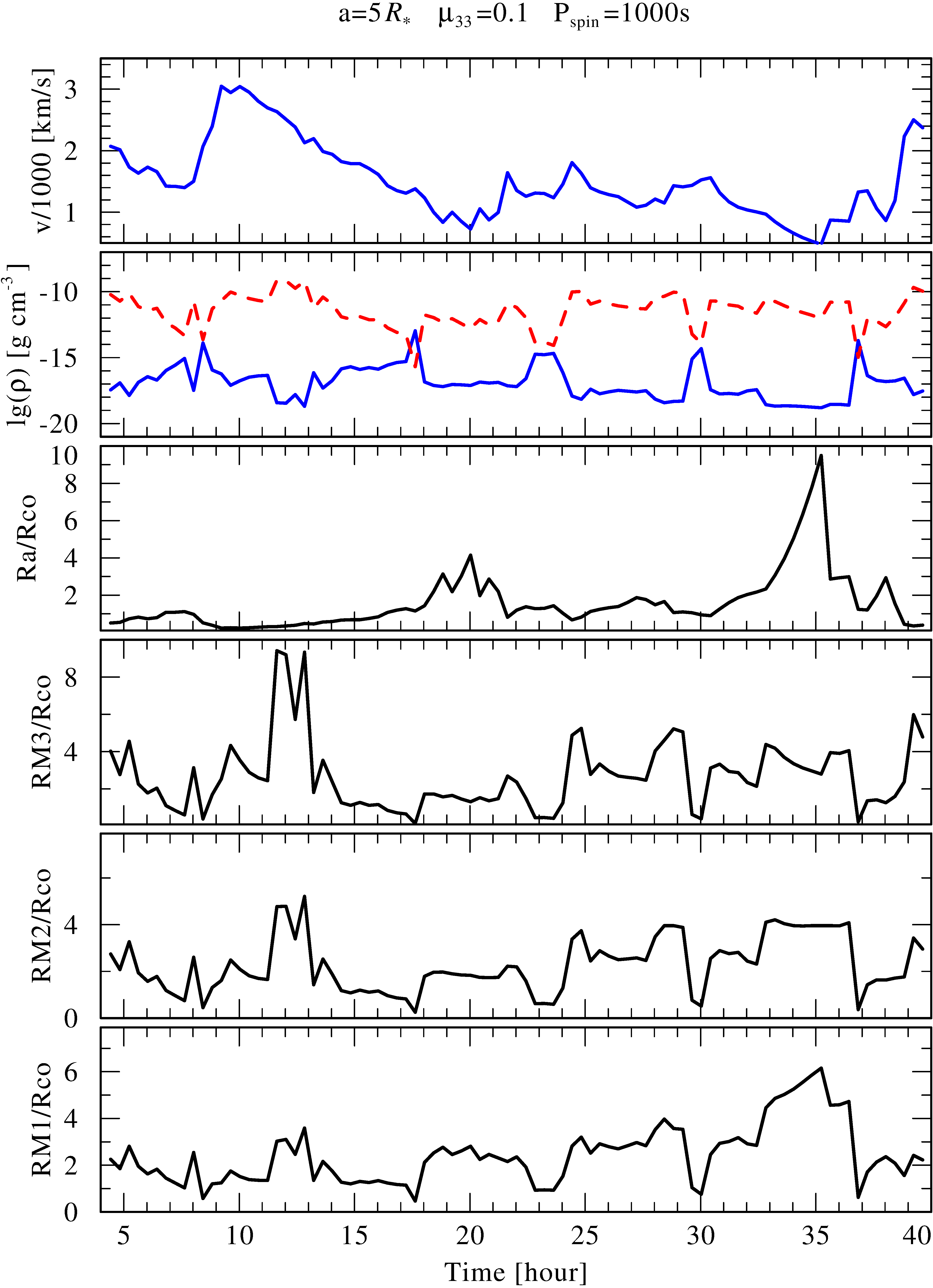}
\includegraphics[scale=0.34]{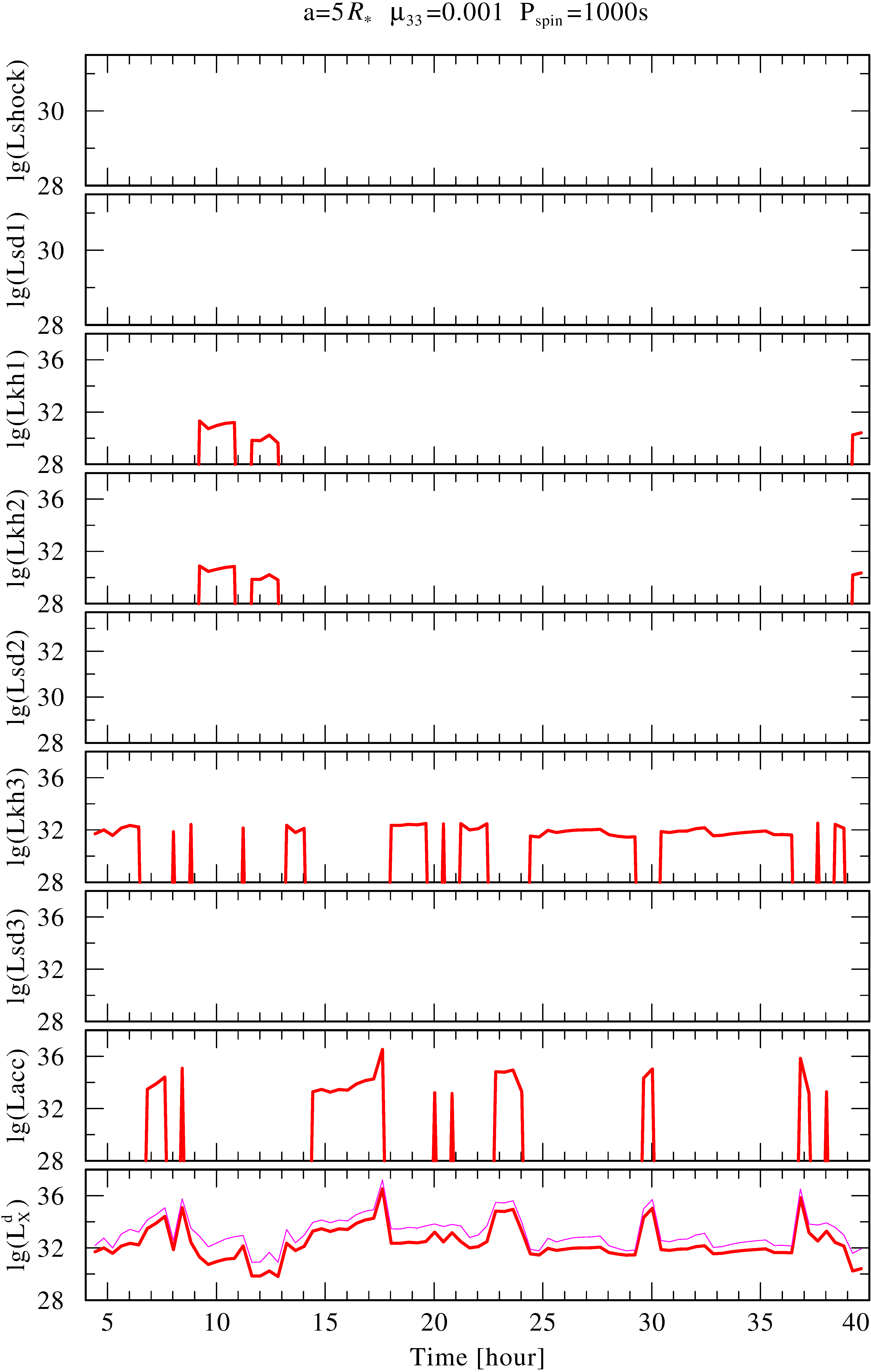}
\includegraphics[scale=0.34]{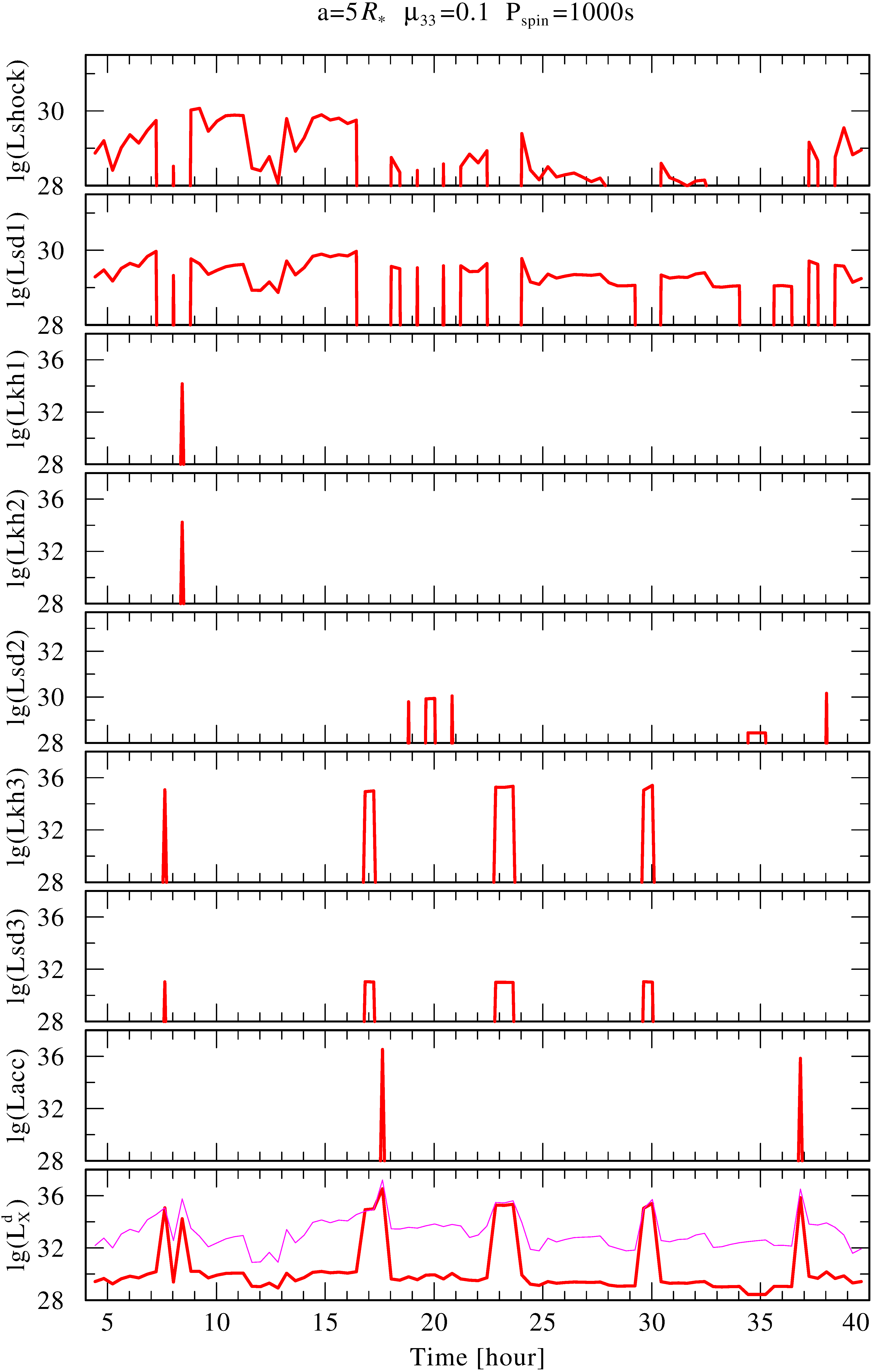}
\caption{Results of the simulations of the accretion onto a NS using an hydrodynamically calculated supergiant clumpy wind model and taking into account the gating accretion mechanisms described in Sect.~\ref{Sec:magnetosphere}. This figure combines Figures~7 and~8 of \citet{2016A&A...589A.102B}. See text for details.}
\label{fig:long3}
\end{figure*}
 
 A first attempt to simulate the transitions between different accretion regimes in SFXTs using an hydrodynamically calculated supergiant clumpy wind model has been presented by \citet{2016A&A...589A.102B}. The authors have shown that the effect of the NS rotation coupled with a strong magnetic field can significantly reduce the average luminosity of a \sghmxb and qualitatively explain the difference between classical systems and SFXTs. This is shown in Fig.~\ref{fig:long3}. 
 
 The system parameters adopted in the simulation are shown on the top of each figure (for the parameters and circular orbits assumed in this work a separation of $5\:R_{\star}$ corresponds to an orbital period of 25.6~days). The top figures both on the left and on the right report the wind velocity and density as a function of time, and all relevant radii to be determined in the gating accretion model (see Sect.~\ref{Sec:magnetosphere}). The bottom figures show the luminosity in each regime that is achieved by the system triggered by the variations of velocity and density in the stellar wind.
 
 The top panel of the top figures shows the instantaneous density of the supergiant wind, while the second panel displays the corresponding density. The red dashed line in these panels represents a critical value of the wind density above which the mass inflow rate toward the compact object becomes large enough to trigger the switch to the direct accretion regime. 
 
In the other panels, $RM1$ corresponds to the magnetospheric radius as defined at the start of Section~\ref{Sec:inhibition-wind},  $RM2$ to the radius in the supersonic propeller regime, $RM3$ to the subsonic propeller regime. 
$L_\mathrm{shock}$ is as above. $L_\mathrm{sd1}$ corresponds to the spin-down luminosity in the super-Keplerian magnetic inhibition regime, $L_\mathrm{sd2}$ and $L_\mathrm{sd3}$ to those in the supersonic and subsonic propeller regimes, respectively.
$L_\mathrm{kh1}$ and $L_\mathrm{kh2}$ are the two somewhat different estimates for the KH-fueled accretion luminosity in the sub-Keplerian magnetic inhibition regime, while $L_\mathrm{kh3}$ is the corresponding luminosity in the subsonic propeller regime.

The bottom panels of the bottom figures show the summed X-ray luminosity (red solid line) compared to the luminosity that a system would have if it was always in the direct accretion regime (solid magenta line). The top and bottom figures on the left differ from the corresponding ones on the right only for the assumed NS magnetic field strength. The figures on the left show a representative case of a classical \sghmxb, where the NS magnetic field as a ``standard'' value close to 10$^{12}$~G and the system in virtually always in the direct accretion regime. The figures on the right show the case where a much stronger NS magnetic field is assumed (``magnetar''-like, 10$^{14}$~G) and how this leads to a very different behaviour in the X-ray domain with a more extreme variability that is closer to what is observed in the SFXTs. With this stronger magnetic field, the velocity and density variations of the stellar wind are able to cause frequent switches among the different accretion regimes due to the fact that the magnetospheric radius, the accretion radius, and the corotation radius are closer to one another.  
 
In a wind-fed system there are also different models to explain the instability of the accretion flow onto the compact object and to interpret the correspondingly induced X-ray variability. As discussed earlier in this section, 
if we assume the specific case of a ``slow'' rotating NS, it can be shown that subsonic (settling) accretion onto the NS magnetosphere occurs at X-ray luminosities 
below $\sim 10^{36}$ erg s$^{-1}$. In this regime, the entry rate of accreting plasma into the NS magnetosphere is determined by plasma cooling. At low accretion rates, the cooling is radiative (inefficient compared to Compton cooling operating at higher $\dot M$), which hampers the development of the RT instability at the magnetospheric boundary. Neglecting the KH instability due to the slow rotation, the mass accretion rate drops down to low values corresponding to luminosities of $\sim 10^{33}-10^{34}$ erg s$^{-1}$, which are comparable to those recorded during the SFXT `low' states. 

It is quite possible that in the low-luminosity states of SFXTs no RT-mediated plasma entry into the NS magnetosphere occurs at all. This could be the case if the plasma cooling time is longer than the time a plasma parcel spends near the magnetosphere because of convection in the magnetospheric shell: $t_\mathrm{cool}>t_\mathrm{conv}\sim t_\mathrm{ff}(R_\mathrm{B})\sim 300-1000$~s. Once this inequality is violated, RT instability can start to develop. Interestingly, the 
rich phenomenology of X-ray flares of SFXTs as derived from the \xmm EXTraS project can naturally be explained by the development of RT instability at low accretion rates \citep{Sidoli2019}.

Additionally, within the context of the settling accretion model, in SFXTs the magnetospheric boundary itself can be made unstable for different reasons. For example, it was conjectured \citep{2014MNRAS.442.2325S} that giant flares in SFXTs are due to a sudden break of the magnetospheric boundary caused by the magnetic field reconnection with the field carried along with stellar wind blobs. This can give rise to short strong outbursts occurring in the dynamical (free-fall) time scale during which the accretion rate onto the NS reaches the maximum possible Bondi value from the surrounding stellar wind. A tentative evidence for the presence of magnetic fields in the OB supergiant in IGR~J11215-5952 was found from ESO-VLT FORS2 spectropolarimetric observations \citep{2018MNRAS.474L..27H}. Another reason for the instability can be due to stellar wind inhomogeneities which can disturb the settling accretion regime and even lead to free-fall Bondi accretion episodes.

So far, only one outburst observed from the SFXTs is difficult to be reconciled with a wind-fed accretion scenario. Independently of the specific assumptions considered for the plasma penetration inside the magnetosphere, the giant outburst observed in 2014 from the SFXT IGR~J17544-2619 \citep{Romano2015giant} reached an unprecedented X-ray peak luminosity of 3$\times$10$^{38}$ \ergs that is virtually impossible to achieve in a wind-fed system due to the limitations on the amount of  material captured by the NS for any reasonable value of the supergiant companion mass loss rate.  \citet{Romano2015giant} suggested that this 
event resulted from the temporary formation of a short-lived accretion disc around the NS hosted in this system. Accretion from an even temporary accretion disc can, indeed, lead to much higher luminosities than those achieved in a wind-fed system due to the larger mass accretion rate that can be transported through the disc by viscosity. As of today, this was the only case in which a disc accretion scenario was adopted for an SFXT, but short-lived accretion discs have also been invoked to explain bright X-ray luminosity states in classical \sghmxbs \citep[see, e.g., the case of OAO~1657-415;][and references therein]{wenrui19}

\subsection{Continuum spectrum}
\label{Sec:continuum}
Soon after the discovery of X-ray binaries, it became clear that Compton scattering in the hot and dense medium close the compact object leads to the shaping of X-ray radiation \citep{DavidsonOstriker:73}.
In neutron-star high-mass X-ray binaries, plasma flowing from the limit of the magnetically dominated region (the magnetosphere) is funneled along the magnetic field lines and then 
falls onto the NS on the magnetic poles forming two or more ``accretion columns''. The accretion column radius depends on the magnetic field strength and on the accretion rate as \citep{Lamb1973} $r_\mathrm{ac} \approx 600 \,L_{37}^{1/7}\, B_{12}^{-2/7}\,\mathrm{m}$, where we have assumed a one solar mass NS with radius of 10\,km, luminosity is expressed in units of $10^{37}\,\ergs$, and the magnetic field in units of $10^{12}$\,G. The spectrum emerging from a hot, dense plasma with a more rarefied medium above is dominated by Compton scattering of some thermal seed photons. At first approximation it has, thus, the spectral shape of an absorbed power-law with an exponential roll-over at high energy. Several empirical functional shapes have been used to describe the spectral energy distribution of these systems \citep[cutoff power law, Fermi-Dirac cutoff, NPEX, etc.; see][for a collection of model shapes]{Coburn2002}. 
However, in all of them, the cutoff energy is indicative of the 
plasma temperature in the accretion column and is of the order of 10\,keV or more. At lower energy, for most systems, photoelectric absorption in the local and Galactic medium prevents the investigation of the spectrum. However, for less absorbed systems, such as Her X-1, additional components due to the accretion disc may appear \citep[e.g.][]{Fuerst:2013}. 

\citet{Basko1976} realized that a high accretion rate would naturally lead to a halt of the infalling material by radiation in the column and identified a critical luminosity at which the formation of a radiatively-induced collisionless shock at some height above the NS surface is unavoidable:

\begin{equation}
L^{\star}\approx 4\times 10^{36}
\frac{r_\mathrm{ac}}{10^{5}\,\mathrm{cm}}
\frac{\sigma_\mathrm{T}}{\sigma_\mathrm{s}}
\frac{10^6\,\mathrm{cm}}{R}\frac{M}{M_\odot}
\,\mathrm{erg\, s^{-1}}\,,
\end{equation}

\noindent 
where $r_\mathrm{ac}$ is the accretion column radius, $\sigma_\mathrm{s}$ is the cross section in the vertical direction, and $M$ is the neutron star mass. At first approximation, below this critical luminosity, the plasma is stopped very close to the surface and radiation can escape vertically forming a ``pencil beam'', while for brighter systems plasma will sink subsonically below the shock and radiation is emitted from the sides of the accretion column in a ``fan beam''. 

In that seminal work, it was noted that a crucial role is played by the value of the opacity due to electron scattering, which is strongly energy dependent in a magnetized plasma, especially at energies comparable to the cyclotron energy in the magnetic field ($E_\mathrm{cyc}\sim 11.57 B_{12}\,\mathrm{keV}$, see Sect.~\ref{Sec:CRSF}). If we indicate with $\sigma_\parallel$ the cross section for electron scattering parallel to the magnetic field lines and with $\sigma_\perp$ the cross section component perpendicular to it, for $E<<E_\mathrm{cyc}$ one finds  $\sigma_\perp \sim \sigma_\mathrm{T}$ and $\sigma_\parallel << \sigma_\perp$ \citep{Canuto1971}. When approaching the cyclotron energy, resonances in the cross section play a 
crucial role and the extraordinary mode polarization dominates with an angle-dependent cross section which can exceed $10^4$ times the Thompson value. Around the cyclotron energy, we thus expect features in the spectrum: the ones that are most known are the cyclotron resonant scattering features, described in Sect.~\ref{Sec:CRSF}, which appear in absorption. However, also the continuum formation is influenced by such a strong energy dependency of the cross section.

\begin{figure}
	\centering
	\includegraphics[width=\linewidth,angle=-90]{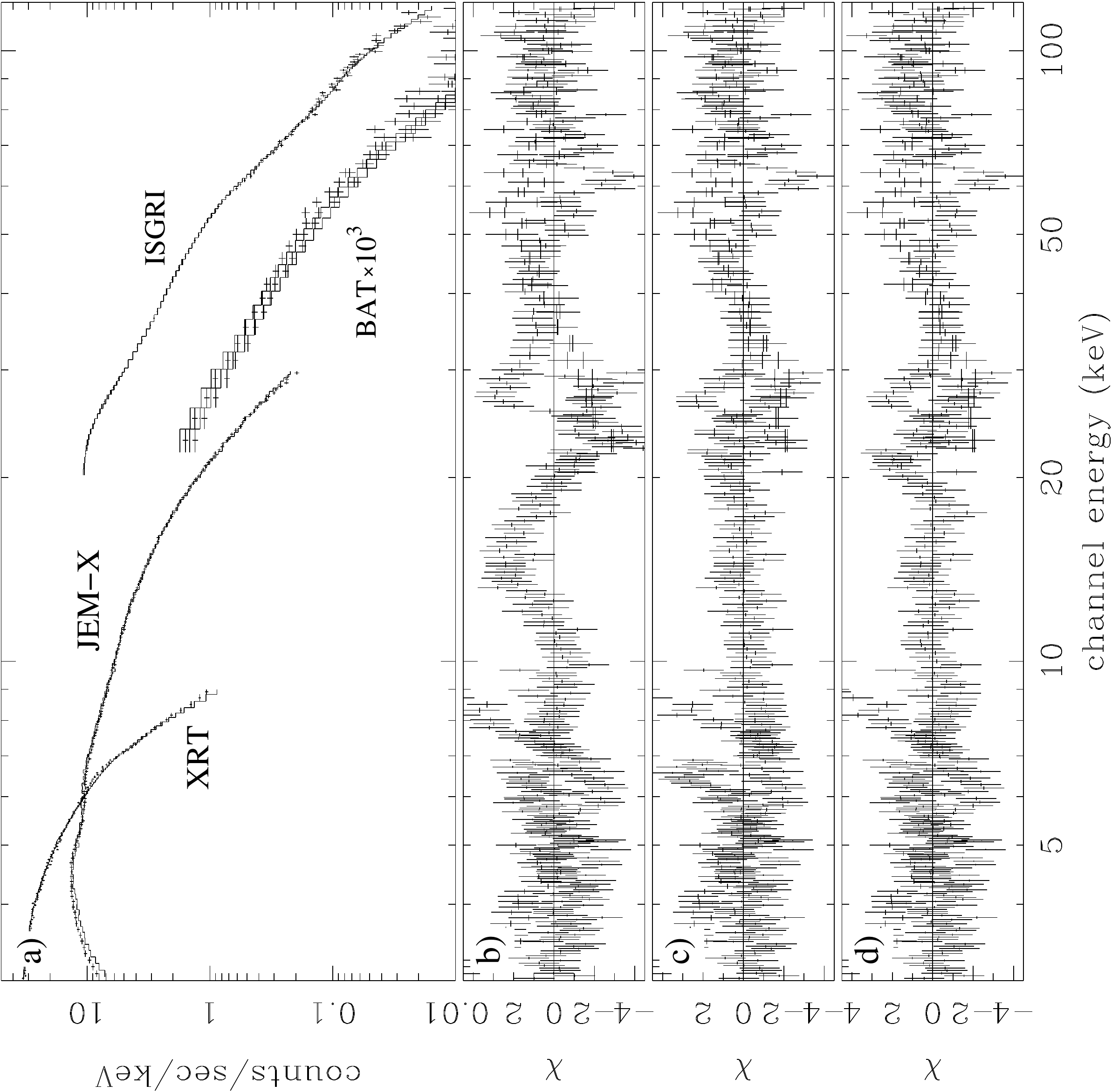}
	\caption{The broad band spectrum of EXO 2030+375 from simultaneous fits 
		of \intg and \swift data of the 2007 giant outburst with \texttt{highecut} (a)
		and residual plots after fitting it without additional features (b),
		adding a ``bump'' around 15 keV (c), or alternatively 
		including two absorption lines at $\sim$10 and $\sim$20 keV (d) \citep[from][]{Klochkov2007}.
		See \citet{Camero-Arranz:2005} for a broadband spectrum from an earlier, normal outburst.}
	\label{spe}
\end{figure}

With the high-sensitivity of the Rossi X-ray Timing Explorer proportional counter array (\xte/PCA), it was noted that there were wiggles appearing around 10\,keV on the top of a smooth continuum. However, it was with the giant outburst of EXO~2030+375 in 2007 that the issue of whether a broad absorption or emission feature is most appropriate became evident.
As depicted in Fig.~\ref{spe}, \citet{Klochkov2007}
showed that during a giant outburst, the continuum of this source could be equally described by adding a broad Gaussian feature centered at $\sim$14 keV (a ``bump'') or by two absorption lines at $\sim$10 and $\sim$20 keV. Spin-phase-resolved analysis revealed a possible absorption feature at much higher energy \citep[$\sim$60\,keV;][]{2008A&A...491..833K}. 
Since then, more cases arose with such a behaviour: particularly relevant is that of 4U~0115+63, for which \citet{Mueller2013}
showed that the simultaneous presence of emission and absorption features is important for the cyclotron line centroid energy determination. However, the emission feature in this source is centered at $\sim$9\,keV, while the absorption line is at about 11\,keV, these values significantly differing between each other (see also \citealt{Ferrigno2009}).

The energy difference between emission and absorption features in 4U~0115+63, but also in EXO~2030+375 and other sources, disfavors the interpretation of cyclotron scattering being responsible for 
both features or at least generating them in the same region around the NS. Moreover, the emission region is intrinsically broad and in general broader than the absorption ones.
Thus, such emission should originate in a region with lower magnetic field and possibly higher optical depth. 

Pioneering works by \citet{Meszaros1985a,Meszaros1985b} described 
the spectrum of magnetized accreting NSs as a function of the pulse phase. They exploited the main interactions between matter and radiation in the column to produce the first energy-dependent beam patterns that could qualitatively reproduce the observed properties of pulse profiles and spectra of pulsating HMXBs. Further refinements were focused on the light bending in the strong gravity regime \citep{Riffert1988,Meszaros1988,Leahy1995}.

A breakthrough in understanding
of the spectral characteristics came from the work by \citet{Becker2005} and \citet{Becker2007} who managed to find an analytical model to describe the spectral continuum of magnetized XRBs. To understand their work, it should be first noted that there are two regions of the system where photons can be originally generated:
the base of the column, where an optically thick thermal mound emits as a black body at 1--2\,keV, and the accretion column in which the flowing plasma, in an optically gray regime, emits bremsstrahlung radiation with a temperature of several keV. However, in presence of a magnetic field with cyclotron energy of the order of the plasma temperature, the collisional excitation of the Landau levels around the magnetic field lines strongly modifies the emission spectrum with a prominent angle-dependent spike at the cyclotron energy produced at the expense of higher-energy photons \citep{Riffert1999}. This spectrum is very complex, but it can be simplified by assuming that the cyclotron emission is a delta function and the bremsstrahlung is not modified.
In this approximation, seed photons are thus of three kinds: black-body from the base of the column, thermal bremsstrahlung and cyclotron emission (delta-function) from the vertical column body.

Seed photons are then up-scattered in the Compton process. In the presence of a strong magnetic field, the Compton cross section is heavily modified because photons have two polarization states 
and these interact differently with the electrons bound to the magnetic field lines. To make the problem viable it can be noted that, below the cyclotron energy, the cross section parallel to the magnetic field is heavily suppressed, while perpendicularly it remains essentially the Thompson one. The effective cross section can thus be treated in the model as an angle-average.
With these assumptions, it is then possible to compute the transfer function for seed photons as they diffuse along the accretion column: which is a Green function \citep{Becker2007}.
The model is substantially analytical and it can be used to fit the spectra of accreting X-ray binaries. This was done by \citet{Ferrigno2009} for 4U~0115+63 using \textit{BeppoSAX} data, where it was necessary to introduce an additional emission component at low energy, and by \citet{Wolff2016} for Her~X-1 using \nustar data.
Model limitations include the assumption of a cylindrical shape of the accretion column, of an analytical plasma velocity profile decoupled from radiation, of a constant magnetic field and electron temperature in the column. \citet{Farinelli2016} managed to relax the assumptions on the vertical dependency on the magnetic field and on the plasma velocity profile; they also introduced a Gaussian profile of the cyclotron emission instead of a delta function. This allowed them to describe the spectra of three X-ray binaries (4U 0115+63, Cen X-3 and Her X-1) without the need for any additional component.

\begin{figure*}[htb]
    \centering
\includegraphics[width=1.0\textwidth]{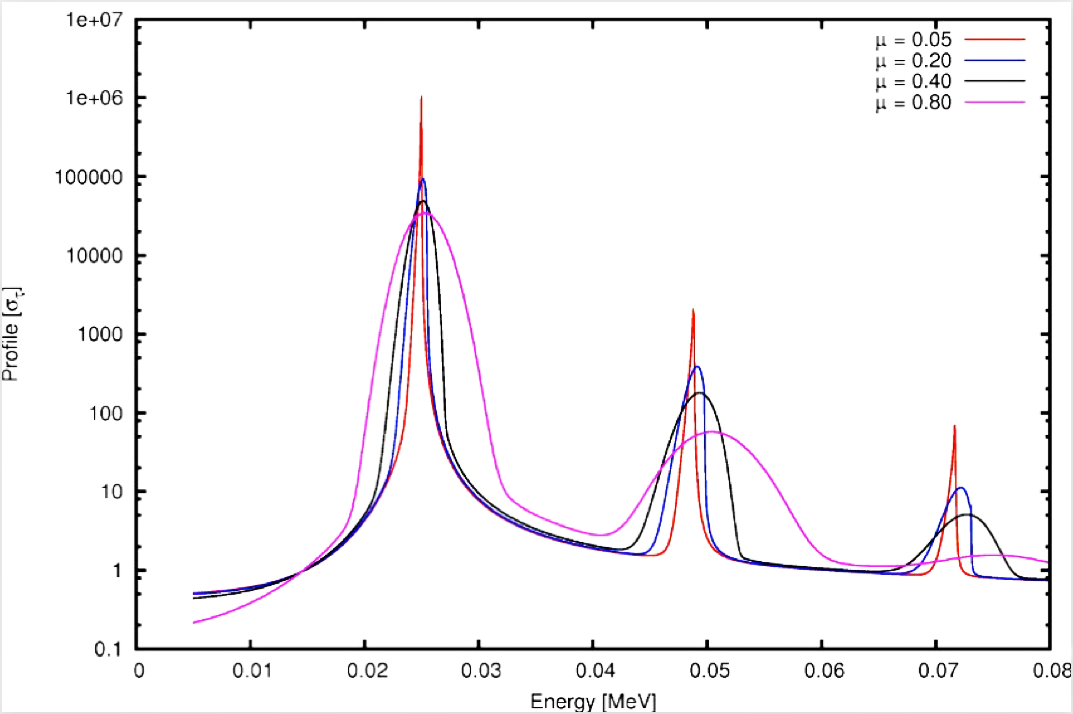}
    \caption{Averaged cyclotron cross-sections in units of the Thomson cross-section calculated by \citet{schwarm17a,schwarm17b}. The different colors indicate different angles $\vartheta$ between the photon path and the magnetic field, where $\mu = \cos \vartheta$.  The calculations were done for a magnetic field strength of $B=2\times10^{12}$\,G and a plasma temperature of $kT=3$\,keV. }
    \label{fig:crsf_crosssection}
\end{figure*}

Despite these theoretical efforts, a comprehensive description of the X-ray spectrum of pulsating HMXBs still eludes our complete understanding, due to the inherent difficulties in treating the coupled MHD problem of a plasma that is emitting near the local Eddington limit. However, with the models proposed so far, it has become clear that the commonly used power law with exponential rollover can only describe the thermal part of the emission, i.e. the Compton upscattering of thermally produced photons. Additional components in the emission are thought to be due to the Compton broadening from collisional excitation of the Landau level (cyclotron emission). Once a continuum is formed, there is a transition from an optically gray regime to free streaming. In this phase, scattering features, mainly in absorption, can be imprinted on the spectrum (see the next section). These features are quite broad 
and sometimes it becomes impossible to disentangle between emission and absorption \citep[see][for recent cases]{Bozzo2017,Ferrigno:2019}.

\subsection{Cyclotron Resonant Scattering Features}
\label{Sec:CRSF}

Electrons moving in a magnetic field are forced onto cyclic paths in the direction perpendicular to the magnetic field. If the magnetic field is strong enough, their cyclotron energy becomes comparable to their rest mass, requiring a quantum mechanical and relativistic treatment \citep[e.g.,][]{1986ApJ...309..362D,Canuto1971, harding1991}. In fields that are strong, the movement of the electrons perpendicular to the magnetic field axis can be described by quantized Landau levels, whose energies are the Eigenvalues of the electron's Hamiltonian and can be approximated when integrating over all angles and polarization as
\begin{equation}
    E_{\mathrm{Landau},n} =  n \frac{\hbar e B}{m_e} 
    \label{eq:e_landau}
\end{equation}

These quantized levels change the cross-section between electron and photons, and hence strongly influence the observed X-ray spectrum. Calculating these cross-sections, however, requires detailed fully relativistic QED-based calculations. The most recent work on this topic was presented by \citet{schwarm17a, schwarm17b}, building on work by, e.g., \citet{sina1996, harding1991, isenberg1998, araya1999, schoenherr2007}.

The cross-sections show two important properties: first of all, they are very strongly peaked close to the Landau level energies, implying that the plasma becomes optically thick at these energies. Secondly, because the energy of the electrons is only quantized in the direction perpendicular to the magnetic field, the angle between the magnetic field and the photon becomes relevant. For large angles, the cross-sections become thermally broadened and shift to higher energies for the harmonic levels, as shown in Fig.~\ref{fig:crsf_crosssection}. 

Because the optical depth is so large at the resonant energies for cyclotron scattering, photons with the exact amount of momentum in the direction perpendicular to the magnetic field cannot escape the line forming region, and we can observe absorption like features in the hard X-ray spectrum. These features are referred to as Cyclotron Resonant Scattering Features (CRSFs) or cyclotron lines for short. The CRSF central energy can directly be related to the magnetic field strength in the line forming region via
\begin{equation}
    E_{\mathrm{CRSF},n} = E_{\mathrm{Landau},n} \frac{1}{1+z}  \approx \frac{n}{1+z} 11.6 \times B_{12}\,\mathrm{keV}
    \label{eq:e_crsf}
\end{equation}
where $B_{12}$ is the magnetic field strength in $10^{12}$\,G and $z$ is the gravitational redshift. 
Equation \ref{eq:e_crsf} is commonly known as the ``12-B-12''-rule and works well for different geometries in the case of $B\lesssim 10^{13}$~G.

CRSFs are the only way to directly measure the magnetic field strength close to the surface of a NS. Currently, 36 CRSF are known. For a recent in-depth review about them and their history, see \citet[][and references therein]{Staubert2019}.

Among the open questions in CRSF research is the fact that model calculations \citep{araya1999,schwarm17a,schwarm17b} tend to predict asymmetrical lines, frequently showing "emission wings" at energies below and above the central energy, while observed features tend to be broad and without a marked asymmetry. 

In order for discrete CRSFs to be observable, the sample magnetic field has to be confined to a very narrow range, indicating a closely confined region within the accretion column, possibly a shock region in the column or close to the poles \citep[and references therein]{Becker2005, Becker2007}. Broad and shallow observed CRSF might be caused by multiple line forming regions contributing \citep{Nishimura:2008}. Another possibility is that CRSFs are formed due to reflection of the downwards beamed radiation from the accretion column on the NS surface around the poles \citep{poutanen13a}.

\subsubsection{Luminosity dependence of the CRSF energy}

It is observationally clearly established that the CRSF energy may change as function of luminosity. The sources exhibiting such behaviour can be divided into two groups: the first, where the centroid energy of CRSF is positively correlated with accretion luminosity, and the second, where an anti-correlation is observed. The sources with detected positive correlation tend to be less bright than the sources with negative correlation \citep{becker12a,2015MNRAS.447.1847M}. Different models which are able to explain the observed CRSF energy behaviour have been proposed.

\begin{figure}
	\centering
	\includegraphics[width=\linewidth,angle=0]{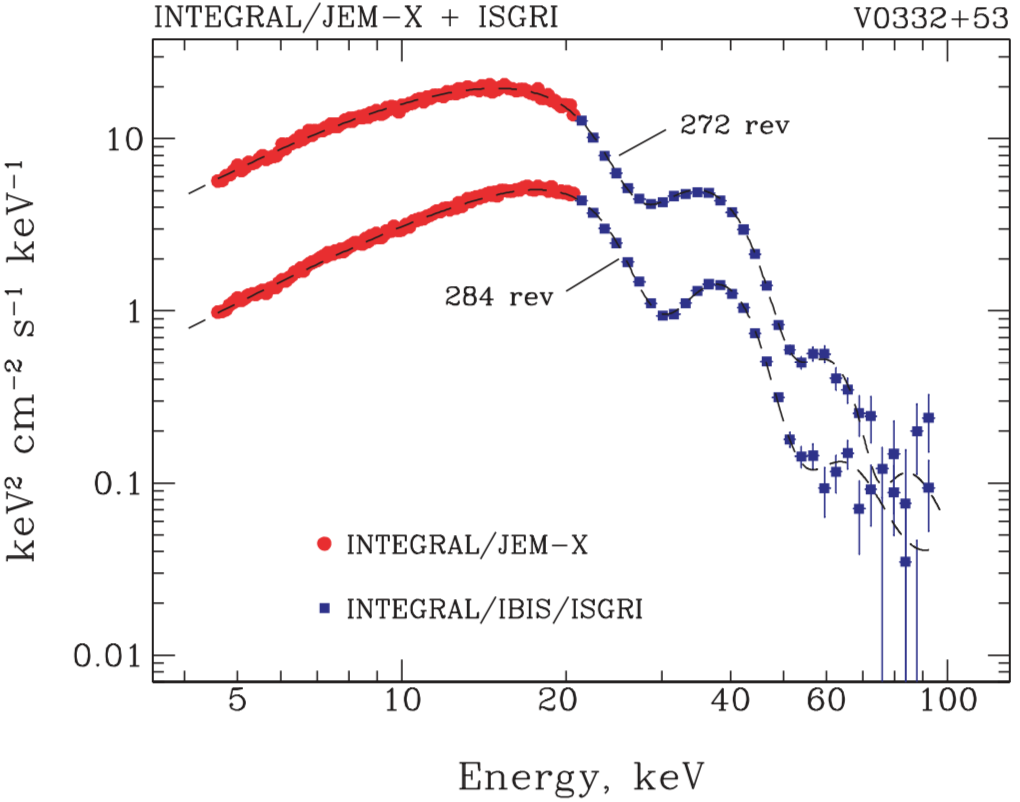}
	\includegraphics[width=\linewidth,angle=0]{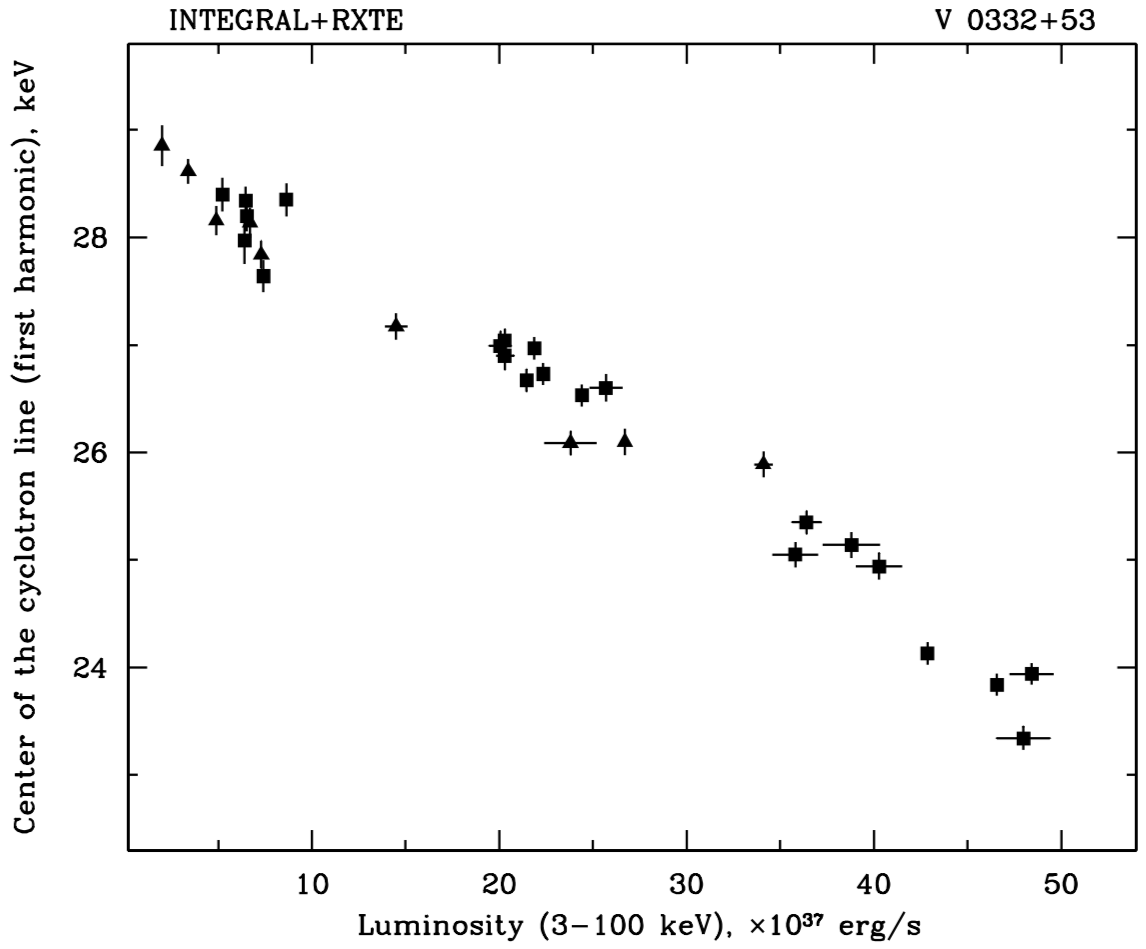}
	\caption{Upper panel: Energy spectra of V0332+53 measured with \intg for two different brightness states, with an immediately visible shift of the cyclotron lines. Lower panel: Dependence of the cyclotron line energy on the source luminosity (3--100 keV);  Triangles and squares mark \intg and \xte results, respectively. (Originally Figures~3 and~4 of \citet{Tsygankov:2006}).}
	\label{fig:cyc_v0332}
\end{figure}

\citet{becker12a} defined three different accretion regimes, depending on luminosity and magnetic field strength. At the lowest luminosities the infalling material is only stopped at the NS surface and the line is formed there. In this case, no change of the line energy with luminosity is expected. At intermediate luminosities a Coulomb-dominated shock is formed in which the line is formed. With increasing luminosity the shock is expected to move closer to the NS surface, sampling higher magnetic field strengths and increasing the observed line energy. At the highest luminosities, above a critical luminosity, the infalling material is decelerated in a radiation dominated shock \citep{1975PASJ...27..311I,Basko1976}. This shock is expected to rise in height with increasing luminosity, decreasing the observed line energy (see Sect.~\ref{Sec:continuum}).

%
In the line formation model of \citet{poutanen13a} a higher luminosity leads to a higher accretion column and a larger fraction of the NS surface being illuminated by this column. A larger area includes regions of a lower magnetic field closer to the NS equator, resulting in a decrease of the observed cyclotron energy with increasing luminosity. 
\citet{Mushtukov:2015c} explain the case of positive correlation for sub-critical luminosities via the Doppler effect of a mildly relativistic falling plasma in the column.

%
%

\subsubsection{Pulse phase dependence of the CRSF profile}

The observed line profiles and energies depend strongly on the angle under which we see the line forming region. Because the rotational axis is typically not aligned with the magnetic field axis (or with our line-of-sight), our viewing angle of the CRSF region changes as function of pulse phase.
The possible changes are very complex and predictions depend strongly on the assumptions about the magnetic field geometry and emission pattern of each column. Energy shifts of the CRSF could for example be explained by different observed relativistic boosting factors of the infalling plasma \citep{fuerst18a}.

In case of a perfectly symmetric accretion geometry, with two opposed accretion columns, a lot of the expected variability as function of pulse phase is considerably suppressed \citep{falknerPhD}. Because of the relativistic light-bending, at most phases both columns are visible and their flux variations as function of the viewing angle almost cancel each other. 

The beaming function, and therefore flux and CRSF profile variations with the pulse phase, can be strongly affected by the material moving from the accretion disc to the NS surface. The influence of magnetospheric accretion flow is expected to be stronger in the case of super-critical accretion, when the gravitationally lensed flux from the accretion columns passes the regions of relatively high density of magnetospheric flow \citep{2018MNRAS.474.5425M}.

\subsubsection{Secular changes of the CRSF energy}

Because accretion is a dynamic process during which the NS gains mass, it is also expected that the accretion geometry might slowly change over time. Because the CRSF is so sensitive to the magnetic field, we might expect a secular change of the CRSF energy, for example if the magnetic field slowly decays. However, expected decay times of magnetic fields are of the order of $10^6$ years \citep[e.g.,][]{bhattacharya92a}, much longer than the history of CRSF science. However, other effects, like screening or burying of the surface magnetic field due to accumulation of the accreted material might occur, but a clear theoretical picture has so far not emerged \citep[see also the discussion in ][]{staubert14a}.



\subsection{X-rays from black hole HMXB systems}
\label{Sec:BH}

The \intg view on Galactic black-hole (BH) binaries is discussed elsewhere in a dedicated chapter, and we refer the reader to this for a more detailed discussion (Motta et al., this volume). Most of these sources are in low-mass X-ray binaries \citep[e.g.,][]{McClintock_Remillard_2006a_book}. Only a few confirmed BH HMXBs are known, but among them is one of the most prominent X-ray sources in the sky, Cygnus X-1, that has been observed by \intg for over 11\,Ms of dead-time corrected exposure \citep{Cangemi_2019a}. Although thought to be wind-accretors, BH HMXBs usually exhibit a stable accretion disc.

Observationally, BHs in XRBs can be found in two main states: \textsl{hard state}, where the total energy output is driven by a hard power law component above 10 keV with an exponential cutoff at a few hundred keV, and \textsl{soft state}, where soft thermal emission from a $\sim$1\,keV accretion disc dominates and a weak, very steep power law component with a cutoff above $\sim$500\,keV may be present. Changing between the states, the source evolves through the hard/soft intermediate state that shows intermediate spectral characteristics \citep[e.g.][]{Fender_2004a,Belloni_2010a}. Radio emission is detected in the hard state but it is strongly suppressed or absent in the soft one; radio jets can be resolved in several sources and radio flares are often associated with state transitions. The states further show distinct X-ray timing characteristics \citep[e.g.][]{Belloni_2010a}. BH X-ray binary states correspond to different accretion/ejection geometries, but in particular the exact origin of the Comptonized hard emission is still controversial \citep{Nowak_2011a,Zdziarski_2014a}. Additionally, sources sometimes show an excess beyond the hard cutoff, the so-called hard tail.

Cygnus X-1 is a key source to understanding BH X-ray binaries: as a HMXB, it is a persistent accretor and thus easy to observe. Additionally, it often transits between the states, crossing the so-called `jet line' where we expect the strongest changes in accretion geometry to take place \citep{Grinberg_2013a}. \intg's unique capabilities at highest energies have contributed in several ways to a better understanding of this system: in particular, the hard tail above 400\,keV in Cygnus X-1 has been shown to be polarized \citep{Laurent_2011a,Jourdain_2012a,Rodriguez_2015a}, hinting at a jet origin for this component. For the spectrum of the hard tail, see also \citet{Walter+Xu:2017}. \citet{Cabanac_2011a} analyzed power spectra and time lags up to $\sim$130\,keV, for the first time assessing the energy-dependence of variability properties at such high energies, giving strict constrains on models that try to reproduce properties of hard X-ray emission.

\subsection{Gamma-ray binaries}
\label{Sec:gamma}

The population of Galactic X-ray sources above 2\,keV is dominated by the X-ray binaries, see e.g. \citet{Grimm2002}. At gamma-ray energies, however, the situation is drastically different. While current Cherenkov telescopes have detected around 80 Galactic sources (see the TeVCat catalogue at \texttt{http://tevcat2.uchicago.edu/}), less than 10 binary systems are regularly observed at TeV energies as non-extended gamma-ray sources \citep{dubus_review13, grlb_cta_ch19}. The properties of PSR B1259$-$63, LS 5039, \lsi, HESS J0632+057 and 1FGL J1018.6$-$5856 are reviewed in \citet{dubus_review13}. Since 2013, three more Galactic binaries, PSR J2032+4127 \citep{PSRJ2032_ATel_TeV}, HESS J1832$-$093 \citep{2016MNRAS.457.1753E,Marti-Devesa+Reimer:2020}, 4FGL J1405.1$-$6119 \citep{Corbet19_4fgl}, and one extragalactic -- LMC P3 \citep{Corbet16:LMCP3}, have been discovered at TeV energies, but still the number of binaries observed in the TeV sky is extremely small; the reason why these systems are able to accelerate particles so efficiently is not known yet. These systems are called gamma-ray binaries as the peak of their spectral energy distribution lies in the gamma-ray range above 1~MeV, sometimes in the GeV--TeV range.

All gamma-ray binaries host compact objects orbiting around massive young stars of O or Be spectral type. This leads to the suggestion that the observed gamma-ray emission is produced as the result of interactions between the relativistic outflow from the compact object and the non-relativistic wind and/or radiation field of the massive companion star. However, neither the nature of the compact object (BH or NS?) nor the geometry (isotropic or anisotropic?) of the relativistic wind from the compact object are fully understood. Only in PSR~B1259$-$63 and PSR~J2032+4127, is the compact object  known to be a young rotation-powered pulsar which produces a relativistic pulsar wind. The interaction of the pulsar wind with the wind of the Be star leads to the observed high energy emission, and in particular to the huge GeV flare observed in PSR B1259$-$63 \citep{Abdo2011,Chernyakova2015,Caliandro:2015,Johnson:2018}.  

In all other cases the source of the high-energy activity of gamma-ray binaries is uncertain. It can be either dissipation of rotation energy of the compact object \citep[e.g.][]{Dubus:2006,Sierpowska-Bartosik+Torres:2008,Torres:2012}, or emission from a jet \citep[e.g.][]{micro1, Zimmermann12}. In these other systems the orbital period is much shorter than in PSR B1259$-$63 and PSR J2032+4127, and the compact object spends most of the time inside the dense wind of the companion star. The optical depth of the wind due to free-free absorption is high enough to suppress most of the radio emission within the orbit, including the pulsed signal of the rotating NS \citep{zdz10}, hampering a direct detection of the possible pulsar.
Super-orbital variability has also been found in at least one of these sources, e.g., see the GeV detection of \lsi in \citet{Ackerman2013lsi}; we refer more about this below.

\citet{Massi2017} tried to deduce the nature of the compact source in \lsi by studying the relation between the X-ray luminosity and the photon index of its X-ray spectrum. It turned out that existing X-ray observations of the system follow the same anti-correlation trend as BH X-ray binaries. 
Under the hypothesis of a microquasar nature for \lsi, they were able to explain the observed radio morphology \citep{micro1} and interpret the observed superorbital period as a beat frequency between orbital and jet-precession periods \citep{Massi2016}. This is further supported by the presence of 55-minute and 2-hour long quasi-periodic oscillations in radio and X-rays, respectively, which are stable over a few days \citep{LSI_qpo_2018}. Conversely, \citet{zdz10} showed that the model in which the compact source is a pulsar allows a natural explanation of the keV\,--\,TeV spectrum of \lsi. These authors argued that the radio source has a complex, varying
morphology, and jet emission is unlikely to dominate the spectrum through the whole orbit. Within this model, the superorbital period of the source is explained as the timescale of the gradual build-up and decay of the disc of the Be star. \citet{Li:2012a} and \citet{2012ApJ...747L..29C} demonstrated the presence of superorbital variability in X-rays, the latter publication showed that a constant time delay between the drifting orbital phases of X-ray and radio flares could be naturally explained if one takes into account the time needed for electrons to reach regions transparent to radio emission. 

Cyclical variations in the mass-loss of the Be star are supported by optical observations confirming the superorbital variability of the Be-star disc \citep{fortuny15}; this is also an explanation for superorbital variability in an alternative flip-flop model  \citep{zamanov01,Torres:2012,Papitto:2012,Ahnen:2016}. This model assumes the compact object in \lsi to be a magnetar and implies a change from a propeller regime at periastron to an ejector regime at apastron. During the periastron the pressure of matter from the Be star outflow compresses and disrupts the magnetosphere of the NS, which leads to the disappearing of the pulsar wind. In this case electrons are accelerated at the propeller shock, which accelerates electrons to lower energies than the inter-wind shock produced by the interaction of a rotationally powered pulsar and the stellar wind of the Be star. A magnetar-like short burst caught from the source supports the flip-flop model and the identification of the compact object in \lsi\ with a NS \citep{Barthelmy2008, Burrows12, Barthelmy2019}. However, RXTE observations of \lsi demonstrated the presence of a few, several second long, flares \citep{2009ApJ...693.1621S}, which were compared by those authors to the flares typically found in the accretion-driven sources; see also \citet{Li:2011a} for a further analysis covering a wider range of orbital cycles. In principle, neither BAT nor RXTE observations can exclude the possibility that the observed flares are coming from another source located close to the line-of-sight \citep{2009ApJ...693.1621S}, although similar observations of instruments with better spatial resolution indicate that it is likely they are from the gamma-ray source \citep{Paredes2007,Rea2010}.
  
Other possible scenarios for the super-orbital modulation in \lsi are related to the precession of the Be star disc \citep{Saha_2016}, or to a non-axisymmetric structure rotating with a period of 1667 days \citep{Xing2017}.

During its orbital motion around the optical companion, the environment of the compact source changes a lot from periastron to apastron. This leads to the observed spectral variability on very different time scales, from hours to the orbital and superorbital periodicities. 
The typical X-ray flux from gamma-ray binaries is at the level of few mCrabs, so it is not possible to study with \intg the spectral variability on short time scales (few hours). Still, \intg is sensitive enough to study the properties of gamma-ray binaries on longer time scales. 

\subsection{Ultraluminous X-ray sources}
\label{Sec:ULX}
Ultraluminous X-ray sources (ULXs) have been defined as a class of extragalactic point-like objects, outside the nucleus of their respective galaxies and with a luminosity exceeding the Eddington limit for a 10~\Msol black hole. Originally often thought to be intermediate-mass black holes with masses $>100$\Msol, further studies rather indicated ``stellar mass'' compact objects accreting at super-Eddington rates for at least most ULXs \citep[e.g.,][]{Sazonov:2014}. The discovery of X-ray pulsations from the ULX M82 X-2 \citep{Bachetti:2014Nature} demonstrated that ULXs can host accreting neutron stars and further examples have been found subsequently. 
Due to their distance, the nature of the mass donor is not well determined for most ULXs and only for NGC~7793 P13 \citep{ATel5552} this has been clearly identified as a high mass star \citep{Motch:2011}. A few more pulsating ULX have been tentatively identified as HMXBs, but, for example, in the case of NGC~300 ULX1, \citet{Heida:2019} reclassified the mass donor as red supergiant.
As noted also in \citet{Bachetti:2014Nature}, very luminous outbursts of BeXRB systems can also reach super-Eddington luminosities and different Be transients like SMC~X-3 \citep{2018MNRAS.479L.134T} or the recently found transient Swift J0243.6+6124 \citep{ATel10809,ATel10812,Wilson-Hodge:2018} are now being labeled as ultraluminous sources, demonstrating a continuum of behaviour from classical accretors to ULXs \citep{2015MNRAS.447.1847M,Kaaret+Feng+Roberts:2017,Grebenev:2017}.

\section{\intg's role in HMXB studies}
\label{Sec:HMXB-INT}

\subsection{Persistent wind-accreting Supergiant HMXBs}
\label{Sec:classic}

Persistent wind-accreting supergiant HMXBS (sgHMXBs) are systems with an early type (O or B-star) supergiant companion, losing large amounts of mass through a stellar wind. The compact objects in these systems accrete from this dense wind, and, while they tend to show a large variability (up to a factor of 100), they are always active and some were among the first HMXBs discovered. \intg contributed to the knowledge about this source class by the merit of long observations of fields containing these sources -- frequently not directly targeting the sources themselves. \intg data in the hard X-ray band have also been important to disentangle intrinsic variations of the X-ray source flux from the effects of absorption when studying X-ray variability. For eclipsing systems, the accumulated long-term data has allowed to refine eclipse parameters and thus derive new constraints on the masses of the binary companions, as detailed in Section~\ref{Sec:Masses}. In the following, we give a few specific examples of \intg results for this source class.

\begin{figure}[htb]
\begin{center}
\includegraphics[width=0.99\linewidth]{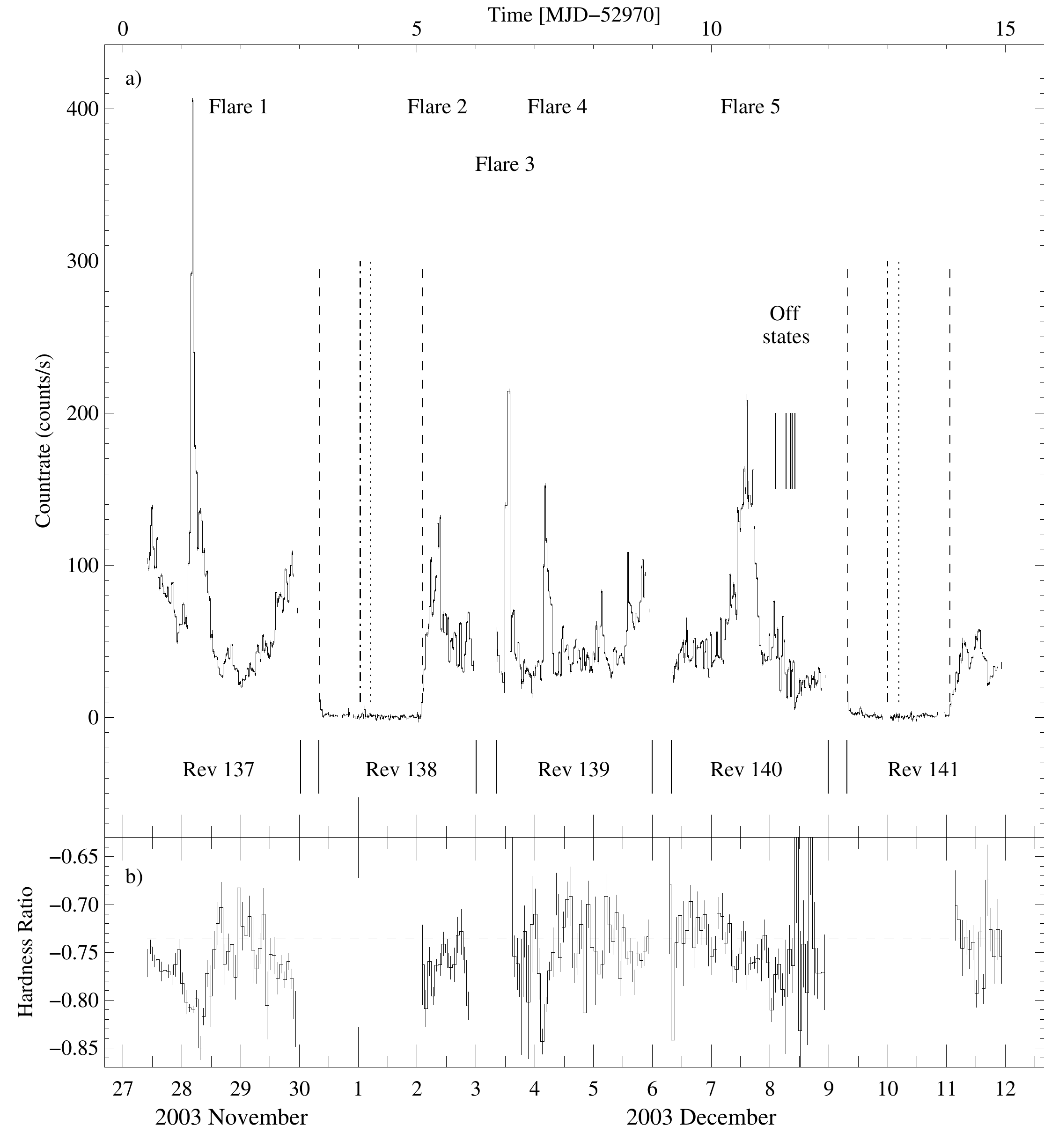}
\caption{Variability of Vela X-1 for the complete Vela region observation from Revolution 137 to 141. a) ISGRI 20--40 keV light curve (time resolution 1 SCW, i.e. $\sim$1800 s) and b) 20--30 keV vs. 40--60 keV hardness ratio. 
Short vertical lines below the X-ray light curve show \intg's perigee passages, during which the instruments are switched off. The long dashed vertical lines show the eclipse ingress and egress times. The dotted vertical line indicates the derived eclipse center, while the dash-dotted line indicates the time of mean longitude T90. Originally Fig.~2 in \citet{kreykenbohm08a}.}
\label{fig:velax1lc}
\end{center}
\end{figure}

The \sghmxb \vela is among the best studied objects in the X-ray sky and often taken as a prototype supergiant X-ray binary in order to study the physics of HMXB or as baseline case for modeling and simulation efforts \citep[see][for an overview]{Kretschmar:2019INT}. 
Long-term X-ray monitor data show on average a clear orbital profile \citep{Fuerst:2010,Falanga2015}, driven by the mean absorption in the dense material present in the system, especially in the accretion and photoionization wakes \citep[and references therein]{grinberg17a}.
Erratic flux variations on timescales from days to minutes have been reported since early deep observations of the system \citep[e.g.,][]{Forman:73,WatsonGriffiths:77,Ogelman:77}. During an extended observation of the Vela region for five consecutive \intg revolutions in November/December 2003 (Fig.~\ref{fig:velax1lc}), covering almost two orbital periods of Vela~X-1, \intg found especially intense flaring, as well as off-states, which during the flux dropped below the detection limit of \intg for 1--2 rotations of the neutron star \citep{Staubert:2004IWS5,kreykenbohm08a}. \citet{Fuerst:2010} found that the \intg flux distribution closely followed a log-normal distribution.   
Early studies of the accretion flow \citep{TaamFryxell88} identified strong time-dependent behaviour as well as indications of a highly asymmetric flow. Other studies \citep[starting with][]{Blondin+90} revealed indeed highly asymmetric structures caused by the photoionization and accretion wakes. Pushing these studies further, \citet{manousakis15a,Manousakis15b} found time-dependent holes in the simulated mass flow, which may explain the off-states observed by \citet{kreykenbohm08a} and others, without requiring intrinsic clumpiness of the wind.

\begin{figure}[ht!]
\begin{center}
\includegraphics[width=0.99\linewidth]{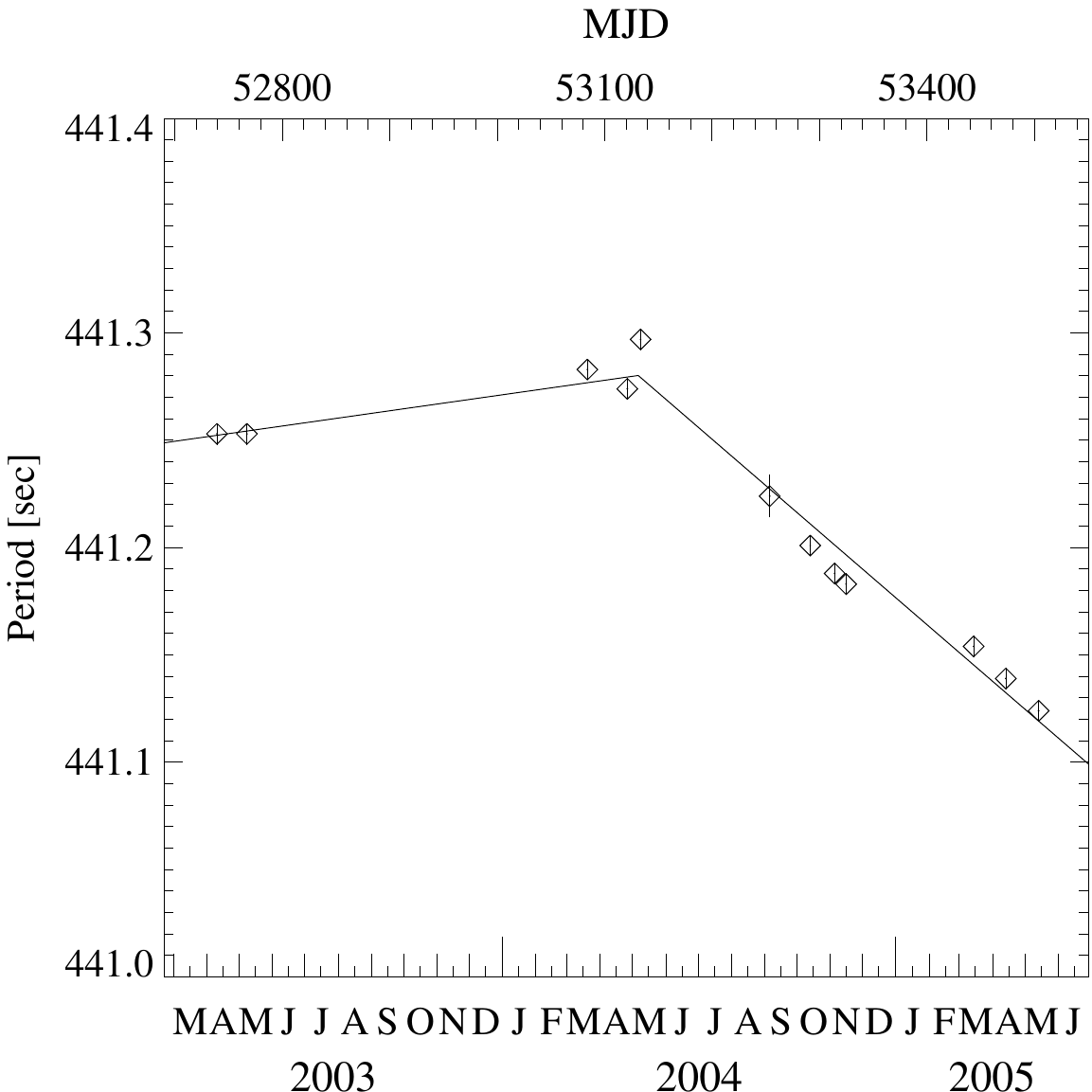}
\caption{A torque reversal of 4U~1907+09 found in \intg observations. Originally Fig.~6 in \citet{Fritz:2006}.}
\label{Fig:4U1907}
\end{center}
\end{figure}

4U~1907+09 is a less studied source, a wind-accreting \sghmxb on a moderately eccentric, close orbit. \intg observations \citep{Fritz:2006} found a clear spin-down after almost 20 years of constant spin-up and confirmed CRSFs at $\sim$19 and $\sim$40~keV, consistent with earlier results (Figure~\ref{Fig:4U1907}). The spin-down trend reversed again later \citep{Sahiner:2012}. Long-term spin period change trends like in this source contrast with the ``random walk'' changes in pulse period observed, e.g., in \vela. This source also shows dips or off-states. Using a \suzaku observation, \citet{Doroshenko:2012} found that the source continues to pulsate in the off-state and that the transitions may be explained by ``gated accretion'' (see Section~\ref{Sec:inhibition-wind}), which might make 4U~1907+09 an interim source between SFXTs and ordinary accreting pulsars.

The somewhat unusual system 2S~0114+650 harbours one of the slowest spinning pulsars ($\Ps\approx 2.65$\,h) in a close orbit ($\Pb\approx 11.6$\,d) around a supergiant star. \intg observations \citep{Bonning+Falanga:2005,Wang:2011} have confirmed the previously observed long-term spin-up trend with shorter variations. Noting stronger short-term variations and super-orbital variations, \citet{Hu:2017} propose the formation of a transient disc (see also Section~\ref{Sec:RLOF}), while \citet{Sanjurjo-Ferrin:2017} consider quasi-spherical settling accretion (see Section~\ref{Sec:magnetosphere}) on a magnetar to explain the observed behaviour.

\subsection{Highly absorbed HMXBs}
\label{Sec:absorbed}

\begin{figure}[htb!]
\begin{center}
\includegraphics[width=0.99\linewidth]{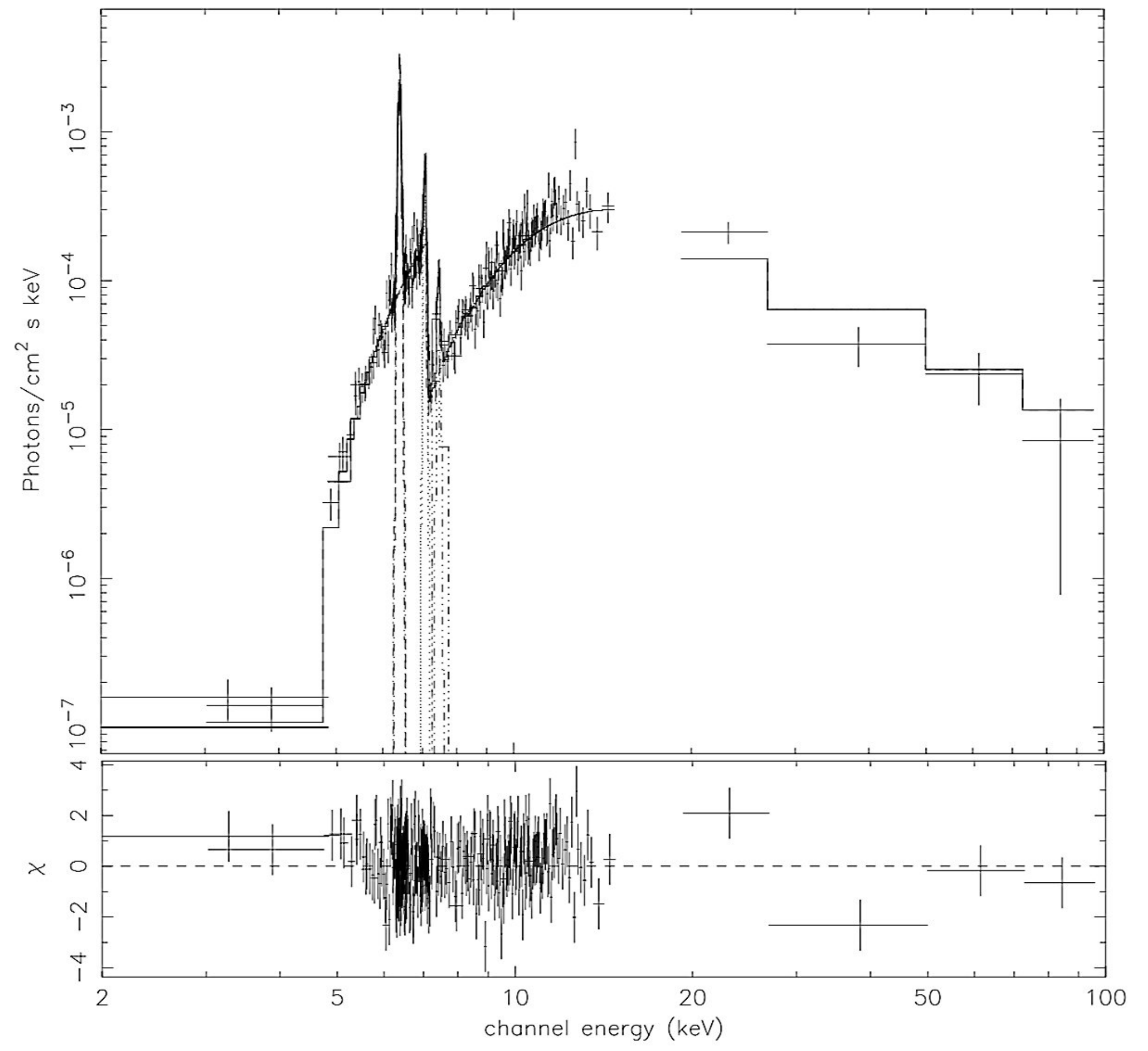}
\caption{XMM-Newton and ISGRI unfolded photon spectra of IGR~J16318--4848, the first new source detected by \intg and an extremely absorbed HMXB. From \citet{Walter03}.}
\label{Fig:IGRJ16318}
\end{center}
\end{figure}

Highly absorbed systems do not define a class themselves: the physics and the nature of the sources are clearly the same as non-absorbed ones. 
It is, however, worth pointing out that due to its unprecedented coverage, and, at that time, the best angular resolution at hard X-rays, we could hope that \intg could see sources otherwise undetected at lower energies, for various reasons: sensitivity, hard spectra, confusion, high absorption, etc.
On 2003 January 29$^{th}$, during the first Galactic Plane Scan of the Norma region (after a few weeks spent on the other side of the Galaxy in the Cygnus region), \intg detected its first such system, and one of the most extreme of all \intg sources (IGRs): IGR J16318$-$4848. This object is indeed the most absorbed HMXB (10 times higher than previously known absorbed systems such as 4U~1700$-$37 or GX~301$-$2) with $N_\mathrm{H}$ in excess of $10^{24}$~cm$^{-2}$ \citep{Walter03}.
This results in a featureless, continuumless spectrum below 4\,--\,5 keV, huge Fe K$_\alpha$, K$_\beta$ and Ni K$_\beta$ lines in the soft X-ray spectrum \citep{Matt03, Walter03}, and variable hard X-ray emission \citep{Barragan:2010IWS8}, see Figure~\ref{Fig:IGRJ16318}. Multi-wavelength follow-ups (mainly from near-to-mid infra-red spectroscopy) have shown the companion to be a peculiar supergiant (a so-called sgB[e]), while the system is enshrouded in a dense cocoon \citep{Filliatre04}. This material also shows up in the X-rays, where the strongly absorbed spectrum can be self-consistently explained with a combination of gas and dust absorber \citep{Ballhausen:2020}. Later observations with the ESO-VLT VISIR instrument suggested that the compact object (of unknown nature) is orbiting within or behind the rim of a torus of matter encircling the supergiant star \citep{Chaty2012}. Recently, \cite{Fortin2020} presented new ESO-VLT X-shooter broad-band spectroscopic observations from optical to near infra-red of this source, and compared it with models made with the POWR code for atmospheres of massive stars.

It was still surprising that the following IGRs (e.g. IGRT~J16320$-$4751, IGR~J16393$-$4643, IGR~J17252$-$3616, IGR~J18325$-$0756) also were   
quite absorbed objects with $N_\mathrm{H}\sim 10^{23}$ cm$^{-2}$ \citep{Rodriguez03,Bodaghee06,Zurita06,walter06,Tomsick08}, more in line, however, with older sources. The strength of the absorption can vary as function of the orbital phase. \citet{Garcia2018} followed the spectral evolution and changes in column density of IGR~J16320$-$4751 along its orbit with a series of \xmm observations, describing the changes in a simple geometrical model.

\subsubsection{Infra-red identification of highly absorbed IGR sources}
\label{Sec:IR-IGR}

After the highly absorbed archetype IGR~J16318$-$4848, \intg found some more sources with similar X-ray spectral properties. Some of these systems seem to be persistent sources, such as IGR~J16320$-$4751 \citep{Rodriguez03}, and some of them are clearly transient, such as IGR~J16358$-$4726 \citep{patel04} and IGR~J16465$-$4507 \citep{Lutovinov2005b}. \citet{kuulkers05} reviewed the X-ray and optical and infra-red (IR) properties of highly absorbed IGR sources and suggested that these are HMXBs containing either a NS or a BH in orbit around a (super)giant donor.
Here we will concentrate on this class of sources and the IR analysis to identify and characterize the counterparts. 

Before \intg, the small number of \sghmxbs was explained by the short-lived supergiant phase. The distribution of this kind of systems was reproduced well by population synthesis models. However, the \intg survey of the Galactic plane and central regions has revealed the existence of more than 900 sources in the energy range 17--200 keV \citep{bird16}, with a location accuracy of 0.5$^\prime$--4$^\prime$, depending on count rate, position in the field of view and exposure. The observing strategy of \intg allowed the detection of new kinds of sources that had been missed in the past due to the very short transient nature (see Section~\ref{Sec:SFXT}) or the very high absorption. A large fraction of these newly discovered sources belongs to the \sghmxb class, resulting in  a substantial increase of its members. Often the highly absorbed sources are not accessible in the optical band due to the high interstellar extinction or would require extremely long exposure times on large optical telescopes.

\begin{figure*}[htb]
  \centering
  \includegraphics[angle=0,width=1.06\columnwidth]{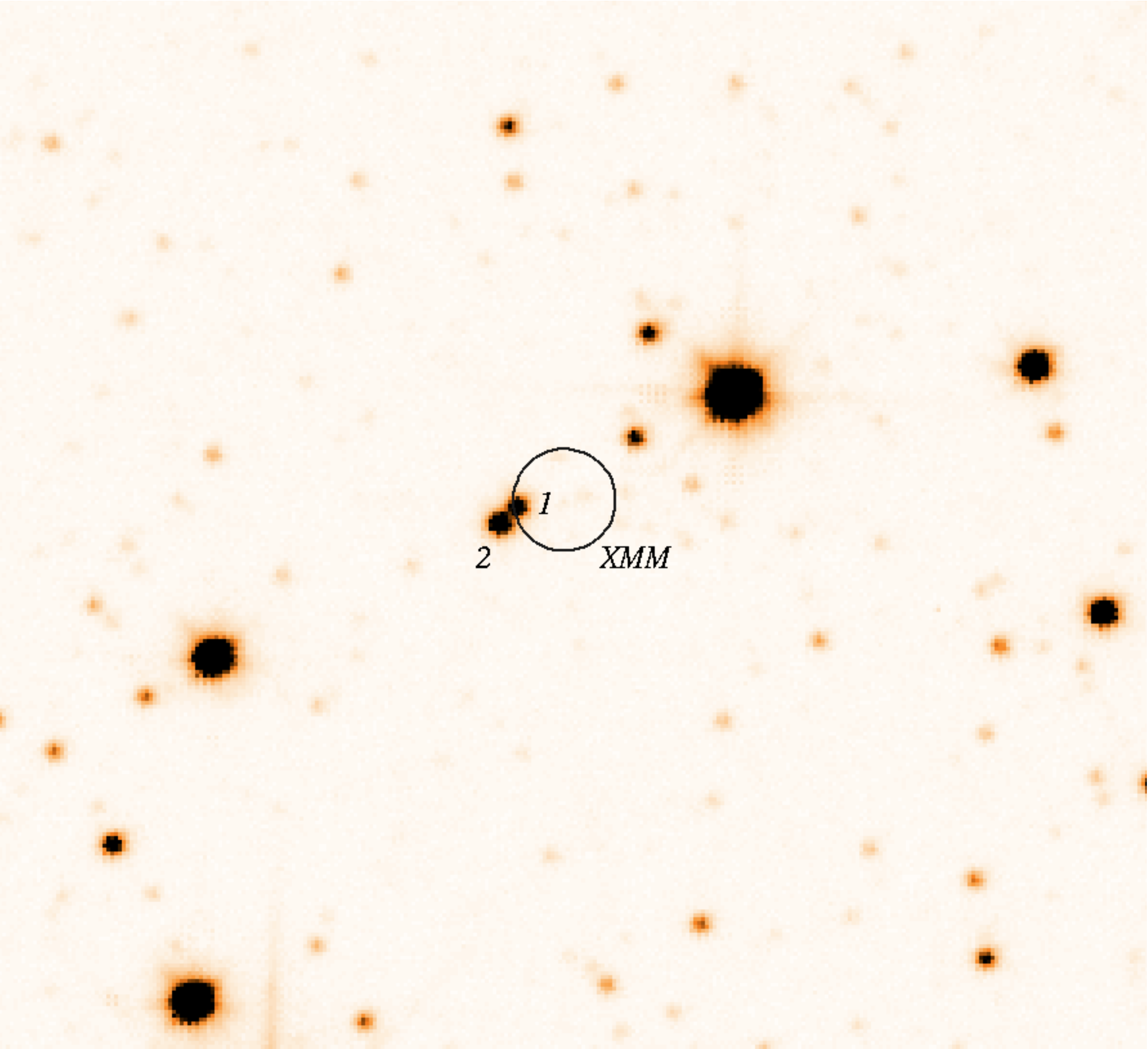}
  \includegraphics[angle=0,width=0.95\columnwidth]{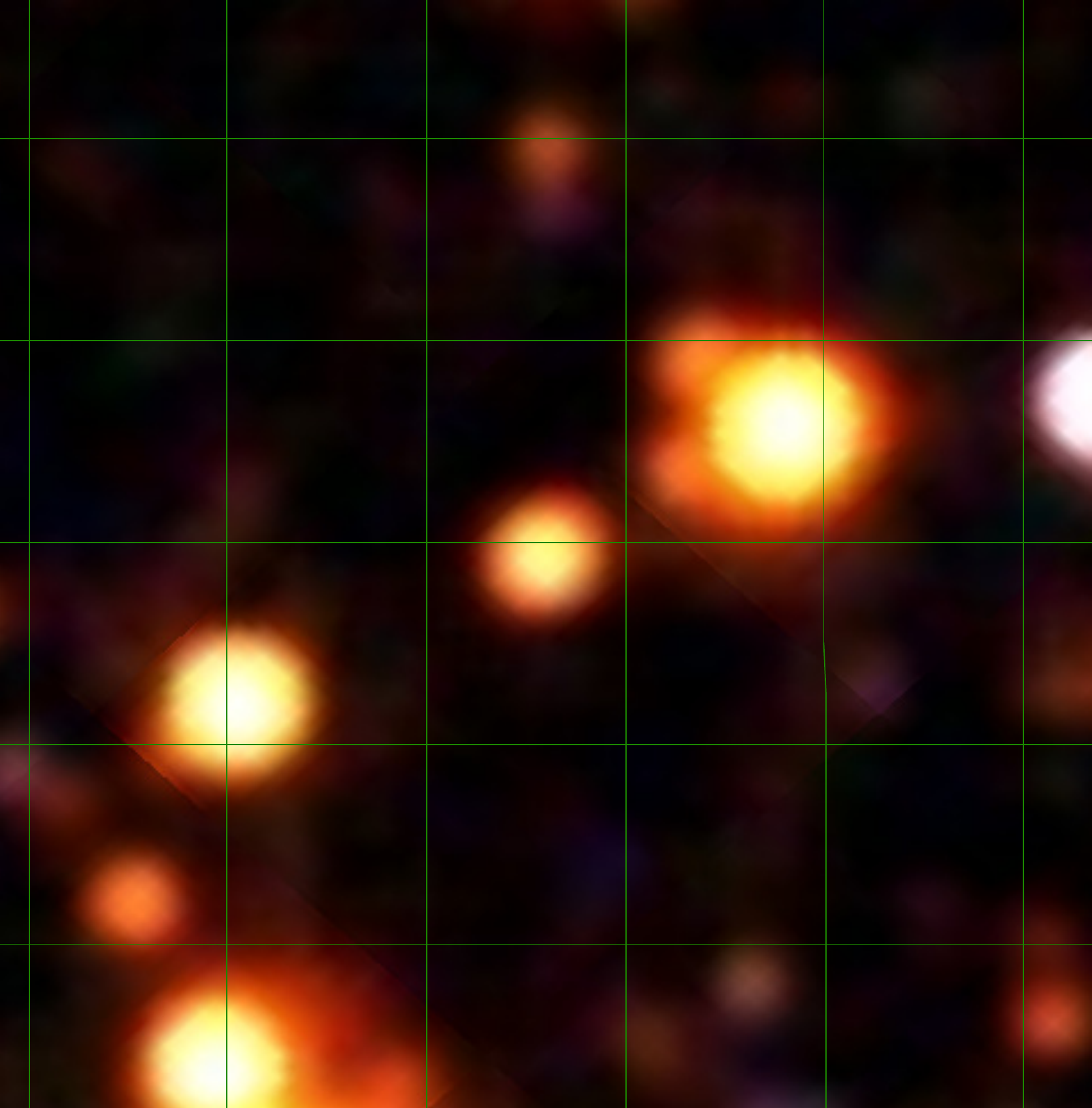}

\caption{\emph{Left panel}: 1.2$^{\prime}$ $\times$1.0$^{\prime}$ \emph{K} finding chart for 2XMM J191043.47$+$091629.4 obtained with the UKIDSS.
The black circle is centred on the \emph{XMM-Newton} position of 2XMM J191043.47$+$091629.4, with the radius indicating the 2.13$^{\prime\prime}$ positional error. 
\emph{Right panel}: 50$^{\prime\prime} \times$50$^{\prime\prime}$ 2MASS coloured map. The images are displayed with north up and east to the left. We note that the two NIR UKIDSS sources appear unresolved in the 2MASS image.}
  \label{fig:IRidentification}
\end{figure*}

Under these circumstances, infrared spectroscopy is an alternative tool to characterize these systems in a multiwavelength context. The detection and study of counterparts in the infrared is possible with the IR instrumentation on a 4-m class telescope. Combining it with available IR photometry and the X-ray properties, the nature of the binary system can be established unambiguously. Follow-up X-ray observations by \xmm \, \citep[e.g.][]{Bodaghee06,Rodriguez06,Bozzo12b}, \chandra \citep[e.g.][]{Tomsick08,Paizis11,Nowak12} or \swift \citep[e.g.][]{Kennea05,Rodriguez09,Pavan11} lead to the reduction of the error circle to a few arcseconds and, consequently, the correct identification of the IR counterpart \citep[e.g.][]{Chaty2008}.
 
To establish or constrain the nature of the companions through IR spectroscopy, IR atlases of \citet{hanson96,hanson98,hanson05} are used for comparing the spectral features present in the IR spectrum.
The IR counterpart of IGR~J19140+0951, 2MASS\,J19140422+0952577 \citep{zand06}, was classified as a B0.5-1 supergiant  \citep{hannikainen07,nespoli08}. Nevertheless,
\citet{Torrejon:2010} showed that the 2MASS companion was formed by two stars. They performed the photometry on images of higher spatial resolution that allowed them to clearly separate both stars. Using their photometry and the spectral type B0.5 Ia, they estimated a distance of 3.6 kpc.
The companion of IGR~J16207$-$5129, 2MASS\,J16204627$-$5130060 \citep{tomsick06}, was classified as a B1 supergiant from
its IR spectrum \citep{nespoli08}, confirming the spectral type derived from optical observations \citep{negueruela07}. \suzaku observations confirmed that this object belongs to the class of absorbed HMXB \citep{bodaghee10}. \citet{pellizza11} obtained optical and IR observations of the field of IGR J16283$-$4838 \citep{rodriguez05} in order to unveil the nature of its counterpart (2MASS\,J16281083$-$4838560). They demonstrated that this source is a highly absorbed HMXB with a blue supergiant companion.

\citet{Torrejon:2010} also investigated the nature of the counterpart to IGR~J18027$-$2016, a HMXB candidate identified as 2MASS\,J16204627$-$5130060 \citep{masetti08}, by combining IR spectra in the $I$, $J$, $H$ and $K$ bands with $JHK$ photometry. They concluded that this IGR source is a persistent X-ray source with a  B1\,Ib companion, i.e. a highly absorbed \sghmxb system \citep[see also][]{pradhan19,aftab16}.

Although \citet{walter06} did not use IR spectroscopy, they studied 10 IGR sources (8 persistent and 2 transient systems), obtained follow-up observations with \xmm and proposed IR counterparts from existing catalogues. They confirmed or demonstrated that 8 out of the 10 sources are intrinsically absorbed and 7 of them are persistent sources. Moreover, they suggested that the companions of these persistent systems are very likely supergiants.

AX~J1910.7+0917, discovered by ASCA \citep{sugizaki01}, was also detected by \intg in the hard X-ray band. \citet{Pavan11} analysed all archival \intg, ASCA, \xmm and \chandra observations around the position of this object. These authors associated the IR counterpart of the source with 2MASS\,J19104360+0916291, but could not firmly establish the nature of the system. However, using IR spectroscopy and new photometric analysis provided by UKIDSS, \citet{rodes13} were able to resolve the 2MASS candidate into two different components (see Fig.\ref{fig:IRidentification}). Their conclusion was that the companion of this source is most likely an early B supergiant located at a distance of $\sim$16.0 kpc, placing it in the Outer arm. This system would also belong to the class of obscured HMXBs containing the slowest pulsar found to date \citep{2017MNRAS.469.3056S}.

Overall,
\intg has increased the number of \sghmxb systems dramatically. The proportion of confirmed \sghmxb related to HMXBs is now around 42\%, and so, almost ten times higher than before \intg \citep{Coleiro2013b}.

\subsection{Masses of eclipsing high-mass X-ray binaries}
\label{Sec:Masses}

\begin{figure}
	\centering
	\includegraphics[width=\linewidth]{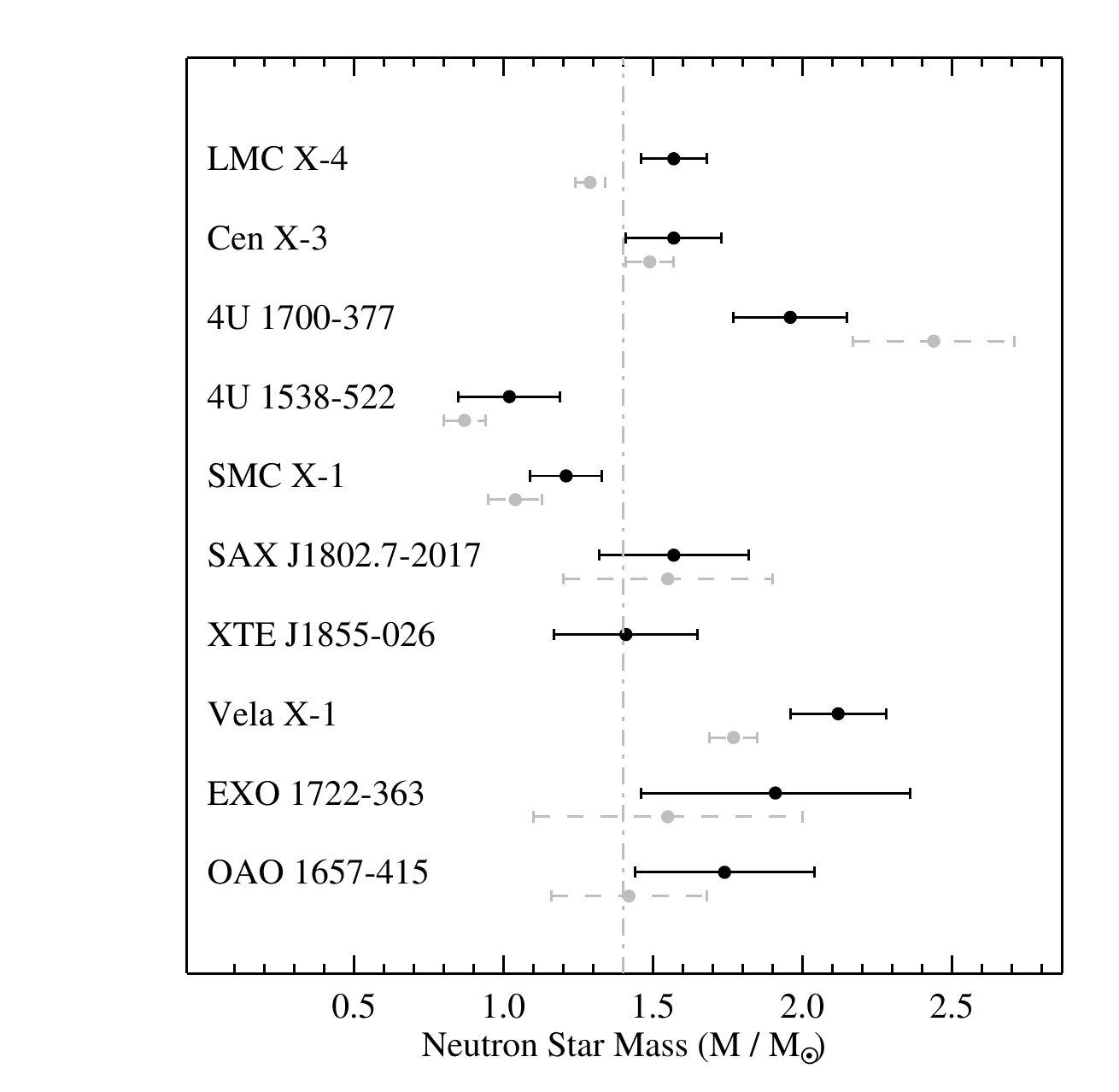}
	\caption{Masses of the ten eclipsing HMXBs. The NS masses determined using 10 years \intg data are shown with solid lines. Values from the literature are represented with dashed lines. The error bars correspond to uncertainties at $1 \sigma$ c.l. The dashed vertical line indicates the canonical NS mass of 1.4 M$_{\odot}$ \citep[see][for more details]{Falanga2015}. }
	\label{fig:masses}
\end{figure}

Only in a few sources among more than one hundred HMXBs is the inclination of the system high enough for the compact star to be periodically occulted along our line of sight by the companion, giving rise to X-ray eclipses. For these eclipsing HMXB systems, LMC~X-4, Cen~X-3, 4U~1700$-$377, 4U~1538$-$522, SMC~X-1, IGR~J18027$-$2016, Vela~X-1, IGR~J17252$-$3616, XTE~J1855$-$026, and OAO~1657$-$415, with the help of more than ten years of monitoring with the \intg and \xte/ASM, the ephemeris, including the duration of the eclipse, orbital decay, and for eccentric systems the angle of periastron and the apsidal advance has been derived \citep{Falanga2015}. Additionally, updated values for the masses of the NS hosted in these ten HMXBs were also provided, as well as the long-term light curves folded on the best determined orbital parameters of the sources. The energy-dependent profile of the X-ray light curve during the eclipse ingress and egress also reveals details of the OB stellar wind structure \citep[see e.g.,][]{White1995}. These light curves reveal complex eclipse ingresses and egresses that are understood mostly as being caused by accretion wakes. These results constitute a database to be used for population and evolutionary studies of HMXBs and for theoretical modeling of long-term accretion in wind-fed X-ray binaries.

Determining the equation of state (EoS) of matter at densities comparable to those inside NSs is one of the most challenging problems of modern physics and can only be addressed based on observations of astrophysical sources. Models proposed in the past years can be tested against observational results, especially by evaluating the highest NS mass that each EoS model is able to sustain \citep[see e.g.,][]{Lattimer2001}. Very soft EoSs predict highest NS masses in the 1.4--1.5\,$M_{\odot}$ range (this occurs when the NS core is made of exotic matter such as kaons, hyperons, and pions), whereas stiff EoSs can reach up to 2.4--2.5\,$M_{\odot}$. More massive NSs can thus provide stronger constraints on the EoS models. As discussed by \citet{Rappaport1983}, eclipsing HMXBs hosting X-ray pulsars provide a means to measure the NS mass and thus place constraints on their EoS. \intg data at high energy band measured semi-eclipse angle smaller than the values reported in the literature, and thus for the ten eclipsing HMXBs NS masses we estimated are generally higher, see Figure~\ref{fig:masses}.

\subsection{Supergiant Fast X-ray Transients}
\label{Sec:SFXT}

\begin{figure*}[htb]
\begin{center}
\includegraphics[width=0.95\textwidth]{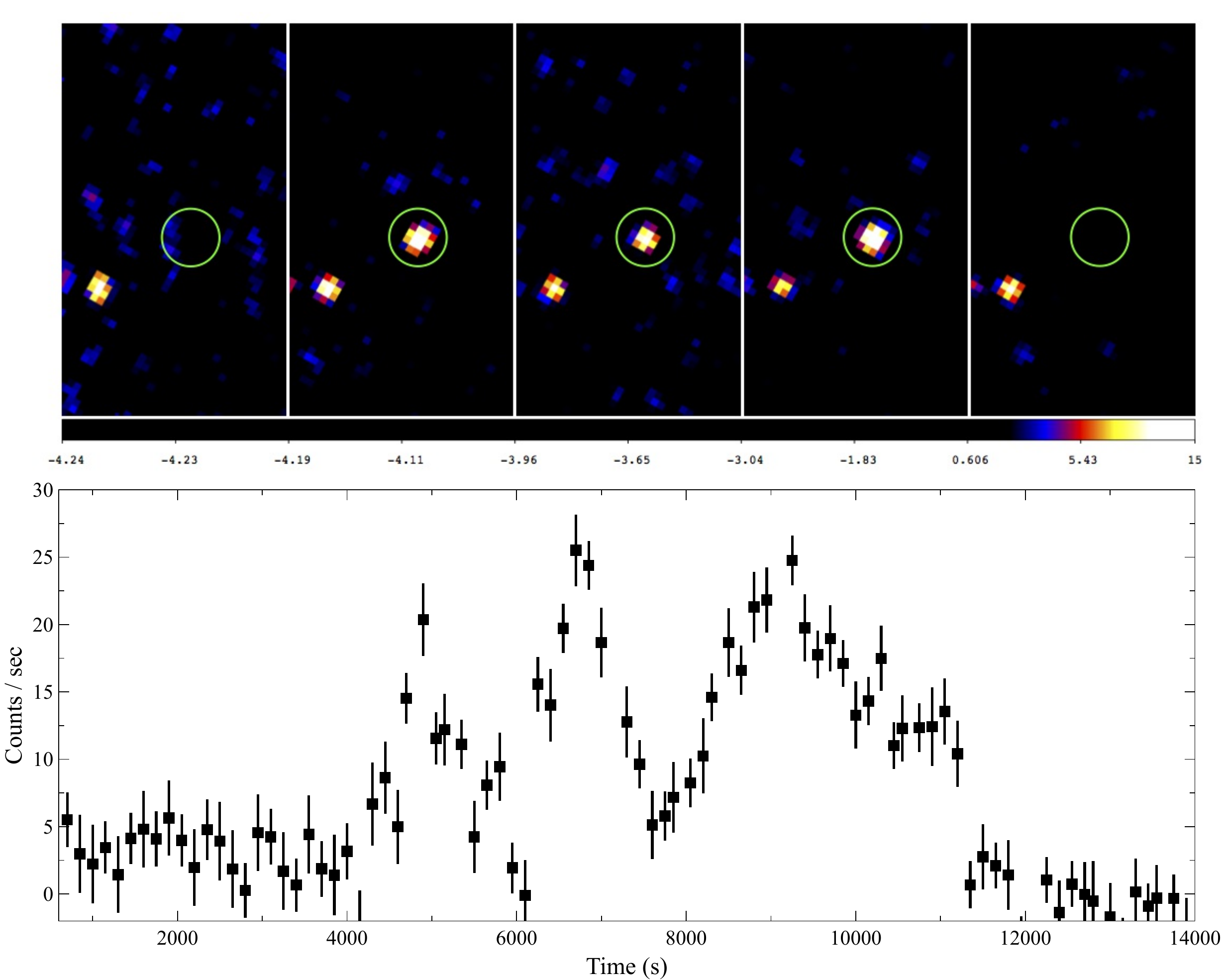}
\caption{Top panel: IBIS/ISGRI pointing-by-pointing image sequence (20--60\,keV) of a fast X-ray outburst from the SFXT XTE~J1739$-$302 (encircled), where each pointing has a duration of about 2000\,s. The colour scale used encodes the difference in sigma from the mean reconstructed flux. A weaker persistent source (1E~1740.7$-$2942) is also visible in the field of view. Bottom panel: the corresponding hard X-ray light curve. See \citet{Sguera2005} for more details.}
\label{fig:sfxt_detection}
\end{center}
\end{figure*}

\intg has proved an excellent tool for the discovery of transients. Its combination of large field of view, fine angular resolution and excellent instantaneous sensitivity, coupled with long exposures as part of regular monitoring of the Galactic Plane, makes it far superior to classical all-sky monitors at this task. In particular, one of the major outcomes of the mission has been the detection of several unidentified fast X-ray transients \citep{Sguera2005, Sguera2006}. Although some of these objects were already known at the time of launch, they had not been studied in depth (Figure~\ref{fig:sfxt_detection}). Several more fast transients were quickly discovered, characterised by strong activity on very short time-scales. Their identification with OB supergiant counterparts \citep{Negueruela2006a} changed our view of the overall HMXB population, by adding a new, distinct class of wind-accreting sources: a totally unanticipated result.

\subsubsection{Discovery and optical identification}


While classical wind-fed supergiant HMXBs are, as described in Sect.~\ref{Sec:classic}, persistent or mildly variable in X-rays, but always detectable around L$_X \sim 10^{36}$ \ergs, these new sources were characterised by their transient nature. Early in the mission, 5 such objects had been identified, namely XTE~J1739$-$302 (IGR~J17391$-$3021; \citealt{Smith2006}), IGR~J17544$-$2619 \citep{ATel190, ATel192, Pellizza2006}, IGR~J16465$-$4507 \citep{ATel329, Chaty2016}, AX~J1845.0$-$0433 and AX~J1841.0$-$0536 \citep{Negueruela2006a}. Given the low duty cycles (see next subsection), previous detections were rare. Only XTE~J1739$-$302 had been characterised with \xte \citep{Smith2006}. An accurate localisation by \chandra allowed the identification of its counterpart, and follow-up optical spectroscopic observations obtained in May 2004 with VLT/FORS1 showed it to be an O8\,Iab(f) star, at a distance of $\sim 2.6$\,kpc \citep{Negueruela2006b}, with interstellar absorption much lower than that implied by X-ray spectral fits. By then, the long stares by \intg during the first few years of operation had already permitted the discovery and characterization of a handful of new systems. Intensive work by different groups \citep[e.g.][]{masetti06a, masetti08, negueruela07} led to the localisation of counterparts to many of these sources, both persistent and transient.

One of the new transient sources, IGR~J17544$-$2619, was unambiguously identified with an optical counterpart thanks to \xmm\ and \chandra\ positions, allowing \cite{Pellizza2006} to take spectra with NTT/EMMI, and identify the counterpart as an O9\,Ib supergiant, at a distance of $\sim3$\,kpc, with a variable amount of absorbing material. Similarly, the counterpart to IGR~J16465$-$4507 was unambiguously identified thanks to the small \xmm error circle, containing only one star. NTT/EMMI spectra of this object taken in 2005 and VLT/X-shooter spectra obtained in 2012 allowed its identification as an early B0.5 to 1\,Ib star with moderate mass loss and very broad lines, indicating a high rotational velocity of 320\,km\,s$^{-1}$, far higher than in most supergiant stars \citep{Chaty2016}. In a few other cases, the infra-red (IR) counterparts of the fast IGR transient sources were identified thanks to the fast repointing and positional capabilities of Swift/XRT \citep[see, e.g.,][and references therein]{romano16}.  

This serves to illustrate that subsequent follow-up optical/infrared spectroscopy in most cases showed that these sources are associated with blue OB supergiant donor stars, just like the majority of \sghmxbs, with a variable amount of absorbing material, though. It was thus proposed to name this new class of \sghmxb as Supergiant Fast X-ray Transients \cite[SFXT;][]{Sguera2005, Negueruela2006a, Smith2006}, because of the fast outbursts and supergiant companions. This new subclass of \sghmxbs, sharing many similarities with other wind-fed \sghmxbs, comprises about 20 sources (including both firm members and candidates), a population size comparable with persistent \sghmxbs.

Although most SFXTs were discovered in the first years of the \intg mission, new members have been uncovered recently \citep[e.g., AX~J1949.8+2534 and IGR~J17503$-$2636;][]{Sguera2017, Hare2019,Ferrigno:2019, Sguera2020}, and two candidate SFXTs have also been discovered in other galaxies \citep{Laycock2014, Vasilopoulos2018}.

\subsubsection{X-ray phenomenology}
   
After the association of SFXTs with early-type supergiant companions, it became evident that these HMXBs are characterized by a remarkable hard X-ray activity: their sporadic high-energy emission caught by \intg only during short duration (a few thousand seconds) flares, put into question the accretion mechanism driving them (see the next subsection), compared with the classical population of supergiant HMXBs (like \vela), known to persistently emit X-rays.

On the timescales typical of SFXT flares ($\sim$2~ks), the detection by IBIS/ISGRI occurs at fluxes greater than a few 10$^{-10}$~\ergscm (18\,--\,50 keV). This implies luminosities above 10$^{35}$\,--\,10$^{36}$~\ergs at the known SFXT distances (see \citealt{sidoli18a} for a systematic analysis of the \intg archival observations of a sample of 58 HMXBs, from 2002 to 2016).
In this bright flaring state, SFXTs spend less than 5\% of the time (their duty cycles at hard X-rays range from 0.01\% in IGR~J17354$-$3255 to 4.6\% in IGR~J18483$-$0311; \citealt{sidoli18a}), and their temporal properties (duration and time interval between consecutive flares, as observed by \intg) show power-law like distributions, possibly indicative of a fractal structure of the supergiant wind matter \citep{sidoli16a}. 

An important observational effort involved several X-ray missions operating at soft X-rays, to investigate the SFXT behaviour outside these outbursts. It consisted of a twofold approach: deep, sensitive pointings at soft X-rays (below 10 keV; from the first studies by \citealt{zand2005, Gonzalez2004, walter06} with \chandra and \xmm, to the most recent \xmm overviews by \citealt{Bozzo2017} and \citealt{Pradhan2018}), alongside long-term monitoring campaigns aimed at characterizing the X-ray properties outside outbursts and at catching new luminous flares (we refer the reader to \cite{Romano2015} and \cite{Bozzo2015} for the most recent reviews 
of the \swift/XRT monitoring of a sample of SFXTs, and to \cite{Smith2012} for the \xte results).

These investigations found that the most extreme SFXTs span a large dynamic range in X-ray luminosity, from quiescence (L$_{X}$$\sim$10$^{32}$~\ergs, e.g. \citealt{Drave2014}) to the peak of the flares (L$_{X}$$\sim$10$^{36}$-10$^{37}$~\ergs, reaching 10$^{38}$~\ergs in IGR~17544$-$2619; \citealt{Romano2015giant}), whereas more typical variability amplitudes are between L$_{X}$$\sim$10$^{33}$-10$^{34}$~\ergs and L$_{X}$$\sim$10$^{36}$~\ergs \citep{sidoli18a}, with a time-averaged X-ray luminosity L$_{X}\lesssim10^{34}$~\ergs \citep{Sidoli2008, Bozzo2015}. 
The comparison of the SFXTs and classical \sghmxbs duty cycles carried out with Swift/XRT highlighted even more the strikingly different behaviours of the two classes of systems in the X-ray domain \citep[see, e.g.,][and references therein]{Bozzo2015}. The SFXT emission is indeed very variable (flaring) at all X-ray luminosities (already evident from the first \xmm observation of the SFXT IGR~J17544$-$2619; \citealt{Gonzalez2004}).

The properties of the lowest luminosity state (L$_{X}\sim10^{32}$~\ergs) of SFXTs remain poorly known, although from \swift/XRT monitoring it can be estimated as the most frequent state in some members of the class (IGR~J08408$-$4503, IGR~J16328$-$4726 and IGR~J17544$-$2619;  \citealt{Romano2015}).

The bright short ($\sim$2~ks duration) flares  belong to longer outbursts, lasting a few days \citep{Romano2007, Rampy2009, Bozzo2011}, and usually occur in an unpredictable way, except for IGR~J11215$-$5259: here, a periodicity in the outburst recurrence was found with \intg \citep{SidoliPM2006}, later refined to $\sim$165~days with \swift/XRT (\citealt{2007A&A...476.1307S}, \citealt{Romano:2009a}) and likely associated with the orbital period of the system.
This is the longest one, while for other SFXTs the orbital cycle ranges from 3.3 days \citep{Jain2009} to 51 days \citep{Drave2010}.
Most periodicities have been discovered from the modulation of their long-term light curve observed with \intg \citep{Bird2009, Zurita2009, Clark2009, Clark2010, Drave2010, Goossens2013}, \swift/BAT \citep{Corbet2006, LaParola2010, Dai2011} and \xte \citep{ATel940,ATel2570}. 
SFXTs display both narrow circular orbits and wide, very eccentric ones, with orbital eccentricities ($e=0.63$ in IGR~J08408$-$4503 and $e>0.8$ in IGR~J11215$-$5952) even higher than in Be/X-ray transients (see the discussion in \citealt{sidoli18a}). In IGR~J17544$-$2619 a high proper motion has been discovered \citep{Maccarone2014}.

In some sources, SFXT flares appear to be more concentrated around periastron (e.g. \citealt{Romano2014:sfxts_cat, Gamen2015}), but they can occur at any orbital phase and it is crucial to remark that not every periastron passage triggers an outburst (except in the case of IGR~J11215$-$5259, where an outburst is present every time the source is observed at the expected times of the periastron passage; e.g. \citealt{Sidoli2020}).
 
Super-orbital periods have been discovered in  IGR~J16418$-$4532 and IGR~J16479$-$4514 \citep{Corbet2013,ATel5131} and, tentatively, in IGR~J08408$-$4503 \citep{Gamen2015}.

Six SFXTs are X-ray pulsars (including the unconfirmed periodicities in IGR~J17544$-$2619 and IGR~J18483$-$0311; \cite{Drave2012,Romano2015giant,ducci13}), with spin periods reaching $\sim$1200~s \citep{walter06}, implying that the accreting compact object is a NS. 
In general, a NS is assumed also in SFXTs where no pulsations have been detected, given the similarity of their broad-band X-ray spectra in outburst with those of accreting pulsars: 
an absorbed, flat power law within 10 keV (with photon index $\Gamma$ between 0 and 1), modified by a high energy cutoff in the range of energies 10--30~keV (e.g. \citealt{walter06, SidoliPM2006, Goetz2007, Filippova2007, Romano2008, Sidoli2009, Zurita2009b, ducci10a}).
The few SFXTs where both spin and orbital periods are detected are spread over a vast area of the Corbet diagram, overlapping (and bridging) persistent \sghmxbs and Be/X-ray transients.

In a few SFXTs a soft X-ray spectral component has been detected during outbursts (as observed in many accreting pulsars; see e.g. \citealt{LaPalombara2006}) and was modeled using a hot blackbody with radius of a few hundred meters, compatible with the NS polar caps (e.g. \citealt{Sidoli2012}). To date, only in the case of the SFXT pulsar IGR~J11215$-$5952 a spectral evolution along the spin phase has been observed \citep{2007A&A...476.1307S}.
In IGR~J17544$-$2619, the most extreme transient of the SFXT class, a spectral evolution was observed in the \xmm plus \nustar spectrum across different luminosity states (from 6$\times$10$^{33}$~\ergs up to the peak of a flare, 1600 times brighter): the combination of a hot (blackbody with $kT\sim$1\,--\,2~keV) thermal component (likely emitted from the surface of the NS), together with a non-thermal, Comptonized component up to $\sim40$~keV, showed evidence of a more prominent contribution from the blackbody at fainter fluxes \citep{Bozzo2016}, similar to what is usually observed in HMXB pulsars.
 
Sometimes the soft excess detected in the same X-ray observation can be accounted for equally well by different models, besides a blackbody: either a partial covering absorption or a ionized absorber \citep{Sidoli2012, Sidoli2017}.
The specific case of IGR~J08408$-$4503 is notable, as this source has the lowest absorbing column density (10$^{21}$~cm$^{-2}$) out of all SFXTs. This allowed \citet{Sidoli2009} to uncover two distinct photon populations (during outburst): one with a temperature of $\sim$0.3~keV, possibly associated with a thermal halo around the NS, and a hotter one at $\sim$1.6~keV, likely from the accretion column. 
A clear soft excess has been observed in this SFXT also in low luminosity states \citep{Sidoli2008, Bozzo2010}, possibly due to thermal X-rays from the donor wind.

The X-ray spectra of SFXTs outside outbursts (at $L_{\mathrm{X}}\sim10^{33}$\,--\,10$^{34}$~\ergs) are usually softer, although still well described by a power law (with photon index $\Gamma\sim1$\,--\,2, \citealt{Sidoli2008, Bozzo2010, Romano2014, Bozzo2017}), implying ongoing accretion.
Sometimes a double component model (non-thermal plus thermal) describes the 1\,--\,10~keV emission better than a single power law even at these low luminosities. 
The soft excess can be accounted for by a blackbody or by a thermal plasma model \citep{Zurita2009b, Bozzo2010, Sidoli2010, Romano2014}. 
This ambiguity in the spectral deconvolution of the soft excess translates into inconclusive results about its nature, at present.
The X-ray spectrum observed during the lowest luminosity state ($\sim10^{32}$ \ergs) can be very soft ($\Gamma\sim6$; \citealt{zand2005}), probably due to thermal X-ray emission from the supergiant wind.

In some sources the local absorbing column density is variable: an increase in the absorption during the rise to the peak of a flare is interpreted as due to and enhancement of the accreting wind matter \citep{Sidoli2009, Bozzo2011, Bozzo2016}, otherwise it can be simply due to the passage of a foreground dense clump \citep{Rampy2009, Boon2016, Bozzo2017}. 
A drop in the column density at the peak of a bright flare is explained with the ionization of the local material  \citep{Bozzo2011, Bozzo2016, Bozzo2017}.

SFXTs have overall lower absorbing column densities than persistent \sghmxbs \citep{Gimenez2015,Pradhan2018}. 
Some exceptions exist, like SAX~J1818.6$-$1703, where $N_{H}\sim5\times10^{23}$~cm$^{-2}$ was observed, similar to what is seen in obscured HMXBs \citep{Boon2016,Bozzo2017}. 
Another diagnostic of the circumsource material is the iron line emission \citep{Bozzo2011, Gimenez2015, Pradhan2018}, contributed by the supergiant wind illuminated by the X-ray source.
The equivalent width (EW) of the 6.4~keV line measured outside eclipses correlates with the absorbing column density and can reach much higher values (EW $>1$~keV) in persistent HMXBs than in SFXTs \citep{Gimenez2015, Pradhan2018}. This is indicative of an accretion environment less dense in SFXTs than in classical \sghmxbs.
Other observations indicate a difference in the donor wind between some members of the two subclasses: in the persistent source Vela X-1 a slower supergiant wind than in the SFXT IGR~J17544$-$2619 has been observed \citep{Gimenez2016}, while a kiloGauss magnetic field has been measured (at $3.8\sigma$) in the supergiant companion of the periodic SFXT IGR~J11215$-$5952 \citep{2018MNRAS.474L..27H}, supporting the hypothesis of the presence of a magnetically focused supergiant wind \citep{2007A&A...476.1307S}. These findings have important implication for the mechanism triggering bright flares (see next subsection).

NS surface magnetic fields, measured from the detection of a CRSF, are elusive: a hint of a low magnetic field ($B\sim10^{11}$~G) has been found in IGR~J18483$-$0311 \citep{Sguera2010}, while a more typical $B\sim10^{12}$~G has been measured with \nustar in IGR~J17544$-$2619 \citep{Bhalerao2015}, although not confirmed by a second \nustar\ observation \citep{Bozzo2016}.
In the 187~s pulsar IGR~J11215$-$5952 only a hint for a CRSF around 17~keV could be found in the \nustar\ spectrum  of a flare \citep{Sidoli2017}, but was not confirmed during a second \nustar\ observation \citep{Sidoli2020}.

As a final remark, we note that the classification of a source as an SFXT or a classical \sghmxb is by no means clear-cut, and there exist several sources displaying a behaviour which is intermediate between the two subclasses (see, e.g., \citealt{walter15a,sidoli18a}). Thus, we should not expect lists of SFXTs or \sghmxbs produced by different groups to agree exactly.

\subsection{Be X-ray Binaries with \intg}
\label{Sec:BeX-INT}
The large field of view and high sensitivity of the instruments on-board \intg allowed this mission to play a leading role in the detection and study of transient sources, and particularly transient X-ray pulsars (XRPs) in Be-binary systems (BeXRBs; see Sect.~\ref{Sec:BeX}). For over 15 years, almost all major outbursts from systems already known were observed, while eight new Be binaries were discovered by \intg, representing an increase in the total number of such sources up to 60 in our Galaxy \citep{walter15a}. Pulsations with periods ranging from $\sim$12 to $\sim$700 s were detected in several new systems: IGR J01583+6713, IGR J11435-6109, IGR J13020-6359, IGR J19294+1816, IGR J22534+6243 \citep{walter15a}, IGR J21343+4738 \citep{reig14b} and IGR J06074+2205 \citep{reig18b}, confirming their pulsar nature.

Observations of bright transient XRPs at different mass accretion rates is essential to understand physical processes at the accretion disc - magnetosphere border and in the vicinity of the NS surface.
A giant (type II) outburst from the BeXRB V~0332+53 starting late in 2004 \citep{2004ATel..349....1S,2005A&A...433L..45K} became the first such an event studied with the \intg observatory in great detail. About 400~ks of exposure time were invested to cover the whole outburst and investigate the properties of the source at very different mass accretion rates. As a result of this monitoring, an unexpected negative correlation of the cyclotron energy (see Sect.~\ref{Sec:CRSF}) with source luminosity was revealed \citep{2006MNRAS.371...19T,2006A&A...451..187M,2010MNRAS.401.1628T,2016A&A...595A..17F}. This discovery led to a surge of interest in the study of cyclotron lines, especially using the \intg observatory in view of its good energy resolution and broad energy coverage \citep[see e.g.,][for the review of spectral properties of XRPs with \intg]{2005AstL...31..729F}. Another pulsar possessing a possible negative correlation of the cyclotron energy with luminosity is 4U~0115+63, where such a trend was suggested by \cite{1998AdSpR..22..987M}. Later, this correlation was confirmed using the \intg and \textit{RXTE} averaged and pulse-amplitude-resolved spectra \citep{2006AdSpR..38.2756N,2004ApJ...610..390M,2007AstL...33..368T,2011A&A...532A.126K}. On the other hand, some authors explained this behaviour of the cyclotron energy in 4U~0115+63 with our poor knowledge of the spectral continuum \citep{Ferrigno2009,Mueller2013,2013AstL...39..375B}. In less bright sources ($L_{\mathrm {X}}\lesssim10^{37}$~erg~s$^{-1}$), a positive correlation of the cyclotron energy with luminosity was discovered using different observations, including some by \intg  \citep{2007A&A...465L..25S,2011PASJ...63S.751Y,2012A&A...542L..28K}. 
For the well known source 1A~0535+262, \intg and \xte observations of the first observable outburst after a long period of quiescence fixed the previously debated magnetic field strength and found no correlation with luminosity over the observed range \citep{ATel601,Caballero:2007}.
These and many other results obtained with different X-ray missions (e.g. \textit{RXTE}, \textit{NuSTAR}, \textit{Suzaku}) substantially improved our knowledge in the field of cyclotron lines formation and evolution (see Sect.~\ref{Sec:CRSF} and the recent review by \citealt{Staubert2019}). 

The good time resolution of the \intg main instruments also permitted studies of temporal properties of emission from XRPs in hard X-rays. In particular, a comprehensive investigation of the pulse profile shapes and pulsed fraction as a function of energy band and flux from the source was performed by \cite{2009AstL...35..433L}. These authors showed that the pulsed fraction systematically increases with energy and has local peculiarities near the cyclotron energy. Phase-resolved spectral analysis also revealed a significant variability of the emission properties over the pulse phase in several sources. Particularly in EXO 2030+375, a hint of the presence of a cyclotron absorption line has been found at $\sim63$~keV in a very narrow phase interval covering less than 10\% of the whole spin period \citep{2008A&A...491..833K}.

In addition to transient sources, the BeXRB family contains a few persistent X-ray pulsars \citep{1999MNRAS.306..100R}. Such objects are characterized by relatively low luminosity ($10^{34}-10^{35}$~erg~s$^{-1}$) and wide orbits ($\Pb\gtrsim200$~days). Due to the high sensitivity of the IBIS telescope some of these systems were detected and studied in the hard energy range with great detail. For instance, RX~J0440.9+4431 and X~Persei were detected up to $\sim120$ keV \citep{2012MNRAS.421.2407T} and $\sim160$ keV \citep{2012MNRAS.423.1978L}, respectively, which is not typical for X-ray pulsars. In the case of X Persei, the high quality of the spectrum allowed \cite{2012A&A...540L...1D} to interpret the broad absorption-like feature around 30~keV not as a cyclotron line \citep{2001ApJ...552..738C}, but as an artificial deficit of photons between two distinct spectral components \citep[see also][]{1998ApJ...509..897D}. The latter interpretation was recently confirmed for two other BeXRBs, GX~304$-$1 and 1A~0535+262, at low accretion rates \citep{2019MNRAS.483L.144T,2019MNRAS.487L..30T}.

\begin{figure}[hbt]
  	\centering
    \includegraphics[width=0.45\textwidth]{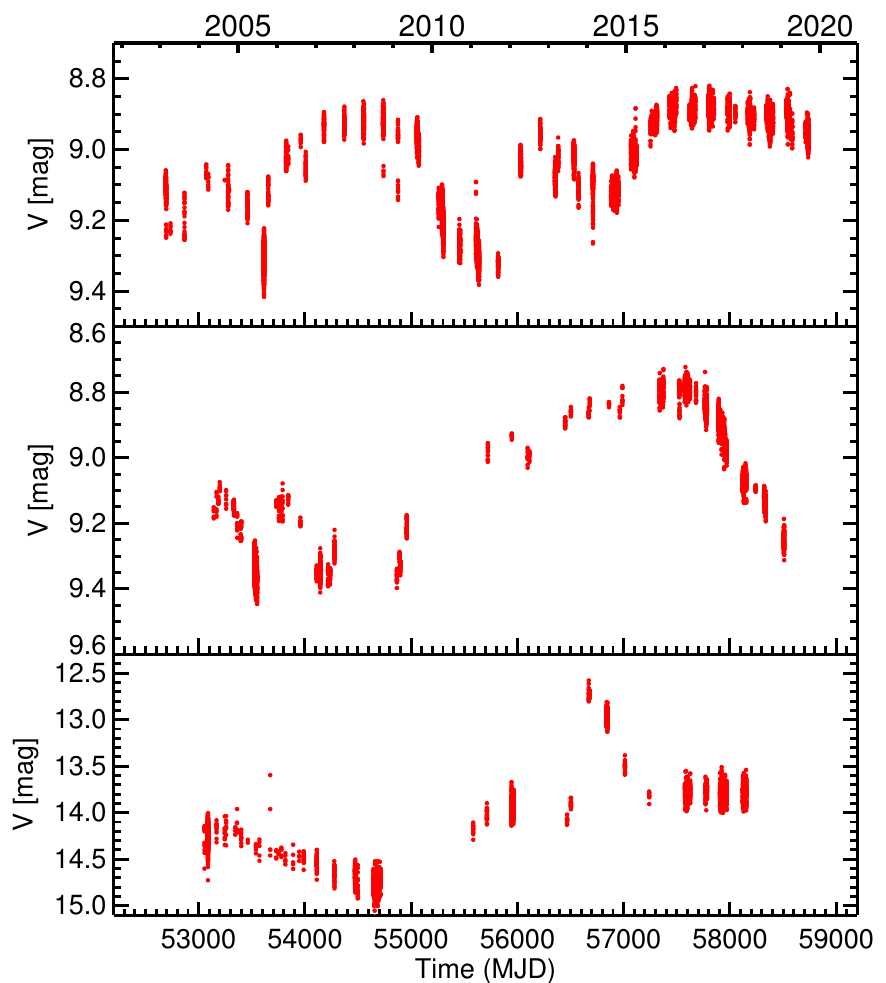}
\caption[OMC light curves of \bexrbs]{OMC light curves of BeXRBs. Top: 1A~0535+262. Middle: H~1145$-$619. Bottom: GX~304$-$1. The apparent dispersion within the observation windows is due to intrinsic low-amplitude variability, typical of Be systems.}
\label{fig:OMC_LC}
\end{figure}

\begin{figure*}[htb]
  	\centering
    \includegraphics[width=0.65\textwidth,angle=-90]{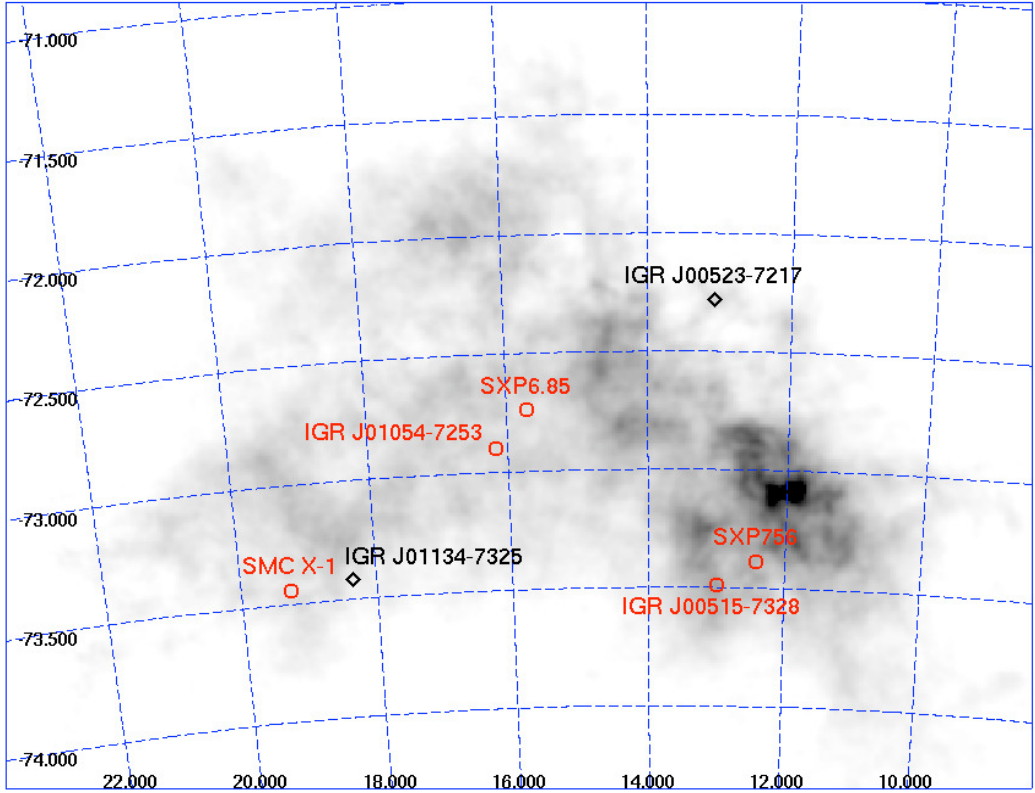}
\caption{Sources detected by IBIS on \intg overplotted on the SMC H I column density map. Figure from  \cite{coe2010}. The sources in red are definite source detections. The two sources in black are candidate sources.}
\label{fig:smc}
\end{figure*}

Another contribution to BeXRB research with \intg is from the Optical Monitoring Camera (OMC), which provides photometry 
in the Johnson $V$--band simultaneously with the high-energy observations \citep{Mas-Hesse:2003}. In this way long-term OMC optical light curves of many HMXBs, along with light curves of many other variable objects \citep{Alfonso-Garzon:2012}, are available through the OMC archive\footnote{\texttt{http://sdc.cab.inta-csic.es/omc/}} \citep{Gutierrez:2004,Domingo:2010}. Long-term optical variability has been observed in the OMC light curves of some BeXRBs, with examples shown in Fig.~\ref{fig:OMC_LC}.

Beyond the Milky Way, \intg has made a significant contribution to identifying the BeXRB population in the nearby Magellanic Clouds. A series of deep observations has led to the discovery and identification of several new systems \citep{mcbride2007, coe2010}. A good example of the power of the \intg wide field of view is shown in this map of objects detected during one set of observations - see Figure \ref{fig:smc}. Identifying all the BeXRB systems in external galaxies like the Magellanic Clouds provides us with a very effective tool for understanding recent star formation and X-ray luminosity functions in these galaxies -- see, for example, \cite{Shtykovskiy2005}.

\subsection{Gamma-ray binary studies with \intg}
\label{Sec:gamma-INT}

\begin{figure}[hbt]
  	\centering
    \includegraphics[width=\columnwidth]{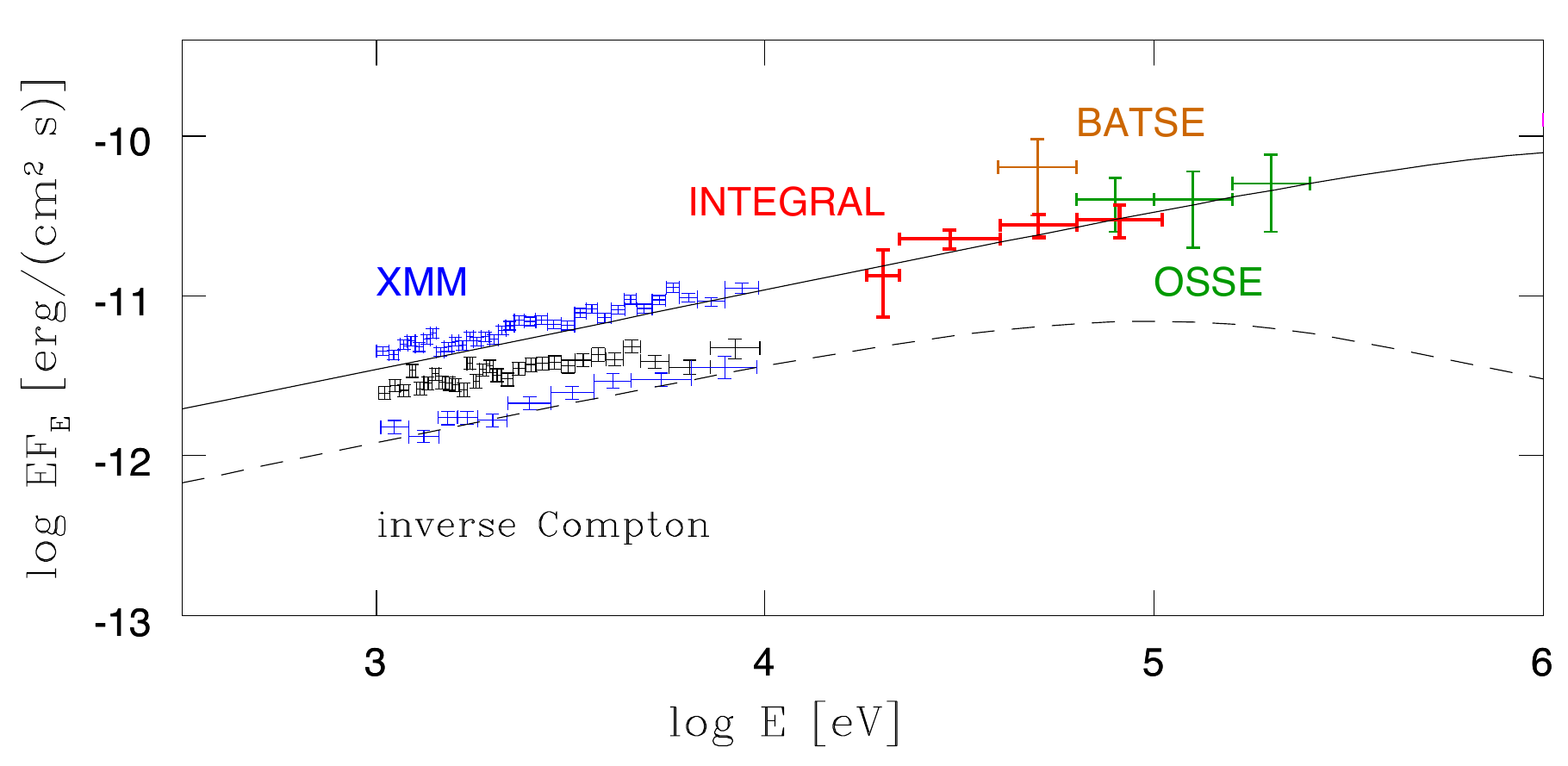}

\caption{Broad-band spectrum of \lsi. This Figure is an adapted extract of Figure 6 in \citet{2006MNRAS.372.1585C}.  The solid (dashed) line shows the model fit within the synchrotron-inverse Compton model for the
high-flux (low-flux) state of the source.}

\label{fig:LSI}
\end{figure}

Joint \intg and \xmm observations of \lsi demonstrated that the overall spectrum of the system in the 0.5--100\,keV energy band is well fit with a featureless power law both at high- and low-flux states, see, e.g., Fig.~\ref{fig:LSI} and  \citet{2006MNRAS.372.1585C,2010MNRAS.408..642Z,2014ApJ...785L..19L}. Non-observation of a cut--off or a break in the spectrum at 10--100\,keV energies, typical of accreting NSs and BHs, favours the scenario in which the compact object is a rotation-powered pulsar. \intg observations demonstrated that in the hard X-rays \lsi is following the overall orbital modulation trend of soft X-rays \citep{2010MNRAS.408..642Z} and also showed hints of a variability similar to the change of the orbital lightcurve on the superorbital time scale observed in the 3--20\,keV range \citep{2012ApJ...747L..29C,2014ApJ...785L..19L}. A joint study of the \lsi spectral variability at hard X-rays with \intg and in the radio band was done by \citet{2012A&A...537A..82Z} and \citet{2014ApJ...785L..19L} to test the possible microquasar nature of the system. \citet{2014ApJ...785L..19L} showed that, for most of the \intg observations, \lsi had a hard spectrum with $\Gamma \sim 1.5$ while the corresponding radio GBI observations (non-simultaneous, but taken at the same orbital and superorbital phase) had a spectral slope $\alpha<0$, which is inconsistent with the predictions of the microquasar model \citep{2012A&A...537A..82Z}. Still, more observations are needed to reach any firm conclusion on the nature of the compact source in the system.

The study of \intg observations of another gamma-ray binary, LS 5039, revealed that that the source significantly emits at hard X-rays (25--200\,keV); this emission varies with the orbit in phase with the very high energy gamma-rays detected with HESS \citep{2009A&A...494L..37H} and is fully anti-correlated with the GeV emission \citep{Abdo:2009:LS5039}. The spectrum at the inferior conjunction is well described by a power law, while at the superior  conjunction the hard X-ray emission is below the sensitivity of \intg. This result indicates that accretion might not be the mechanism for the production of the hard emission in the system,  since, in this case, one would expect a rather sharp flux maximum near periastron \citep{2009A&A...494L..37H}.  An investigation of the orbital light curve in a broad energy range from classical X-ray to TeV energies revealed a transition energy range of tens to hundreds MeV, in which the dominant emission moves from around apastron region to around periastron \citep{Chang:2016}.

A similar behaviour is observed in 1FGL~J1018.6$-$5856, another gamma-ray binary with a period of 16.5\,days, discovered blindly in Fermi-LAT data \citep{LAT2012}. \citet{Li:2011b} have carried out an analysis of all \intg data available at that time. The total effective exposure extracted on the source amounted to 5.78\,Ms, and led to a credible detection of the source at hard X-rays (18--40\,keV). The count rate of the source is very low, but hints at an anti-correlation with the 100\,MeV--200\,GeV emission detected by Fermi-LAT. 

In 2004 \intg performed the first imaging observations of PSR~B1259$-$63 in the hard X-ray range ($>20$\,keV) \citep{2004A&A...426L..33S}, which allowed separating the emission of PSR B1259$-$63 from the emission of the nearby pulsar 2RXP J130159.6$-$635806 \citep{2005MNRAS.364..455C}, which may have contaminated previous hard X-ray observations of this system. Unambiguous measurement of the non-cutoff power law spectrum in the 20--200\,keV energy range served as invaluable input for further broad band spectral modeling. \intg also traced the hard X-ray behaviour of PSR~B1259$-$63 between 55 and 70 days after its 2014 periastron passage \citep{Chernyakova2015}.

\begin{figure*}
	\centering
	\includegraphics[width=\textwidth,angle=0]{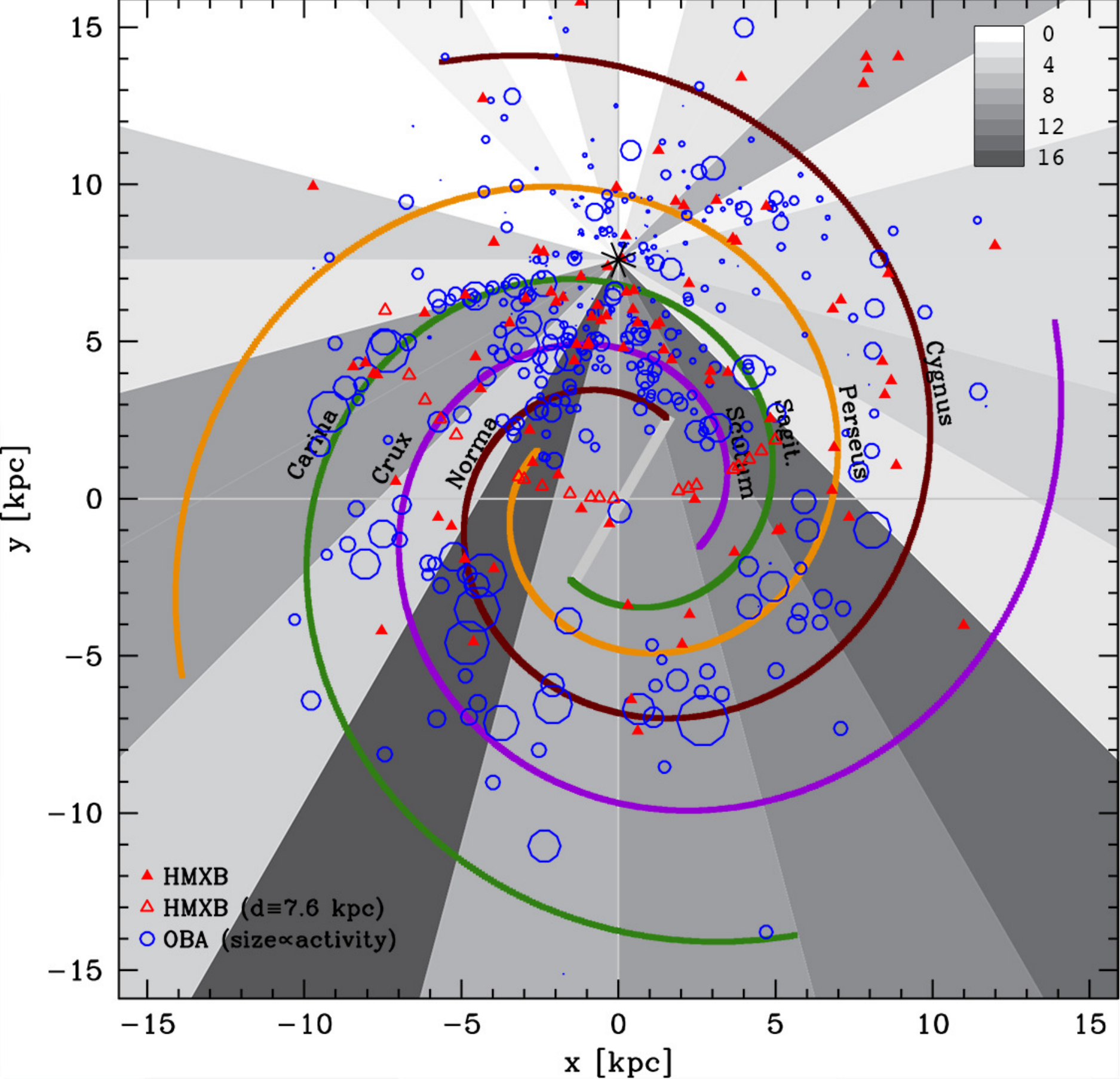}
	\caption{The distribution of \intg-IBIS-detected HMXBs is shown for an observer situated outside the Milky Way. The closed triangles represent 91 HMXBs whose distances are known. The open triangles denote 21 HMXBs whose distances are not known, so they are placed at the galactocentric distance of 7.6\,kpc used in the spiral arm model of \citet{vallee2002}. The largest circles indicate the most active sites of massive star formation \citep{russeil2003}. The shaded bands in the background indicate the number of HMXBs per bin of 15$^{\circ}$ in Galactic longitude, as viewed from the Sun (star symbol). This is an update of the study by \citet{Bodaghee12a}.}
	\label{fig:gal_model}
\end{figure*}

\begin{figure*}[hbt]
	\centering
	\includegraphics[angle=0,width=\linewidth,clip]{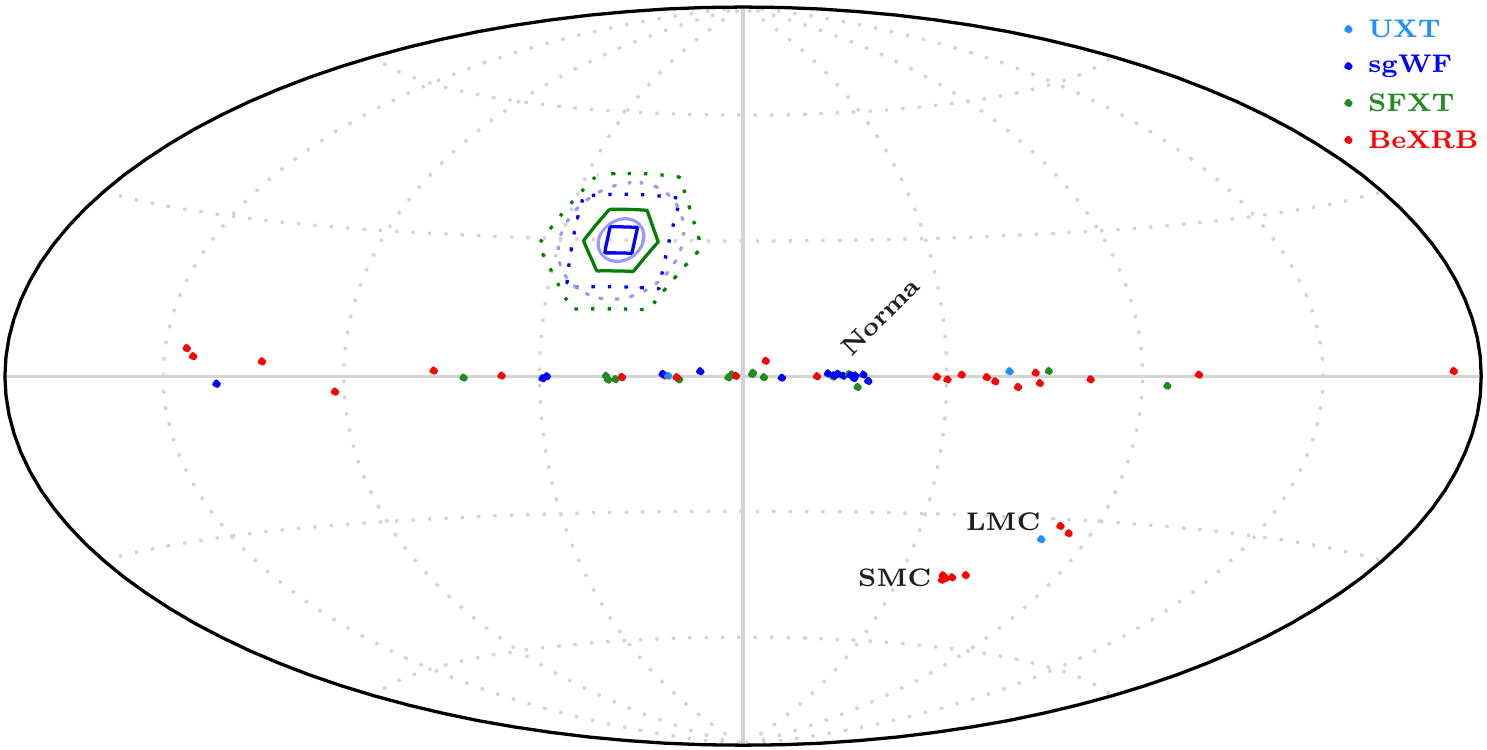} 
	\caption{Sky distribution in Galactic coordinates of HMXBs for which \intg played a significant role (see Table~\ref{tab:obs}). The overlay contours indicate the fully-coded and maximum fields of view of the \intg instruments.}
	\label{fig:gal_fov}
\end{figure*}

\section{Population overview and distribution in the Galaxy (and beyond)}
\label{Sec:Pop}

All-sky surveys by \intg-IBIS have lowered the sensitivity limit to $2.2 \times 10^{-12}$\,erg\,cm$^{-2}$\,s$^{-1}$ for a 5$\sigma$ source detection above 20\,keV \citep{Krivonos17}. This is equivalent to an X-ray luminosity of $2 \times 10^{35}$\,erg\,s$^{-1}$ for a source located at a distance of 20\,kpc, i.e., at the far side of the Milky Way. As a result, \intg doubled the number of HMXBs detected in hard X-rays in the Galaxy, and tripled the number of those with supergiant donor stars \citep{walter15a}. These HMXBs feature column densities in excess of $10^{22}$ or even $10^{23}$\,cm$^{-2}$. The nature of the compact object in these new systems is known, or suspected, to be a NS in almost every case. Many of them exhibit long pulsation periods expected from wind accretors. Therefore, the catalog of \intg-detected HMXBs provides a large (in number), uniform (in exposure), and nearly complete (in luminosity) population for statistical analysis.

\begin{figure*}[t]
\begin{center}
\includegraphics[angle=0,width=\textwidth]{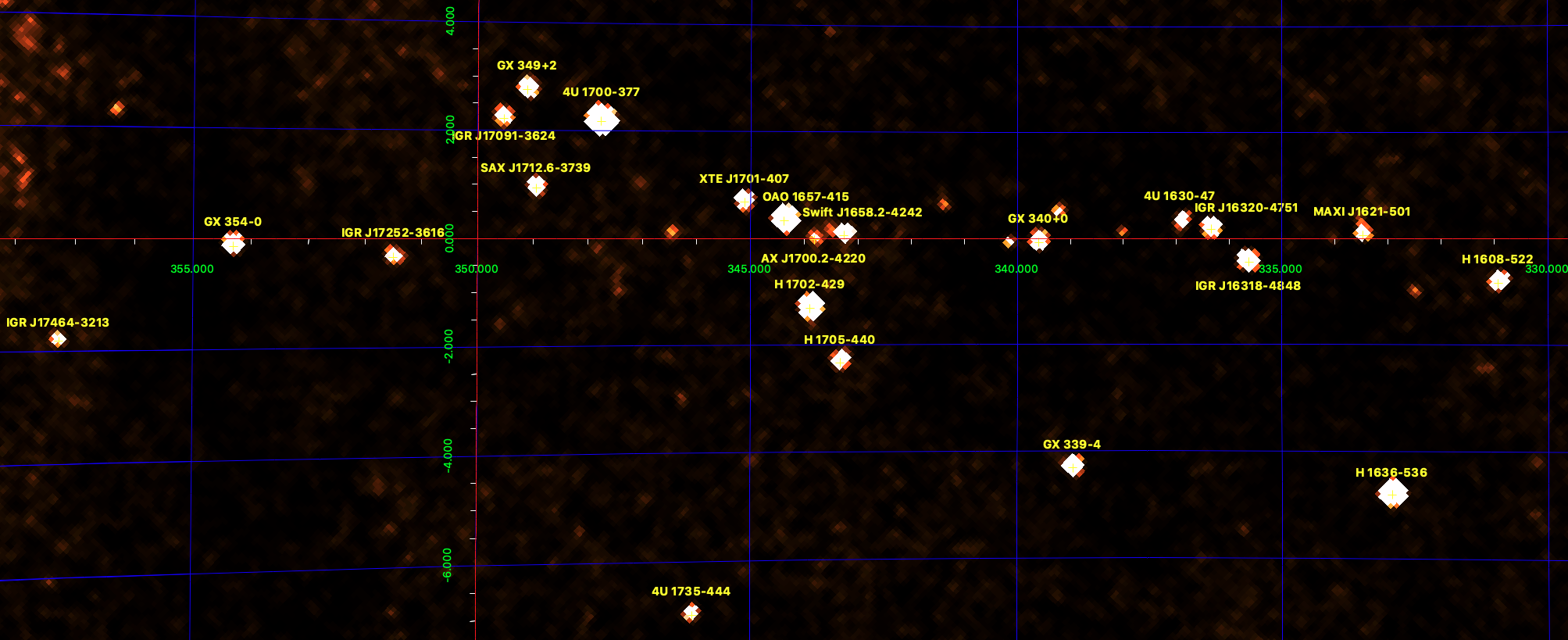}
\caption{Deep mosaic image in the 28--80~keV energy range of the Norma region close to the tangent direction, a zone in the Galaxy rich in hard X-ray sources. The mosaic includes 770 individual \intg pointings between 2016 and 2020.}
\label{fig:Norma}
\end{center}
\end{figure*}

\subsection{Spatial Distributions}
\label{Sec:Pop:Spatial}

Figure~\ref{fig:gal_model} presents the Milky Way as viewed by an outside observer with the locations of 112 HMXBs detected by \intg-IBIS, as well as active sites of massive star formation \citep{russeil2003}. There are now 91 HMXBs whose distances, as reported in the literature, place them somewhere within our Galaxy. The previous version of this map had 79 such objects \citep{Bodaghee12a}. Distances to 47 HMXBs were either refined or collected for the first time thanks to measurements of the optical counterpart as part of the 2nd data release from the \emph{Gaia} mission \citep{gaia2018,bailer2018}. From our perspective within the Galaxy, the direction tangent to the Norma spiral arm continues to feature the highest concentration of HMXBs (the darkest band in Fig.\,\ref{fig:gal_model}). That particular wedge of 15$^{\circ}$ in longitude contains 16 HMXBs which has motivated X-ray surveys by \chandra\ and \nustar\ \citep{Fornasini2014,Fornasini:2016PhD,Fornasini2017} to uncover its faint HMXB population. Figure~\ref{fig:Norma} shows a deep mosaic of this region.

The spatial distribution of HMXBs traces the recent history of massive star formation in the host galaxy. This is because only a few tens of Myr are thought to elapse between the birth of a massive stellar binary in an OB Association (OBA), and the supernova phase that leaves behind a compact object \citep{Schaller92}. Studies of longitudinal distributions of HMXBs consistently show that they are most abundant near sites where the formation of massive stars is most intense, i.e., towards the tangents to the Galactic Spiral Arms \citep{Grimm2002,Dean05, Lutovinov2005b, Bodaghee07}. The surface density of HMXBs is largest for galactocentric distances of 2--8\,kpc, which is again consistent with the distribution of OBAs \citep{lutovinov2013}. 

Nevertheless, significant differences emerge between the HMXB and OBA populations when comparing their physical locations within the Galaxy. The scale height of the HMXB population is larger than that of the OBA population: $\sim$90\,pc and $\sim$30\,pc, respectively \citep{lutovinov2013}. On average, a HMXB is located 0.3$\pm$0.1\,kpc from the nearest OBA. Including this offset in the two-point cross-correlation function increases the significance of the clustering between HMXBs and OBAs, which indicates that the HMXBs have moved away from their parent OBAs and from the Spiral Arms as well. Assuming typical HMXB ages, this corresponds to an average migration velocity of 100$\pm$50\,km\,s$^{-1}$ \citep{Bodaghee12a}. \citet{Coleiro2013a}, after computing the distance and absorption to 46 HMXBs by fitting the optical and infrared SED of the donor star, could associate them with their likely parent OBAs, deriving a clustering size between HMXB and OBA of 0.3$\pm$0.05\,kpc, with a characteristic inter-cluster distance of 1.7$\pm$0.3\,kpc. With these accurate HMXB distances, by taking into account the Galactic spiral arm rotation and assuming a regular value of kick velocity of 100\,km\,s$^{-1}$, \citet{Coleiro2013a} constrained age and maximum migration distance of 13 HMXBs: 9 \bexrbs and 4 \sghmxbs.

These velocities are higher than expected from recoil due to anisotropic mass loss from the primary to the secondary \citep{Blaauw61}, or via dynamical ejection and outflows from the cluster \citep{Poveda67,Pflamm10}. Instead, the velocity range is consistent with values expected for a natal kick acquired during the formation of the NS, which can be significant in an asymmetrical supernova \citep[e.g.,][]{Shklovskii70}. Empirical evidence of migration velocities in HMXBs can help determining the characteristic timescale between the formation of the NS and the X-ray emission phase, which is still poorly understood, as well as constraining models of type II supernovae. Unfortunately, there are only a handful of HMXBs whose proper motions are known well enough to enable the measurement of a kick velocity away from a specific OBA \citep[e.g.,][]{Ankay01,Ribo02,Mirabel04}. Fortunately, the evolutionary history of massive binary stars is imprinted as an offset in the spatial distribution of HMXBs relative to their birthplaces, and this offset yields an average velocity for the population.

\begin{figure}[ht!]
\centering
\includegraphics[width=0.95\columnwidth]{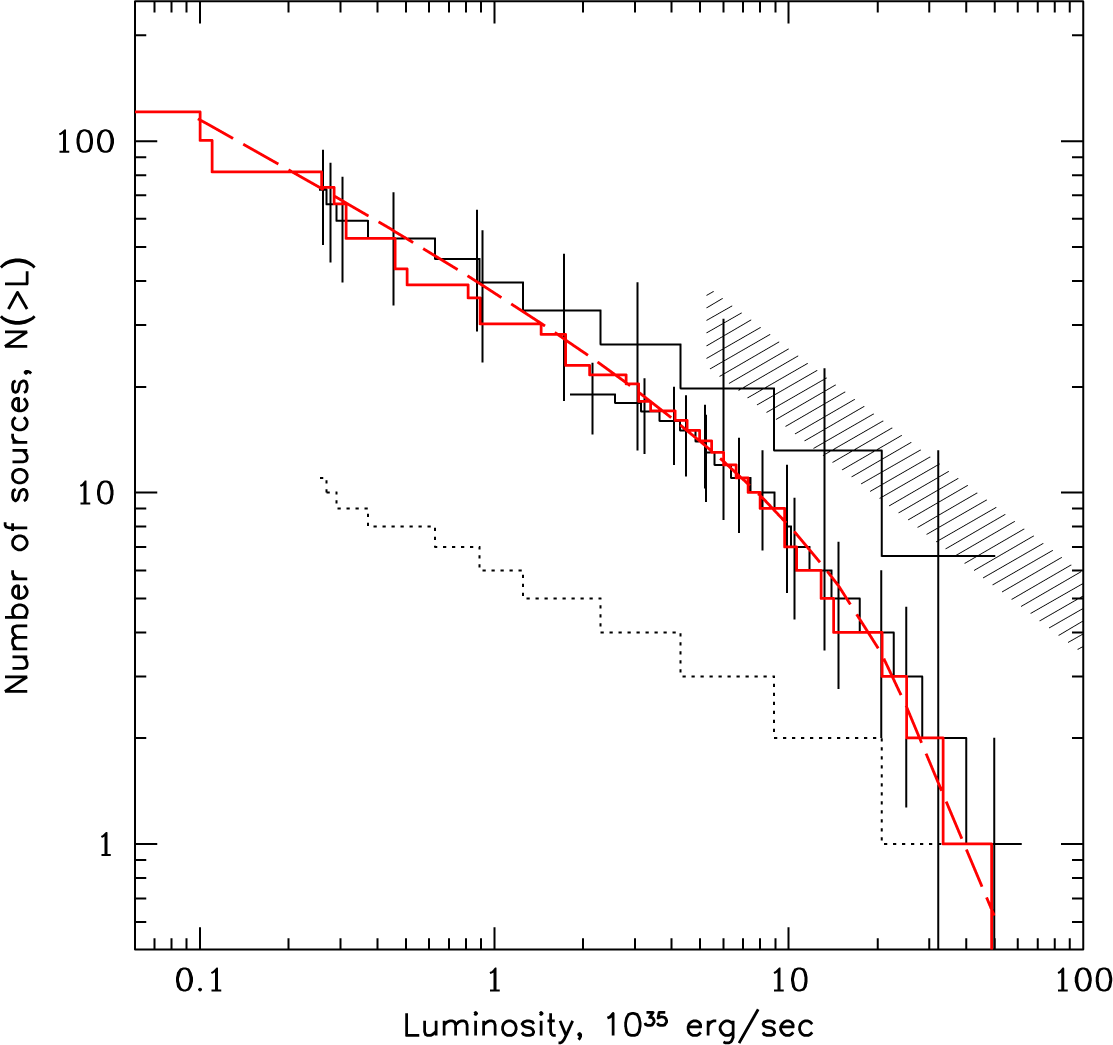}
\includegraphics[width=0.95\columnwidth]{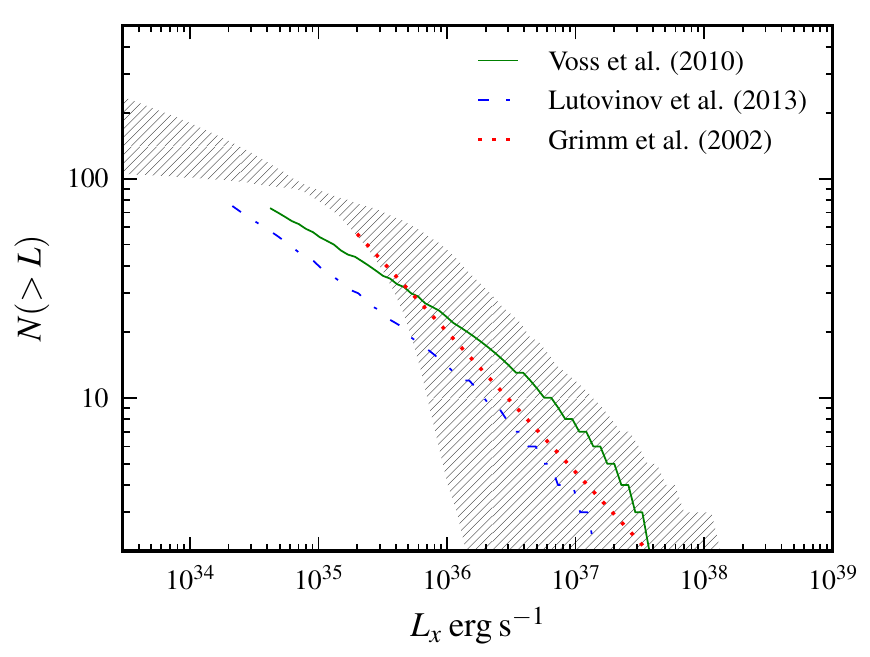}
\caption{Top panel: luminosity function for persistently-emitting HMXBs in the Milky Way (red histogram) fit with a broken power law (dashed curve) as presented in \citet{lutovinov2013}. Volume-limited samples are shown as black histograms. The shaded region corresponds to the luminosity function from \xte-ASM data \citep{Grimm2002}. 
Bottom panel: the same luminosity function (shaded area) as reconstructed through Monte Carlo modeling by \citet{Doroshenko14}. For comparison,  the results by \citet{lutovinov2013} and \citet{Voss2010} are shown.
}
\label{fig:lumin_func}
\end{figure}

\subsection{Luminosity Functions}

A useful diagnostic of HMXB populations is the luminosity function 
$N(>L)$, which is the number distribution of sources with luminosities greater than $L$, plotted as a function of $L$, as shown in Fig.\,\ref{fig:lumin_func}. The shape of the HMXB luminosity function depends on the recent star-formation rate of the host galaxy \citep[e.g.,][]{Grimm2002}, and on the relations governing the mass accretion rate and luminosity in wind-accreting systems \citep{Postnov2003}. 

Luminosity functions are relatively easy to build for HMXBs in other galaxies since the entire population is located at the well-known distance of the host galaxy. Measuring the distances of HMXBs in the Milky Way, however, is more challenging given that the Galactic veil of dust and gas leads to optical extinction of the stellar companion, and photoelectric absorption of the soft X-rays. So not only are many of the distances unknown but also, for those objects whose distances are known, the sensitivity of the survey in each region of the Milky Way must also be considered.

Figure\,\ref{fig:lumin_func} presents the HMXB luminosity function with \intg-IBIS data \citep{lutovinov2013}. A single power law does not adequately fit the luminosity function over the range of luminosities sampled ($10^{34}$--$10^{37}$\,\ergs), given that a break in the slope around $10^{36}$\,\ergs and a flattening at lower luminosities are evident. That becomes even more clear if the spatial distribution of early type stars is taken into the account. This is illustrated by an independent reconstruction of the HMXB luminosity function using the same data-set by \citet{Doroshenko14} and shown in Figure~\ref{fig:lumin_func}, where the impact of distance uncertainties to individual objects is minimized through Monte Carlo modeling under the assumption that the spatial distribution of HMXBs is roughly known.

These results could imply that the IBIS surveys may be missing a significant number of faint HMXBs. However, a flattening at the faint end was also observed in the luminosity function of HMXBs in the more uniform \swift-BAT survey \citep{Voss2010}, as well as in the luminosity function of HMXBs in the Small Magellanic Cloud \citep{Shtykovskiy2005}.
Therefore, an adjustment may be necessary in the universal luminosity function which is expected to follow a single power law $dN/dL \propto L^{-\alpha}$ with $\alpha = 1.6\pm 0.1$ for a wide range of luminosities ($10^{35}$--$10^{40}$\,\ergs). Surveys by \intg-IBIS and other facilities such as {\it eROSITA} will continue to probe the faint end of the luminosity function, helping to clarify the recent history of massive star formation and the relative contribution of HMXBs to the total X-ray luminosity of galaxies.

\subsection{Cumulative Luminosity Distributions}

\begin{figure*}
\centering
\includegraphics[width=1.0\linewidth]{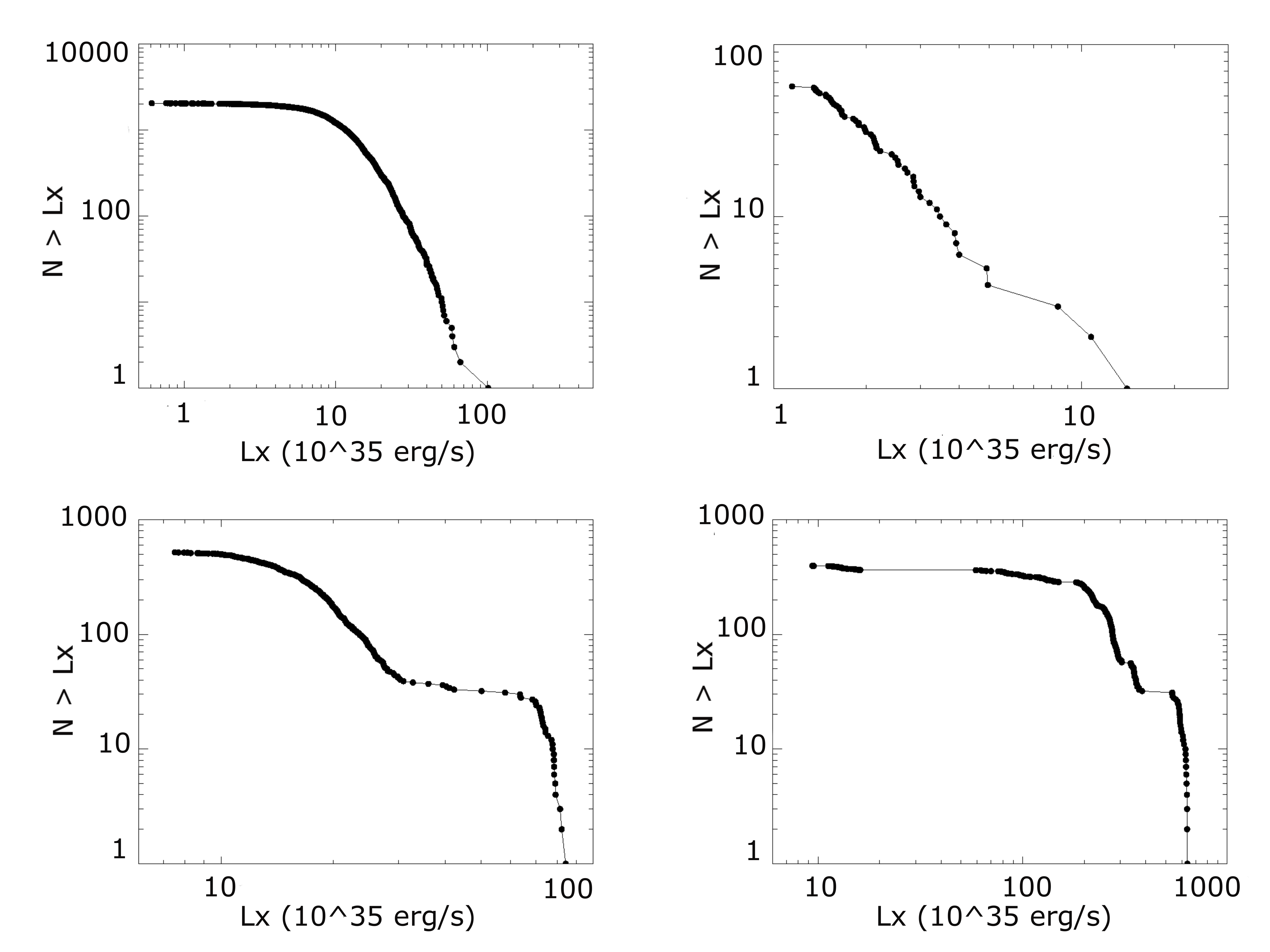}
\caption{Cumulative luminosity distributions (18--50\,keV) of four representative sources considering fourteen years of \intg IBIS/ISGRI data.  Each point refers to a source detection in a $\sim$2\,ks \intg pointing. The highest y-axis values indicate the duty cycles in the given band, while the range of variability can be seen on the x-axis. From left to right, top to bottom: Vela~X$-$1 (persistent \sghmxb),  SAX~J1818.6$-$1703 (SFXT) and the two BeXRB transients SAX~J2103.5$+$4545 and EXO~0331$+$530 (the bimodal shape can be explained by the occurrence of normal and giant outbursts). Adapted from \cite{sidoli18a}.}  \label{fig:CLD_4sources}
\end{figure*}

A long-term  systematic analysis of \intg\ data (fourteen year span) has been performed by \citet{paizis14} and \citet{sidoli18a} who investigated the hard X-ray properties of 58 HMXBs. The sample, about half of the total number of HMXBs known in our Galaxy, comprises persistent and transient systems, including BeXRBs, \sghmxbs and SFXTs hosting a NS or a BH. Light-curves of 2\,ks time bins were used to derive hard X-ray (18--50\,keV) cumulative luminosity distributions (CLD) for all the sources. This approach leads to a full quantitative characterization of the hard X-ray luminosity distributions of the single sources, displaying in a concise way the different phenomenologies and patterns at play, the sources' duty cycles, range of variability and the time spent in each luminosity state. Figure~\ref{fig:CLD_4sources} shows the CLDs of four representative sources. 

In \citet{sidoli18a}, the phenomenology observed with \intg is juxtaposed with other known source properties in order to obtain a quantitative overview of the main subclasses of accreting massive binaries as they tend to cluster in the different parameter spaces explored.

With respect to \citet{lutovinov2013} who studied the luminosity and spatial properties of persistent HMXBs in our Galaxy with \intg, \citet{sidoli18a}, while considering transient sources as well,  focused on the bright luminosity end. Indeed, the selection criteria chosen (source detection in a single 2\,ks pointing) resulted in a flux-limited sample with a sensitivity of a few 10$^{-10}$ \ergscm, about one order of magnitude worse than what considered by \citet{lutovinov2013}.

\begin{figure}
\centering
\includegraphics[width=1.05\columnwidth]{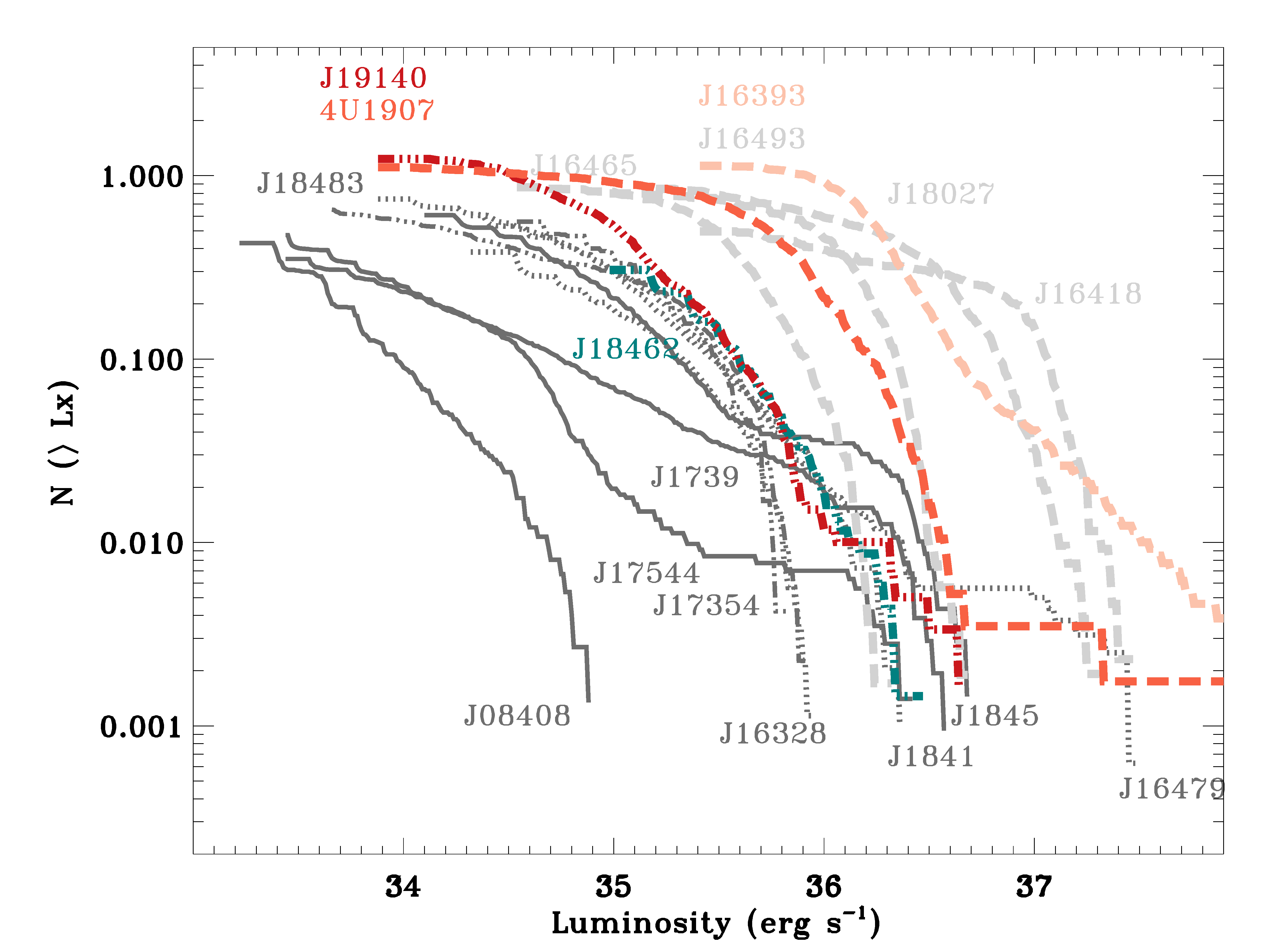}
\caption{Cumulative luminosity distributions (2--10\,keV) of several SFXTs and classical \sghmxbs derived from the long term monitoring campaigns of these objects performed with Swift/XRT \citep[][Romano et al. 2020, in preparation]{Bozzo2015}. The figure is adapted from \citet{Bozzo2015} and includes both data previously published (gray lines) and newly obtained data (Romano et al. 2020, in preparation). Solid lines are used for SFXTs and dashed lines for classical \sghmxbs. The typical ``step-like'' shape of the SFXT cumulative luminosity distributions can be clearly seen in all cases.}  \label{fig_xrtcumul}
\end{figure}

Motivated by the intriguing \intg results, the study of the HMXB cumulative luminosity distributions presented by \citet{paizis14} was extended in the soft X-ray energy domain (2--10 keV) by \citet{Bozzo2015}. These authors used Swift/XRT data collected during the long-term monitoring campaign of several classical \sghmxbs and SFXTs, which regularly span a significant fraction of the orbital phases of many revolutions of these systems (providing a few 100~ks of effective exposure per source), in order to provide an efficient probe of their different luminosity states. The advantage of the XRT data is that the instrument is endowed with a high sensitivity and can extend the cumulative luminosity functions down to fluxes as low as $\sim$10$^{-13}$ \ergscm, complementing the \intg curves. We show in Fig.~\ref{fig_xrtcumul} an improved version of the fig.~5 of \citet{Bozzo2015}, which includes the results of the most recent and partly still ongoing XRT observation campaigns (the gray curves are those already published previously, whereas the other ones will be reported with all details in Romano et al. 2020, in preparation). In agreement with the results found by \intg, the XRT observations revealed that the cumulative distributions of the SFXTs have a more complex shape, with multiple ``steps'' (as opposed to the single knee curve of the classical systems) marking the presence of different accretion states. These findings have been interpreted by \citet{Bozzo2015} in terms of the different accretion regimes introduced in Sect.~\ref{Sec:magnetosphere}.

\section{Future perspective}
\label{Sec:Future}

\intg has now been confirmed to operate at least until end 2020, with a possible further extension up to the end of 2022. Other extensions will be decided following pending reviews of the mission status and operations, but are formally possible until 2029, when a planned maneuver will force the satellite to a controlled re-entry in the Earth atmosphere. As emphasized multiple times in this review, the unique combination of sensitivity, pointing strategy, and large field of view of the \intg instruments will surely lead in the years to come to the discovery of new transient HMXBs, including SFXTs. The most recent results in these fields, as summarized in this review, prove that each newly discovered IGR source offers new perspectives to understand the physics of HMXBs and often poses new theoretical/observational challenges. The community is thus certainly looking forward to more years of successful \intg operations. 

Following the success of the \intg mission, a number of different high energy facilities, covering from the softer ($\lesssim$0.5~keV) to the harder X-rays (up to the MeV range) are being planned or already in the timeline. These will certainly provide a dramatic step forward in our understanding of all classes of HMXBs. 

The \Athena mission \citep[see, e.g.,][]{athena} is set to fly in the early 2030s and, especially thanks to the X-IFU instrument \citep{barret18}, it will open the possibility of performing high resolution spectroscopy of HMXBs even at relatively low fluxes. These observations will be able to probe for the first time, among others, the physical processes occurring during the accretion of single clumps from the stellar winds onto the compact objects hosted in classical \sghmxbs and SFXTs \citep{bozzo15athena}. This is something that today is only possible by exploiting the grating spectrometers on-board \chandra and \xmm for the uniquely bright HMXB Cyg~X-1 \citep{hirsch19}. High-resolution spectroscopy of wind-fed HMXBs with X-IFU (with a typical accuracy of a few eV, depending on the specific energy) will also permit us to investigate in unprecedented detail the physics of the interaction between the X-rays from the compact object and material in the stellar wind, finally providing a solid measure of the perturbed and unperturbed stellar wind. This is the key also to likely distinguish between the scenarios proposed to explain the different behaviours of classical \sghmxbs and SFXTs. 

The eXTP mission \citep{zhang19} is planned to be at present the next X-ray facility harboring a large field-of-view instrument \citep[WFM;][]{galvez19} that will cover the full sky every $\sim$2~days looking for X-ray transient sources in the 2--50\,keV energy band. The instrument is based on a similar coded-mask technique as \intg/IBIS, but will complement the \intg discoveries by monitoring the sky in the softer energy band. The eXTP/WFM will provide a good energy resolution of 300\,eV on the entire band and a time resolution as accurate as a few $\mu$s. An on-board system will detect and broadcast to the ground impulsive events, as for example the outbursts from SFXTs. The instrument will thus be capable to discover many more transient/flaring HMXBs\footnote{See, e.g., \citet{bozzo13loft}. These authors estimated the number of SFXT outbursts for the WFM instrument on-board the LOFT mission \citep{feroci2012}, which has a similar design as the eXTP/WFM.}, as well as flares/outbursts from already known sources, collecting for each of these events a uniquely rich amount of useful data that will be provided to the community a few hours after the detection. The eXTP/WFM will also provide daily monitoring data for hundreds of sources with good timing and spectral resolution, that can be used to verify the evolution of the spectral states, as well as the spin period and/or the timing state of the different HMXBs (including possible evolution over time of the cyclotron line features and orbital/super-orbital modulations). Furthermore, the suite of instruments on-board eXTP \citep{zhang19} will be able to provide an unprecedentedly high statistics, good energy resolution, and high timing accuracy data in a broad energy range (2--50\,keV) to perform a combined timing-spectroscopy-polarimetry analysis of many bright HMXBs\footnote{Preliminary measurements of the X-ray polarimetry in HMXBs \citep{kallman15} could be already expected thanks to the NASA mission IXPE \citep{ixpe}, currently planned for a launch in 2021.}. This will open new perspectives to resolve changes of the cyclotron line parameters as a function of the system physical conditions (geometry, luminosity, line of sight) and understand intimate details of the physics of accretion in wind- / disc-fed systems, as well as aspects of the microphysics of plasma penetration in the NS magnetosphere in a never accessed before way\citep{zand19, santangelo19}.

Improvements over the already exciting results that will be provided by eXTP are expected from its successor STROBE-X \citep{ray19}, which will provide yet more effective area to increase the statistics and optimize the background in a more extended energy range (0.5--50\,keV). STROBE-X is a mission concept proposed for the NASA decadal survey\footnote{See  \url{https://www.nrl.navy.mil/space/strobe-x-headed-decadal-survey}}.

Flares and outbursts from Galactic and extra-Galactic HMXBs are expected to be among the easiest targets for detection by the next generation of wide-field X-ray facilities using Lobster-eye telescopes \citep{angel79}, as those planned onboard the Einstein Probe \citep{ep} and the THESEUS mission \citep{amati18, stratta18}. The former mission is led by China and is planned to be launched as early as 2021, while THESEUS is one of the three candidates competing for a launch opportunity in 2032 within the context of the ESA M5 call\footnote{See \url{https://www.esa.int/Our_Activities/Space_Science/ESA_selects_three_new_mission_concepts_for_study}.}. The advantage of Lobster-eye telescopes is to provide a large field of view (up to 60$\times$60\,deg in the currently planned configurations) and a low background, achieving a sensitivity able to discover sources as faint as $\sim$10$^{-10}$~erg~s$^{-1}$ (0.3--6\,keV) already in a few tens of seconds. This matches well the range of luminosity of the flares and outbursts observed from HMXBs. THESEUS will also be endowed with an infrared telescope (IRT) which will be automatically re-pointed toward the sources of interest, and could be used to identify the massive companions of HMXBs. 

In the gamma-ray domain, the AMEGO\footnote{But see also the similar mission concept e-ASTROGAM \citep{astrogam}.} mission concept \citep{amego} is expected in a farther future to provide new insights about the emissions from HMXBs in the MeV energy range, distinguishing between the possibilities that such emission is due to accretion phenomena or to the presence of jets \citep{astrogamscience}.  

At even higher energies, as summarized in Sect.~\ref{Sec:gamma}, during the last decade an increasing number of HMXBs has been  detected in the MeV--GeV energy range (HE), as well as in the TeV energy range (VHE). This proved the existence of an emerging new population of ``gamma-ray binaries'' and/or ``gamma-ray loud binaries'' \citep{dubus_review13,grlb_cta_ch19}, which rapidly became subject of major interest. Similarly to these general classes, in principle SFXTs, as well as other flaring \sghmxbs, could be able to produce HE and VHE emission as well since they are characterized by the same ingredients in terms of a compact object and a massive early-type companion star. However, the detection of such emission is nontrivial for current generations of HE and VHE instruments, since it should likely be in the form of unpredictable flares having short duration, small duty cycle and relatively low flux. To date, only few hints have been reported in the literature on SFXTs as best candidate counterparts of unidentified transient HE sources, merely based on circumstantial evidences \citep{Sguera2009a, Sguera2009b, Sguera2011}. No detection of VHE emission from other \sghmxbs has been reported so far. Hopefully, this situation will change in the near future thanks to a new generation of ambitious facilities under constructions, e.g. the Cherenkov Telescope Array, which will offer an order of magnitude improvement in the VHE domain in terms of sensitivity and survey speed compared to existing facilities, hence holding the promise of performing well as a transient detection factory \citep{ctas}.  We expect that future studies will likely shed new light by detection or not-detection SFXTs, as well as 
other \sghmxbs, as HE and VHE transient emitters. If confirmed, this would open the investigation to a completely unexplored energy window, allowing a deep study of the most extreme physical mechanisms at work,  which could be tested on very short timescales, not usually investigated. All this could eventually add a further extreme characteristic to the class of SFXTs, as well as to other subclasses of HMXBs in general.\\

The \intg mission has provided an important impulse to the study of HMXBs: its large field of view together with its broad spectral range, sensitivity and angular resolution has allowed us to peer through the obscured as well as fast transient X-/gamma-ray sky. Our knowledge has grown broader and deeper in the ever-changing source population that traces the star-forming arms of our Galaxy. \intg plays an important role in connecting the soft X-rays to the very high-energy range of the electromagnetic spectrum (MeV to TeV), enriching our view of the physical processes involved. It serves as an important springboard for the missions and the knowledge to come.

\section*{Acknowledgments}

Based on observations with \intg, an ESA project with instruments and a science data center funded by ESA member states (especially the PI countries: Denmark, France, Germany, Italy, Spain, and Switzerland), the Czech Republic, and Poland and with the participation of Russia and the USA. The \intg teams in the participating countries acknowledge the continuous support from their space agencies and funding organizations: the Italian Space Agency ASI (via different agreements including the latest one, 2019-35HH, and the ASI-INAF agreement 2017-14-H.0), the French Centre national d'{\'e}tudes spatiales (CNES), the Russian Foundation for Basic Research (KP, 19-02-00790), the Russian Science Foundation (ST, VD, AL; 19-12-00423), the Spanish State Research Agency (via different grants including 
ESP2017-85691-P, ESP2017-87676-C5-1-R and Unidad de Excelencia Mar{\'{\i}}a de Maeztu -- CAB MDM-2017-0737). IN is partially supported by the Spanish Government under grant PGC2018-093741-B-C21/C22 (MICIU/AEI/FEDER, UE). LD acknowledges grant 50 OG 1902.
 
\clearpage 
\onecolumn
\appendix

\section{HMXBs discovered or identified with  \intg}
\begin{table*}[ht!]
\caption{HMXB discovered or identified with \intg. Under System type, sgWF stands for wind-accreting \sghmxb, BeXRB for Be X-ray binary, SFXT for Supergiant Fast X-ray Transient and UXT for Unidentified X-ray Transient. Reference numbers are expanded below the table. Distance values are taken from \citet{bailer2018} for many sources. For sources in the SMC we use the ``canonical distance modulus'' proposed by \citet{Graczyk:2014}, which is consistent, e.g., with \citet{deGrijs:2015}. For sources in the LMC we use the recent value by \citet{Pietrzynski:2019}}
\label{tab:obs}

\renewcommand\arraystretch{1.1}
\footnotesize
\begin{tabular}{@{}l@{\enspace}r@{\enspace}rrrrlll}\hline\hline
\multicolumn{1}{l}{Source} & \multicolumn{1}{c}{R.A.} & \multicolumn{1}{c}{Decl.} & 
\multicolumn{1}{c}{\Ps}  & \multicolumn{1}{c}{\Pb} & \multicolumn{1}{c}{distance} & 
\multicolumn{1}{c}{Companion} & \multicolumn{1}{c}{System} & References \\  
\multicolumn{1}{l}{name}   & \multicolumn{1}{c}{deg}  & \multicolumn{1}{c}{deg} &
\multicolumn{1}{c}{s}      & \multicolumn{1}{c}{days} & \multicolumn{1}{c}{kpc} &
\multicolumn{1}{c}{type}   &\multicolumn{1}{c}{type}  & \\ \hline
\href{http://simbad.u-strasbg.fr/simbad/sim-basic?Ident=IGR+J00370%2B6122}{IGR\,J00370+6122} &   9.286 &    61.386 &    346 & 15.67 &          3.3 & BN0.5\,II-III / BN0.7\,Ib   & sgWF   & \tbsmartref{A281}{Reig05}{iZd07}{Hai20} \\
\href{http://simbad.u-strasbg.fr/simbad/sim-basic?Ident=IGR+J00515-7328}{IGR\,J00515$-$7328} &  13.003 & $-$73.490 & --     & --     &    $60\pm2$ & O8\,V / B0\,III & \bexrb & \tbsmartref{Coe10}{Mas13} \\ 
\href{http://simbad.u-strasbg.fr/simbad/sim-basic?Ident=IGR+J00569-7226}{IGR\,J00569$-$7226} &  14.259 & $-$72.432 &   5.05 & 17.13 &     $60\pm2$ & B0.2\,Ve        & \bexrb & \tbsmartref{A5547}{A5557}{C15b}{Ev06} \\
\href{http://simbad.u-strasbg.fr/simbad/sim-basic?Ident=IGR+J01054-7253}{IGR\,J01054$-$7253} &  16.173 & $-$72.901 &  11.48 &  36.3 &     $60\pm2$ & O9.5-B0\,IV-V   & \bexrb & \tbsmartref{Kah99}{A2079}{A2088}{Tow11a} \\
\href{http://simbad.u-strasbg.fr/simbad/sim-basic?Ident=IGR+J01217-7257}{IGR\,J01217$-$7257} &  20.419 & $-$72.959 & --     & --     &     $60\pm2$ & B0-5\,(II)e    & \bexrb  & \tbsmartref{A5806}{Ev04} \\
\href{http://simbad.u-strasbg.fr/simbad/sim-basic?Ident=IGR+J01363%2B6610}{IGR\,J01363+6610} &  24.060 &    66.188 &  --    &   160 &            2 & B1\,Ve          & \bexrb & \tbsmartref{A275}{A3079}{Tom11} \\
\href{http://simbad.u-strasbg.fr/simbad/sim-basic?Ident=IGR+J01572-7259}{IGR\,J01572$-$7259} &  29.318 & $-$72.976 &  11.58 &  35.6 &     $60\pm2$ & Be              & \bexrb & \tbsmartref{A1882}{Seg13a} \\
\href{http://simbad.u-strasbg.fr/simbad/sim-basic?Ident=IGR+J01583%2B6713}{IGR\,J01583+6713} &  29.576 &    67.224 &  469.2 &    -- &          4.0 & B2\,IVe         & \bexrb & \tbsmartref{A672}{A681}{Kau08} \\
\href{http://simbad.u-strasbg.fr/simbad/sim-basic?Ident=IGR+J05007-7047}{IGR\,J05007$-$7047} &  75.203 & $-$70.775 & --     & 30.77 &     $49.6\pm0.6$ & --          & UXT    & \tbsmartref{Saz05}{A572}{A2594}{A2597} \\
\href{http://simbad.u-strasbg.fr/simbad/sim-basic?Ident=IGR+J05305-6559}{IGR\,J05305$-$6559} &  82.462 & $-$66.041 &   13.7 & --    &     $49.6\pm0.6$ & B0.7Ve      & \bexrb & \tbsmartref{Kri10}{Gre13} \\
\href{http://simbad.u-strasbg.fr/simbad/sim-basic?Ident=IGR+J05414-6858}{IGR\,J05414$-$6858} &  85.377 & $-$69.008 &   4.42 &  19.9 &     $49.6\pm0.6$ & B0-1\,IIIe  & \bexrb & \tbsmartref{A2695}{A2696}{A3537}{Stu12} \\
\href{http://simbad.u-strasbg.fr/simbad/sim-basic?Ident=IGR+J06074%2B2205}{IGR\,J06074+2205} &  91.850 &    22.083 &  373.2 & --    &          4.5 & B0.5\,Ve        & \bexrb & \tbsmartref{A223}{A959}{RZ18b}   \\ 
\href{http://simbad.u-strasbg.fr/simbad/sim-basic?Ident=IGR+J08262-3736}{IGR\,J08262$-$3736} & 126.557 & $-$37.620 &        &       &          6.1 & OB\,V           & \bexrb & \tbsmartref{Bird10}{Mas10}{Boz12b} \\ 
\href{http://simbad.u-strasbg.fr/simbad/sim-basic?Ident=IGR+J08408-4503}{IGR\,J08408$-$4503} & 130.199 & $-$45.058 & --     &  9.54 &          2.7 & O8.5\,Ib-II(f)p & SFXT   & \tbsmartref{A813}{Gtz07}{Duc19b}{Hai20}     \\
\href{http://simbad.u-strasbg.fr/simbad/sim-basic?Ident=IGR+J10100-5655}{IGR\,J10100$-$5655} & 152.529 & $-$56.914 & --     & --    & --           & B0.5\,Ve or B0\,Ivpe & \bexrb & \tbsmartref{A684}{Col13b} \\ 
\href{http://simbad.u-strasbg.fr/simbad/sim-basic?Ident=IGR+J11215-5952}{IGR\,J11215$-$5952} & 170.445 & $-$59.863 &  186.8 &   165 & $6.5^{+1.4} _{-1.0}$ & B0.5\,Ia & SFXT  & \tbsmartref{A469}{Swa07}{Sid07}{Hai20}  \\
\href{http://simbad.u-strasbg.fr/simbad/sim-basic?Ident=IGR+J11305-6256}{IGR\,J11305$-$6256} & 172.779 & $-$62.945 & --     & --    &            3 & B0\,IIIne       & \bexrb & \tbsmartref{A278}{Mas06a}{Hai20} \\ 
\href{http://simbad.u-strasbg.fr/simbad/sim-basic?Ident=IGR+J11435-6109}{IGR\,J11435$-$6109} & 176.031 & $-$61.106 &  161.8 & 52.46 &      $8\pm2$ & B0.5\,Ve        & \bexrb & \tbsmartref{A350}{Col13b} \\ 
\href{http://simbad.u-strasbg.fr/simbad/sim-basic?Ident=1ES+1210-646}{1ES\,1210-646}    & 183.312 & $-$64.879 & --     &  6.71 &          2.8 & B5\,V           & \bexrb & \tbsmartref{Mas09}{Mas10}{A1861} \\
\href{http://simbad.u-strasbg.fr/simbad/sim-basic?Ident=IGR+J12341-6143}{IGR\,J12341$-$6143} & 188.467 & $-$61.796 & --     & --    & --           & --              & UXT    & \tbsmartref{Bird06}{Sgu20} \\
\href{http://simbad.u-strasbg.fr/simbad/sim-basic?Ident=IGR+J13020-6359}{IGR\,J13020$-$6359} & 195.495 & $-$63.969 &    643 & --    & 7 or 0.8--2.3 & B0-6\,Ve       & \bexrb & \tbsmartref{Bird06}{Mas06a}{Kri15}{For18} \\
\href{http://simbad.u-strasbg.fr/simbad/sim-basic?Ident=IGR+J13186-6257}{IGR\,J13186$-$6257} & 199.652 & $-$62.946 & --     & 19.99 &     0.7--5.4 & B0-6\,Ve        & \bexrb & \tbsmartref{Bird07}{Dai11}{For18} \\ 
\href{http://simbad.u-strasbg.fr/simbad/sim-basic?Ident=IGR+J14059-6116}{IGR\,J14059$-$6116} & 211.310 & $-$61.308 & --     & 13.71 &          7.7 & O6.5\,III       & \bexrb & \tbsmartref{Lan17}{Cor19} \\
\href{http://simbad.u-strasbg.fr/simbad/sim-basic?Ident=IGR+J14331-6112}{IGR\,J14331$-$6112} & 218.273 & $-$61.221 & --     & --    &     $\sim$10 & early B\,III or mid B\,V & \bexrb & \tbsmartref{Mas08}{Col13b} \\
\href{http://simbad.u-strasbg.fr/simbad/sim-basic?Ident=IGR+J14488-5942}{IGR\,J14488$-$5942} & 222.180 & $-$59.705 & --     & 49.63 &  --          & Be              & \bexrb & \tbsmartref{Bird10}{A2598} \\
\href{http://simbad.u-strasbg.fr/simbad/sim-basic?Ident=IGR+J16195-4945}{IGR\,J16195$-$4945} & 244.884 & $-$49.742 & --     & 3.95 &             5 & ON9.7\,Iab      & sgWF   & \tbsmartref{A229}{Tom06}{Col13b}{Cus16} \\ 
\href{http://simbad.u-strasbg.fr/simbad/sim-basic?Ident=IGR+J16207-5129}{IGR\,J16207$-$5129} & 245.193 & $-$51.502 & --     & 9.73  &     1.8--4.1 & B0\,I / B1\,Ia  & sgWF   & \tbsmartref{A229}{A783}{Mas06b}{Nes08} \\
\href{http://simbad.u-strasbg.fr/simbad/sim-basic?Ident=IGR+J16283-4838}{IGR\,J16283$-$4838} & 247.034 & $-$48.652 & --     & 287.6$\pm$1.7 & 18$\pm$4 & sgBe        & sgWF    & \tbsmartref{A456}{pel2011}{cus2013} \\
\href{http://simbad.u-strasbg.fr/simbad/sim-basic?Ident=IGR+J16318-4848}{IGR\,J16318$-$4848} & 247.951 & $-$48.817 & --     &    80 &  $3.6\pm2.6$ & B0-5\,sgBe        & sgWF   & \tbsmartref{IA8063}{Wal03}{Jain09} \\
\href{http://simbad.u-strasbg.fr/simbad/sim-basic?Ident=IGR+J16320-4751}{AX\,J1631.9$-$4752} & 248.007 & $-$47.875 &   1303 &  8.99 &          3.5 & O8\,I            & sgWF   & \tbsmartref{Sug01}{Rod03}{A649}{Lut05a}{Rod06} \\ 
\href{http://simbad.u-strasbg.fr/simbad/sim-basic?Ident=IGR+J16328-4726}{IGR\,J16328$-$4726} & 248.158 & $-$47.395 & --     & 10.08 &  $7.2\pm0.3$ & O8\,Iafpe         & SFXT   & \tbsmartref{A2075}{Fio10}{A2588}{Per15} \\
\href{http://simbad.u-strasbg.fr/simbad/sim-basic?Ident=IGR+J16374-5043}{IGR\,J16374$-$5043} & 249.306 & $-$50.725 & --     & --    &          3.1 & --             & SFXT   & \tbsmartref{A2809}{Sgu20} \\
\href{http://simbad.u-strasbg.fr/simbad/sim-basic?Ident=AX+J163904-4642}{AX\,J1639.0$-$4642} & 249.772 & $-$46.704 &    912 & 3.69(4.24) &    10.6 & B\,IV-V(OB)$^\dag$ & sgWF   & \tbsmartref{Sug01}{Bod06}{Tho06}{Cor10}{Bod12b} \\ 
\href{http://simbad.u-strasbg.fr/simbad/sim-basic?Ident=IGR+J16418-4532}{IGR\,J16418$-$4532} & 250.462 & $-$45.540 &   1246 &  3.74 &           13 & O8.5\,I        & SFXT  & \tbsmartref{A224}{A779}{Cha08}{Sid12}{Dra13} \\
\href{http://simbad.u-strasbg.fr/simbad/sim-basic?Ident=IGR+J16465-4507}{IGR\,J16465$-$4507} & 251.647 & $-$45.118 &    228 & 30.32 &  $9.5\pm5.7$ & B0.5I/O9.5Ia & SFXT & \tbsmartref{A329}{A429}{Rom08}{Clk10} \\
\href{http://simbad.u-strasbg.fr/simbad/sim-basic?Ident=IGR+J16479-4514}{IGR\,J16479$-$4514} & 252.027 & $-$45.202 & --     &  3.32 &  $\sim$4.9/$\sim$4.5  & O8.5Ib         & SFXT   & \tbsmartref{A176}{Cha08}{Rah08}{Col15} \\
\href{http://simbad.u-strasbg.fr/simbad/sim-basic?Ident=IGR+J16493-4348}{IGR\,J16493$-$4348} & 252.362 & $-$43.819 &  1093  &  6.78 &   16.0($>$6) & B0.5Ib    & sgWF   & \tbsmartref{A457}{Hill08}{A2599}{Nes10b}\\
\href{http://simbad.u-strasbg.fr/simbad/sim-basic?Ident=AX+J1700.2-4220}{AX\,J1700.2$-$4220} & 255.100 & $-$42.337 &   54.2 & 44.12 &           10 &  B2e            & \bexrb & \tbsmartref{Sug01}{A783}{A2564} \\
\href{http://simbad.u-strasbg.fr/simbad/sim-basic?Ident=IGR+J17200-3116}{IGR\,J17200$-$3116} & 260.027 & $-$31.288 &  328.2 & --    & --           & --             & \bexrb & \tbsmartref{A229}{A3205} \\
\href{http://simbad.u-strasbg.fr/simbad/sim-basic?Ident=IGR+J17252-3616}{EXO\,1722$-$363} & 261.297 & $-$36.283 & 413.7  &  9.72 & 5.3--8.7$^\ddag$ & O8.5I          & sgWF   & \tbsmartref{War88}{Taw89}{A229}{Tho07}{Msn09} \\ 
\href{http://simbad.u-strasbg.fr/simbad/sim-basic?Ident=IGR+J17354-3255}{IGR\,J17354$-$3255} & 263.839 & $-$32.937 & --     &  8.45 &          8.5 & O8.5\,Iab(f) or O9\,Iab   & SFXT   & \tbsmartref{A874}{A2596}{Sgu11}{Col13b} \\ 
\href{http://simbad.u-strasbg.fr/simbad/sim-basic?Ident=IGR+J17375-3022}{IGR\,J17375$-$3022} & 264.391 & $-$30.388 & --     & --    &           26$^\star$ & --             & SFXT   & \tbsmartref{A1781}{A1783}{Sgu20} \\
\href{http://simbad.u-strasbg.fr/simbad/sim-basic?Ident=IGR+J17391-3021}{XTE\,J1739$-$302} & 264.775 & $-$30.358 & --     & 12.87 or 51.47 &          2.7 & O8Iab(f)       & SFXT   &  \tbsmartref{I6757}{A181}{Rom09b}{Dra10} \\
\href{http://simbad.u-strasbg.fr/simbad/sim-basic?Ident=AX+J1749.1-2733}{AX\,J1749.1$-$2733} & 267.287 & $-$27.554 &    132 & 185.5 & $14.5\pm1.5$ & --             & \bexrb & \tbsmartref{Sak02}{GrSu07}{Kar08}{Zur08} \\
\href{http://simbad.u-strasbg.fr/simbad/sim-basic?Ident=IGR+J17503-2636}{IGR\,J17503$-$2636} & 267.575 & $-$26.605 & --     & --    & --           & --             & SFXT   & \tbsmartref{Fer19} \\
\href{http://simbad.u-strasbg.fr/simbad/sim-basic?Ident=IGR+J17544-2619}{IGR\,J17544$-$2619} & 268.560 & $-$26.317 & 11.58 or 71.49 & 4.93  &  3.6 & O9Ib           & SFXT   & \tbsmartref{A190}{Dra12}{Rom15g} \\
\hline
\multicolumn{9}{l}{$^{\dag}$~In doubt as \tbsmartref{Bod12b}\ found a Chandra position inconsistent with the proposed counterpart.}\\ 
\multicolumn{9}{l}{$^{\ddag}$~Various existing distance estimates falling withing the quoted range \tbsmartref{Tho07}.}\\
\multicolumn{9}{l}{$^{\star}$~Tentative determination}\\ 
\multicolumn{9}{r}{\textsl{continued on next page \ldots}} \\   
\end{tabular}
\renewcommand\arraystretch{1.0}
\vspace*{-5mm}
\end{table*}
\clearpage

\setcounter{table}{0}

\begin{table*}[t!]
\caption{\textsl{(continued from previous page)} HMXB discovered or identified with \intg.}  
\label{tab:obs2}

\footnotesize
\renewcommand\arraystretch{1.1}
\begin{tabular}{@{}l@{\enspace}r@{\enspace}rrrrlll}\hline\hline
\multicolumn{1}{l}{Source} & \multicolumn{1}{c}{R.A.} & \multicolumn{1}{c}{Decl.} & 
\multicolumn{1}{c}{\Ps}  & \multicolumn{1}{c}{\Pb } & \multicolumn{1}{c}{distance} & 
\multicolumn{1}{c}{Companion} & \multicolumn{1}{c}{System} & References \\  
\multicolumn{1}{l}{name}   & \multicolumn{1}{c}{deg}  & \multicolumn{1}{c}{deg} &
\multicolumn{1}{c}{s}      & \multicolumn{1}{c}{days} & \multicolumn{1}{c}{kpc} &
\multicolumn{1}{c}{type}   &\multicolumn{1}{c}{type}  & \\ \hline
\href{http://simbad.u-strasbg.fr/simbad/sim-basic?Ident=IGR+J18027-2016}{IGR\,J18027$-$2016} & 270.666 & $-$20.283 &   139.6 & 4.57(4.47) & 12.4$\pm$0.1 & B1\,Ib       & sgWF   & \tbsmartref{Rev04}{Jain09} \\
\href{http://simbad.u-strasbg.fr/simbad/sim-basic?Ident=IGR+J18179-1621}{IGR\,J18179$-$1621} & 274.466 & $-$16.359 &   11.82 & --         & --           & --           & \bexrb & \tbsmartref{A3947}{A3949}{Boz12a}{Li12b}\\
\href{http://simbad.u-strasbg.fr/simbad/sim-basic?Ident=SAX+J1818.6-1703}{SAX\,J1818.6$-$1703}& 274.662 & $-$17.052 & --      &         30 &  $2.1\pm0.1$ & B0.5\,Iab    & SFXT   & \tbsmartref{I6840}{GrSu05}{Sgu05}{Bird09}{Tor10} \\
\href{http://simbad.u-strasbg.fr/simbad/sim-basic?Ident=AX+J1820.5-1434}{AX\,J1820.5$-$1434} & 275.125 & $-$14.573 &   152.3 &    54 or 111 &          4.7 & O9.5-B0\,Ve  & UXT    & \tbsmartref{Kin98}{Kaur10}{Seg13b}{Cor17}  \\
\href{http://simbad.u-strasbg.fr/simbad/sim-basic?Ident=IGR+J18214-1318}{IGR\,J18214$-$1318} & 275.332 & $-$13.311 & --      & --         &      8$\pm$2 & OB           & sgWF   & \tbsmartref{Bird06}{But09} \\ 
\href{http://simbad.u-strasbg.fr/simbad/sim-basic?Ident=IGR+J18219-1347}{IGR\,J18219$-$1347} & 275.478 & $-$13.791 & --      &      72.44 &  --            & --           & \bexrb? & \tbsmartref{Kri10}{Kar12}{LaP13} \\ 
\href{http://simbad.u-strasbg.fr/simbad/sim-basic?Ident=IGR+J18406-0539}{IGR\,J18406$-$0539} & 280.196 &  $-$5.681 & --      & --         &          1.1 & B5\,V        & \bexrb & \tbsmartref{Mol04}{Mas06a} \\ 
\href{http://simbad.u-strasbg.fr/simbad/sim-basic?Ident=AX+J1841.0-0536}{AX\,J1841.0$-$0536} & 280.252 &  $-$5.596 & --      &  6.45$^{\star}$ &     6.9 & B0\,I        & SFXT   & \tbsmartref{Bam01}{A340}{Sgu09a}{GG14} \\ 
\href{http://simbad.u-strasbg.fr/simbad/sim-basic?Ident=AX+J1845.0-0433}{AX\,J1845.0$-$0433} & 281.259 &  $-$4.565 & --      &       5.72 &  6.4$\pm$0.7 & O9.5\,I      & SFXT   & \tbsmartref{Yam95}{Coe96}{Sgu07a}{Goo13}{Col13b} \\ 
\href{http://simbad.u-strasbg.fr/simbad/sim-basic?Ident=IGR+J18462-0223}{IGR\,J18462$-$0223} & 281.565 &  $-$2.392 &     997 & $\sim$2.13? & $\sim$11?   & --           & SFXT   & \tbsmartref{A1319}{Sgu13} \\
\href{http://simbad.u-strasbg.fr/simbad/sim-basic?Ident=IGR+J18483-0311}{IGR\,J18483$-$0311} & 282.075 &  $-$3.183 &   21.05 &      18.55 &         2.8  & B0.5-1Iab    & SFXT   & \tbsmartref{A157}{A940}{Sgu07a}{Jain09} \\
\href{http://simbad.u-strasbg.fr/simbad/sim-basic?Ident=AX+J1910.7%2B0917}{AX\,J1910.7+0917} & 287.682 &     9.275 &   36200$\pm$110 & --         &           16 & early B\,I   & sgWF   & \tbsmartref{Sug01}{RR13}{Isr16} \\ 
\href{http://simbad.u-strasbg.fr/simbad/sim-basic?Ident=IGR+J19140%2B0951}{IGR\,J19140+0951} & 288.526 &     9.885 &    5900 &      13.56 &          3.6 & B0.5Ia       & sgWF   & \tbsmartref{Tor10}{Sid16b} \\ 
\href{http://simbad.u-strasbg.fr/simbad/sim-basic?Ident=IGR+J19294%2B1816}{IGR\,J19294+1816} & 292.483 &    18.311 &   12.44 &     116.2  &    2.9 or 11 & B1\,Ve       & \bexrb & \tbsmartref{Boz11a}{RR18}{Mlc20} \\
\href{http://simbad.u-strasbg.fr/simbad/sim-basic?Ident=AX+J1949.8%2B2534}{AX\,J1949.8+2534} & 297.481 &    25.567 & --      & --         &       7--8.8 & early B\,Ia  & SFXT   & \tbsmartref{Sug01}{Sgu17}{Hare19} \\ 
\href{http://simbad.u-strasbg.fr/simbad/sim-basic?Ident=IGR+J20006%2B3210}{IGR\,J20006+3210} & 300.091 &    32.190 & $\sim$890 or 1056 & -- & $\sim$8    & early B\,V or mid B\,III & \bexrb & \tbsmartref{Mas08}{Mor09}{Pra13} \\
\href{http://simbad.u-strasbg.fr/simbad/sim-basic?Ident=IGR+J21343%2B4738}{IGR\,J21343+4738} & 323.625 &    47.614 & 320.35 & --          &          8.5 & B1--1.5\,III--V & \bexrb & \tbsmartref{Rei14a}{Rei14b} \\ 
\href{http://simbad.u-strasbg.fr/simbad/sim-basic?Ident=IGR+J22534%2B6243}{IGR\,J22534+6243} & 343.480 &    62.727 &   46.67& $>$100        &         4--5 & B0-1\,III-Ve & \bexrb &  \tbsmartref{Kri12}{A4240}{A4248}{Esp13}{Mas13} \\ 
\hline
\multicolumn{9}{l}{$^{\star}$~Tentative determination}\\ 
\end{tabular}
\renewcommand\arraystretch{1.0}
\vspace*{-3mm}
\end{table*}

\noindent\textbf{References used in Table~\ref{tab:obs}: }\\
\setcounter{tabref}{0}
\begin{small}
\noindent%
\tbrefcite{A281}{ATel281}; \tbrefcite{Reig05}{Reig:2005}; \tbrefcite{iZd07}{intZand:2007}; \tbrefcite{Hai20}{Hainich:2020}; 
\tbrefcite{Coe10}{coe2010}; \tbrefcite{Mas13}{Masetti:2013}; 
\tbrefcite{A5547}{ATel5547}; \tbrefcite{A5557}{ATel5557}; \tbrefcite{C15b}{coe2015b}; \tbrefcite{Ev06}{Evans:2006};
\tbrefcite{Kah99}{Kahabka:1999SMC}; \tbrefcite{A2079}{ATel2079}; \tbrefcite{A2088}{ATel2088}; \tbrefcite{Tow11a}{Townsend:2011a};   
\tbrefcite{A5806}{ATel5806}; \tbrefcite{Ev04}{Evans:2004_2dF};  
\tbrefcite{A275}{ATel275}; \tbrefcite{A3079}{ATel3079}; \tbrefcite{Tom11}{Tomsick:2011}; 
\tbrefcite{A1882}{ATel1882}; \tbrefcite{Seg13a}{Segreto:2013a};  
\tbrefcite{A672}{ATel672}; \tbrefcite{A681}{ATel681}; \tbrefcite{Kau08}{Kaur:2008} 
\tbrefcite{Saz05}{Sazonov:2005}; \tbrefcite{A572}{ATel572}; \tbrefcite{A2594}{ATel2594}; \tbrefcite{A2597}{ATel2597}; 
\tbrefcite{Kri10}{Krivonos:2010Survey}; \tbrefcite{Gre13}{Grebenev:2013LMC}; 
\tbrefcite{A2695}{ATel2695}; \tbrefcite{A2696}{ATel2696}; \tbrefcite{A3537}{ATel3537}; \tbrefcite{Stu12}{Sturm:2012}; 
\tbrefcite{A223}{ATel223}; \tbrefcite{A959}{ATel959}; \tbrefcite{RZ18b}{reig18b}; 
\tbrefcite{A813}{ATel813}; \tbrefcite{Gtz07}{Goetz2007}; \tbrefcite{Duc19b}{Ducci:2019IGRJ08408}; 
\tbrefcite{Bird10}{Bird:2010}; \tbrefcite{Mas10}{Masetti:2010}; \tbrefcite{Boz12b}{Bozzo12b}; 
\tbrefcite{A684}{ATel684}; \tbrefcite{Col13b}{Coleiro2013b}; 
\tbrefcite{A469}{ATel469}; \tbrefcite{Swa07}{Swank:2007}; \tbrefcite{Sid07}{2007A&A...476.1307S};  
\tbrefcite{A278}{ATel278}; \tbrefcite{Mas06a}{masetti06a}; 
\tbrefcite{A350}{ATel350}; 
\tbrefcite{Mas09}{Masetti:2009}; \tbrefcite{A1861}{ATel1861};  
\tbrefcite{Bird06}{Bird:2006}; \tbrefcite{Sgu20}{Sguera2020}; 
\tbrefcite{Kri15}{Krivonos:2015}; \tbrefcite{For18}{Fortin2018};  
\tbrefcite{Bird07}{Bird:2007}; \tbrefcite{Dai11}{Dai2011}; 
\tbrefcite{Lan17}{Landi:2017}; \tbrefcite{Cor19}{Corbet19_4fgl}; 
\tbrefcite{Mas08}{masetti08}; 
\tbrefcite{A2598}{ATel2598}; 
\tbrefcite{A810}{ATel810};  
\tbrefcite{A229}{ATel229}; \tbrefcite{Tom06}{tomsick06}; \tbrefcite{Cus16}{Cusumano:2016}; 
\tbrefcite{A783}{ATel783}; \tbrefcite{Mas06b}{masetti06b}; \tbrefcite{Nes08}{nespoli08}; 
\tbrefcite{A456}{ATel456}; \tbrefcite{pel2011}{pellizza11}; \tbrefcite{cus2013}{Cusumano:2013};   
\tbrefcite{IA8063}{IAUC8063}; \tbrefcite{Wal03}{Walter03}; \tbrefcite{Jain09}{Jain:2009RAA};  
\tbrefcite{Sug01}{sugizaki01}; \tbrefcite{Rod03}{Rodriguez03}; \tbrefcite{A649}{ATel649}; \tbrefcite{Lut05a}{Lutovinov2005a}; \tbrefcite{Rod06}{Rodriguez06}; 
\tbrefcite{A2075}{ATel2075}; \tbrefcite{Fio10}{Fiocchi:2010}; \tbrefcite{A2588}{ATel2588}; \tbrefcite{Per15}{Persi:2015}; 
\tbrefcite{IA8097}{IAUC8097}; 
\tbrefcite{A2809}{ATel2809}; 
\tbrefcite{Bod06}{Bodaghee06}; \tbrefcite{Tho06}{Thompson:2006}; \tbrefcite{Cor10}{ATel2570}; \tbrefcite{Bod12b}{Bodaghee12b}; 
\tbrefcite{A224}{ATel224}; \tbrefcite{A779}{ATel779};\tbrefcite{Cha08}{Chaty2008}; \tbrefcite{Sid12}{Sidoli2012}; \tbrefcite{Dra13}{Drave:2013}; 
\tbrefcite{A329}{ATel329}; \tbrefcite{A429}{ATel429}; \tbrefcite{Rom08}{Romano2008}; \tbrefcite{Clk10}{Clark2010}; 
\tbrefcite{A176}{ATel176}; \tbrefcite{Rah08}{Rahoui:2008}; \tbrefcite{Col15}{Coley:2015}; 
\tbrefcite{A457}{ATel457}; \tbrefcite{Hill08}{Hill:2008};  \tbrefcite{A2599}{ATel2599}; \tbrefcite{Nes10b}{Nespoli:2010b};   
\tbrefcite{A2564}{ATel2564}; 
\tbrefcite{A3205}{ATel3205}; 
\tbrefcite{War88}{Warwick:1988}; \tbrefcite{Taw89}{Tawara:1989}; \tbrefcite{Tho07}{Thompson:2007}; \tbrefcite{Msn09}{Mason:2009}; 
\tbrefcite{A874}{ATel874}; \tbrefcite{A2596}{ATel2596}; \tbrefcite{Sgu11}{Sguera2011}; 
\tbrefcite{A1781}{ATel1781}; \tbrefcite{A1783}{ATel1783}; 
\tbrefcite{I6757}{IAUC6757}; \tbrefcite{A181}{ATel181}; \tbrefcite{Rom09b}{Romano:2009b}; \tbrefcite{Dra10}{Drave2010}; 
\tbrefcite{Sak02}{Sakano:2002}; \tbrefcite{GrSu07}{2007AstL...33..149G}; \tbrefcite{Kar08}{Karasev:2008}; \tbrefcite{Zur08}{ZuritaHeras+Chaty:2008};  
\tbrefcite{Fer19}{Ferrigno:2019}; 
\tbrefcite{A190}{ATel190}; \tbrefcite{Clk09}{Clark2009}; \tbrefcite{Dra12}{Drave2012}; \tbrefcite{Rom15g}{Romano2015giant}; 
\tbrefcite{Rev04}{Revnivtsev:2004Survey}; 
\tbrefcite{A3947}{ATel3947}; \tbrefcite{A3949}{ATel3949}; \tbrefcite{Boz12a}{Bozzo12a}; \tbrefcite{Li12b}{Li:2012b};  
\tbrefcite{I6840}{IAUC6840}; \tbrefcite{GrSu05}{2005AstL...31..672G}; \tbrefcite{Sgu05}{Sguera2005}; \tbrefcite{Bird09}{Bird2009}; \tbrefcite{Tor10}{Torrejon:2010}; 
\tbrefcite{Kin98}{Kinugasa:1998}; \tbrefcite{Kaur10}{Kaur:2010}; \tbrefcite{Seg13b}{Segreto:2013b}; \tbrefcite{Cor17}{Corbet:2017};  
\tbrefcite{But09}{Butler:2009}; 
\tbrefcite{Kar12}{Karasev:2012}; \tbrefcite{LaP13}{LaParola:2013}; 
\tbrefcite{Mol04}{Molkov:2004AstL}; 
\tbrefcite{Bam01}{Bamba:2001}; \tbrefcite{A340}{ATel340}; \tbrefcite{Sgu09a}{Sguera2009a}; \tbrefcite{GG14}{Gonzalez-Galan:PhD}; 
\tbrefcite{Yam95}{Yamauchi:1995}; \tbrefcite{Coe96}{Coe:1996}; \tbrefcite{Sgu07a}{Sguera2007a}; \tbrefcite{Goo13}{Goossens2013}; 
\tbrefcite{A1319}{ATel1319}; \tbrefcite{GrSu10}{2010AstL...36..533G} \tbrefcite{Sgu13}{Sguera2013}; 
\tbrefcite{A157}{ATel157}; \tbrefcite{A940}{ATel940}; \tbrefcite{Sgu07b}{Sguera2007b};  
\tbrefcite{RR13}{rodes13}; \tbrefcite{Isr16}{Israel2016}; 
\tbrefcite{Sid16b}{sidoli16b}; 
\tbrefcite{Sgu17}{Sguera2017}; \tbrefcite{Hare19}{Hare2019};  
\tbrefcite{Boz11a}{Bozzo:2011a} ; \tbrefcite{RR18}{rodes18}; \tbrefcite{Mlc20}{Malacaria:2020}; 
\tbrefcite{Mor09}{Morris:2009}; \tbrefcite{Pra13}{Pradhan:2013}; 
\tbrefcite{Rei14a}{reig14a}; \tbrefcite{Rei14b}{reig14b}  
\tbrefcite{Kri12}{Krinovos:2012Survey}; \tbrefcite{A4240}{ATel4240}; \tbrefcite{A4248}{ATel4248}; \tbrefcite{Esp13}{Esposito:2013}; 
\end{small}
\twocolumn

\section*{References}
\bibliographystyle{elsarticle-harv} 
\bibliography{hmxb}

\end{document}